\documentclass[12pt,letterpaper]{article}
\usepackage{jheppub}

\usepackage{graphicx}
\usepackage{bbm}
\usepackage{amsmath}
\usepackage{amssymb}
\usepackage{mathtools}
\usepackage{amsthm}
\usepackage{enumitem}

\usepackage[labelformat=simple]{subcaption}

\usepackage[usenames, dvipsnames, table]{xcolor}

\newcommand{\ept}[1]{\exp\left(#1\right)}

\def\be{\begin{equation}}
\def\ee{\end{equation}}

\newtheoremstyle{named}{0.75\baselineskip}{0.75\baselineskip}{\itshape}{}{\bfseries}{.}{.5em}{#3}
\theoremstyle{named}
\newtheorem*{namedconjecture}{Conjecture}

\newcommand{\Iota}{\mathrm{I}}
\newcommand{\Mu}{\mathrm{M}}

\newcommand{\tmop}[1]{\ensuremath{\operatorname{#1}}}

\newcommand{\halfBPS}{$\frac{1}{2}$-BPS}

\definecolor{SatColor}{rgb}{1,1,0.5}

\begin{document}

\begin{titlepage}

{\flushright {ACFI-T24-05} \\ }

\setcounter{page}{1} \baselineskip=15.5pt \thispagestyle{empty}

\bigskip\

\vspace{2cm}
\begin{center}
{\LARGE \bfseries A Distance Conjecture for Branes}

 \end{center}
\vspace{0.5cm}

\begin{center}
{\fontsize{14}{30}\selectfont Muldrow Etheredge$^{a,b}$, Ben Heidenreich$^a$, and Tom Rudelius$^c$}
\end{center}

\begin{center}
\vspace{0.25 cm}
\textsl{$^a$Department of Physics, University of Massachusetts, Amherst, MA 01003, USA}\\
\textsl{$^b$Kavli Institute for Theoretical Physics, University of California, Santa Barbara, CA 93106, USA}\\
\textsl{$^c$Department of Mathematical Sciences, Durham University, Durham, DH1 3LE, UK}

\vspace{0.25cm}

\end{center}
\vspace{1cm}
\noindent

We use branes to generalize the Distance Conjecture. We conjecture that in any infinite-distance limit in the moduli space of a $d$-dimensional quantum gravity theory, among the set of particle towers and fundamental branes with at most $p_\text{max}\leq d-2$ spacetime dimensions, at least one has mass/tension decreasing exponentially $T\sim \exp(-\alpha\Delta)$ with the moduli space distance $\Delta$ at a rate of at least $\alpha\geq 1/\sqrt{d-p_\text{max}-1}$. Since $p_\text{max}$ can vary, this represents multiple conditions, where the Sharpened Distance Conjecture is the $p_\text{max}=1$ case. This conjecture is a necessary condition imposed on higher-dimensional theories in order for the Sharpened Distance Conjecture to hold in lower-dimensional theories. We test our conjecture in theories with maximal and half-maximal supersymmetry in diverse dimensions, finding that it is satisfied and often saturated. In some cases where it is saturated---most notably, heterotic string theory in 10 dimensions---we argue that novel, low-tension non-supersymmetric branes must exist. We also identify patterns relating the rates at which various brane tensions vary in infinite-distance limits and relate these tensions to the species scale.
\vspace{.9cm}

\bigskip
\noindent\today

\end{titlepage}

\setcounter{tocdepth}{2}

\hrule
\tableofcontents

\bigskip\medskip
\hrule
\bigskip\bigskip

\section{Introduction\label{s.introduction}}

In asymptotic limits of moduli spaces of quantum gravity theories (QGTs), the Distance Conjecture of Ooguri and Vafa \cite{Ooguri:2006in} predicts towers of exponentially light particles. Such towers have been well studied \cite{ Klaewer:2016kiy, Baume:2016psm, Blumenhagen:2017cxt, Grimm:2018cpv, Heidenreich:2018kpg, Corvilain:2018lgw, Blumenhagen:2018nts, Grimm:2018ohb, Lee:2018urn, Buratti:2018xjt, Font:2019cxq, Marchesano:2019ifh, Joshi:2019nzi, Palti:2019pca, Baume:2019sry, Erkinger:2019umg, Lee:2019wij, Lee:2019xtm, Gendler:2020dfp, Calderon-Infante:2020dhm, Klaewer:2020lfg, Lanza:2020qmt, vanBeest:2021lhn, Lanza:2021udy,  Etheredge:2022opl, Castellano:2023jjt, Calderon-Infante:2023ler, Rudelius:2023mjy, Etheredge:2023odp, Castellano:2023stg, Etheredge:2024tok}, and the Distance Conjecture has been refined and sharpened in multiple ways. For instance, the Sharpened Distance Conjecture (Sharpened DC) \cite{Etheredge:2022opl} places a precise bound on the exponential scaling at which towers become light. The Sharpened DC requires that, when traveling an asymptotic distance $\Delta$ towards an infinite distance limit in the moduli space of a $d$-dimensional theory, at least one tower of particles becomes exponentially light as least as quickly as:
\begin{align}
	m_n\sim e^{-\alpha \Delta},\qquad \alpha \geq \frac 1{\sqrt{d-2}}, \label{eqn:SDCbound}
\end{align}
where $\Delta$ is measured in $d$-dimensional Planck units.

The Sharpened DC bound~\eqref{eqn:SDCbound} is sharp, as it is saturated in the $g_s \to 0$ limit of a perturbative string theory when holding all moduli besides the dilaton fixed, i.e., in an ``emergent string limit''~\cite{Lee:2019xtm,Lee:2019wij}. In fact, there is strong evidence~\cite{Etheredge:2022opl} that the Sharpened DC can \emph{only} be saturated in such limits.

In order for the Sharpened Distance Conjecture to hold upon compactification, branes must exist in the higher dimensional theory. This is because, in the limits in moduli space in which cycles of the compactification manifold become small, typically the only objects that become light are either Kaluza-Klein (KK) monopoles or wrapped branes. In order for the Sharpened Distance Conjecture to hold, the branes of the higher-dimensional theory must satisfy some further constraints.

In this paper, we identify multiple constraints on the asymptotic moduli-dependence of branes, which we call the Brane Distance Conjecture (Brane DC). Our conjecture builds on earlier studies of the asymptotic moduli dependence of branes \cite{Palti:2017elp, Font:2019cxq, Lanza:2020qmt, Herraez:2020tih, Alvarez-Garcia:2021pxo, Etheredge:2023zjk} by a) placing precise, often-saturated bounds on the exponential scaling of brane tensions in asymptotic limits of moduli space, which b) generalize the Sharpened DC, and c) are closely linked to the Emergent String Conjecture \cite{Lee:2019wij}.

\begin{namedconjecture}[Brane Distance Conjecture (Brane DC)]
	In any infinite-distance limit in the moduli space of a $d$-dimensional quantum gravity theory, among the set of particle towers and fundamental branes\footnote{Here and in what follows, a ``brane'' will always mean an object with more than one spacetime dimension. For a discussion of the meaning of ``fundamental'', see, e.g.,~\cite{Reece:2018zvv}.} with at most $p_\text{max}\leq d-2$ spacetime dimensions, at least one has mass/tension $T$ decreasing exponentially at least as quickly as
	\begin{align}
		T&\sim \exp\left( -\alpha \Delta \right),  & \alpha &\geq \frac{1}{\sqrt{d-p_\text{max}-1}}, \label{e.BDCbound}
	\end{align}
	where $\Delta$ is the proper distance traveled along the infinite-distance geodesic.
\end{namedconjecture}

This is really a collection of conjectures, one for each $p_\text{max} \in \{1,2,\ldots,d-2\}$. The Sharpened DC \cite{Etheredge:2022opl}, requiring a particle tower with mass decreasing exponentially at a rate $\alpha \ge \frac{1}{\sqrt{d-2}}$, is equivalent to the $p_\text{max}=1$ Brane DC. As $p_\text{max}$ increases, the lower bound on the rate $\alpha \ge  \frac{1}{\sqrt{d-p_\text{max}-1}}$ becomes stronger, requiring the existence of fundamental branes except in directions where the Sharpened DC is satisfied with ample room to spare. Note that the bound depends only on the combination $d-p_\text{max}$, which is the minimum codimension of the branes that may satisfy the conjecture.

Just like the Sharpened DC, we will argue that the Brane DC can only be saturated in certain, specific circumstances. To describe these circumstances, let us first define the ``$\alpha$-vector'' for a $(p-1)$-brane to be
\begin{align}
	\vec \alpha_p \equiv - \vec{\nabla} \log T_p(\phi),
	\label{eqn:alphadef}
\end{align}
where $T_p(\phi)$ is the tension of the brane as a function of the moduli and the derivative is taken with respect to the moduli with the $d$-dimensional Planck scale held fixed. We will argue that
\begin{namedconjecture}[Brane DC Saturation Condition]
The Brane DC bound~\eqref{e.BDCbound} cannot be saturated for $1<p_{\rm max}<d-3$. Along an infinite-distance geodesic where the bound is saturated for $p_{\rm max} = 1$ or $p_{\rm max} = d-3$,\footnote{As we will see in examples, the $p_{\rm max} = d-2$ Brane DC can also be saturated, but the consequences of this are presently less clear.} a fundamental $p_{\rm max}$-brane becomes light with $\vec{\alpha}_{p_{\rm max}+1}$ asymptotically tangent to the geodesic with magnitude\footnote{Here, the length of an $\alpha$-vector is determined by contraction with the metric on moduli space, i.e., $|\vec{\alpha}|^2 = \alpha_i g^{ij} \alpha_j$.}
\begin{equation}
|\vec{\alpha}_{p_{\rm max}+1}| = \frac{2}{\sqrt{d-p_{\rm max} -1}} .
\end{equation}
\end{namedconjecture}
The required brane is a string (dimension-2 brane) for $p_{\rm max} = 1$ or a ``vortex'' (codimension-2 brane) for $p_{\rm max} = d-3$, in either case having one spacetime dimension more than the limit $p \le p_{\rm max}$ for the branes satisfying the Brane DC. The conditions on $\vec{\alpha}_{p_{\rm max}+1}$ imply that the brane becomes exponentially low tension in this limit at a rate \emph{twice} the Brane DC requirement for the $p \le p_{\rm max}$ branes, and that the geodesic points in the direction in which the brane tension decreases the quickest (i.e., the geodesic asymptotically approaches a $\log T_p(\phi)$ gradient flow~\cite{Etheredge:2023usk, Etheredge:2023zjk, Ruehle:2024}.) Note that the saturation condition for $p_{\rm max} = 1$ is simply a restatement of the conditions that apply to an emergent string limit.

In what follows, we present several lines of evidence in support of the Brane DC and its saturation condition. 
First and foremost, we argue that under mild assumptions, the Brane DC in $d$ dimensions follows from the Sharpened DC in $d-p_{\rm max}+1$ dimensions after compactification on a torus. As a result, the compelling evidence for the Sharpened DC in $d \geq 4$ dimensions given in \cite{Etheredge:2022opl} translates immediately to evidence for the Brane DC with $p_{\rm max} \leq d-3$.

Further evidence comes from examples in string theory. In maximal supergravity examples in dimension $d$, with $10 \geq d \geq 3$, we show that the Brane DC is satisfied for $p_\text{max}\in\{1,\dots,d-2\}$, and also provide a novel test of the Sharpened DC in 3d maximal supergravity. In theories with less supersymmetry, the Brane DC frequently requires non-BPS branes. In several such examples we identify the required non-BPS branes in each infinite distance limit, showing that they have exactly the moduli-dependence required by the Brane DC.

In some other cases, our conjectures require the existence of novel, non-BPS branes whose identities are as yet unknown but whose properties we can constrain. For instance, some apparent saturations of the $p_\text{max}=d-3$ Brane DC in heterotic string theories have no known corresponding low-tension vortex, suggesting either that a novel vortex exists, or that a higher-codimension brane exists with a tension that decreases rapidly enough to avoid saturating the $p_\text{max}=d-3$ Brane DC.

Coupled with this analysis, we also observe numerous repeating patterns in the structure of brane $\alpha$-vectors. Firstly, the $\alpha$-vector of a fundamental $(p-1)$-brane in $d$ dimensions generally has length
\begin{align}
	\vec \alpha_p^2=2-\frac{p(d-p-2)}{d-2}\,. \label{eqn:taxRule1}
\end{align}
Surprisingly, this formula applies to numerous, disparate types of fundamental branes regardless of their origins, and even applies to some non-BPS branes (in an appropriate duality frame.) We show that it holds when the brane fully wrapped on a torus is dual to a KK tower for the decompactification of a single dimension, suggesting a close connection to the Emergent String Conjecture \cite{Lee:2019wij}.

Moreover, the $\alpha$-vectors of certain ``nearby'' pairs of fundamental $(p-1)$ and $(q-1)$-branes satisfy:
\begin{align}
\vec \alpha_p\cdot \vec \alpha_q=1-\frac{q(d-p-2)}{d-2} ,\qquad \text{for $q \le p$}. \label{eqn:taxRule2}
\end{align}
We show that this holds when the KK towers dual to the two wrapped branes can be made light at the same time, decompactifying two dimensions in the dual description.

We also observe that brane $\alpha$-vectors often lie on a regular lattice and that $\vec{\alpha}_p/p$ typically lies on the boundary of the ``species polytope''~\cite{Calderon-Infante:2023ler, Etheredge:2024tok}. All of these observations suggest the need for more work on the behavior of low tension branes in infinite distance limits. For instance, although we lack a systematic understanding of these patterns, the equations~\eqref{eqn:taxRule1} and \eqref{eqn:taxRule2} greatly resemble and can be related to the taxonomy rules of \cite{Etheredge:2024tok} constraining the $\alpha$-vectors of particle towers. We defer further consideration of such ``brane taxonomy'' to future work~\cite{Etheredge:BraneTaxonomy}.

An outline of this paper is as follows. In Section \ref{s.dimred}, we discuss how the Brane DC is required by the Sharpened DC under dimensional reduction, given our assumptions. In Section \ref{s.examples}, we test the Brane DC in maximal and half-maximal supergravity examples in diverse dimensions. In Section \ref{s.observations}, we discuss how tension scales compare with the species scale, and we also discuss how further constraints on branes can be obtained from the Emergent String Conjecture. We conclude with a summary of results and discussions of future directions. In Appendices \ref{s.formulas}, \ref{s.cod2branes}, and \ref{s.taxonomy}, we summarize useful reduction formulas, study codimension-2 branes in maximal supergravity, and further discuss constraining the asymptotic decay of brane tensions.

\section{The Brane Distance Conjecture from dimensional reduction} \label{s.dimred}

In this section we argue that under mild assumptions the Brane DC with $p_{\rm max} \leq d-3$ follows from the Sharpened DC upon dimensional reduction. We also discuss the more-subtle $p_{\rm max} = d-2$ case as well as the conditions under which the Brane DC can be saturated.

\subsection{Dimensional reduction argument} \label{subsec:dimred}

Consider a $D$-dimensional Quantum Gravity Theory ($\mathrm{QGT}_D$), which can be compactified on an $n$-torus $T^n$ to produce a $(d=D-n)$-dimensional Quantum Gravity Theory ($\mathrm{QGT}_d$). The moduli of $\mathrm{QGT}_d$ are those of $\mathrm{QGT}_D$ plus the metric moduli of the torus and the axions arising from the $k$-form gauge fields of $\mathrm{QGT}_D$ reduced along $k$ cycles. Let us focus on a slice of this moduli space $\mathcal{M}_{d}$ where the axions are held fixed, so that
\begin{equation}
\mathcal{M}_{d}^{\text{slice}} = \mathcal{M}_{D} \times \mathcal{M}_{T^n} \,,
\end{equation}
where $\mathcal{M}_D$ is the moduli space of $\mathrm{QGT}_D$ and $\mathcal{M}_{T^n}$ is the metric moduli space of the torus. Note that geodesics within $\mathcal{M}_{d}^{\text{slice}}$ are geodesics within $\mathcal{M}_{d}$, hence the infinite-distance limits of $\mathcal{M}_{d}^{\text{slice}}$ are a subset of those of $\mathcal{M}_{d}$.

Now consider taking an infinite-distance limit within $\mathcal{M}_D$ while holding the size and shape of $T^n$ fixed. Assuming $d\ge 4$ for reasons to be discussed later, the following types of particle towers in $\mathrm{QGT}_d$ may become light in this limit:
\begin{enumerate}
	\item Particle towers of $\mathrm{QGT}_D$, \label{case:tower}
	\item Towers of winding modes of a $\mathrm{QGT}_D$ fundamental brane fully wrapped on $T^n$, \label{case:wind}
	\item Towers of oscillation modes of a $\mathrm{QGT}_d$ fundamental brane that arises from a $\mathrm{QGT}_D$ fundamental brane partially wrapped on $T^n$. \label{case:osc}
\end{enumerate}
Here a $(p-1)$-brane is ``fully'' wrapped if its $(p-1)$ spatial directions wrap a $(p-1)$ cycle on $T^n$, which is possible only if $p-1\leq n = D-d$. Partially wrapped branes give rise to branes rather than particles in $\mathrm{QGT}_d$, but these branes may have oscillation modes generating particle towers (case~\ref{case:osc} above), analogous to perturbative fundamental strings.\footnote{Unwrapped branes may also have towers of oscillation modes, but these are particle towers in $\mathrm{QGT}_D$ (case~\ref{case:tower}) so they do not need to be accounted for separately.}

In each of the above cases, the $\mathrm{QGT}_d$ particle tower becomes light in $d$-dimensional Planck units if and only if the $\mathrm{QGT}_D$ particle tower or fundamental brane from which it arises becomes light / low-tension in $D$-dimensional Planck units.\footnote{Note that $d$-dimensional Planck units differ from $D$-dimensional Planck units by a fixed factor since the torus is held at fixed size.} There are numerous other possible $\mathrm{QGT}_d$ particle towers besides those listed above, such as towers of KK modes, KK monopoles, wrapped branes carrying compact momentum, etc., but these other towers all have a contribution to their mass that depends on the size of $T^n$, and they can only become parametrically light when $T^n$ is either parametrically large or small (depending on the specific tower). Thus, we do not need to consider these towers in the limit in question.

We now apply the Sharpened DC to this infinite distance limit of $\mathrm{QGT}_d$. Assume for now that the Sharpened DC is satisfied by a tower of type~\ref{case:tower} or~\ref{case:wind} enumerated above. Since the mass of a fully wrapped brane is proportional to its tension (times a factor that depends on the fixed volume of the $T^n$ cycle that it wraps), the particle tower or fundamental brane of $\mathrm{QGT}_D$ must therefore have a mass or tension that decreases exponentially at a rate
\begin{equation}
\alpha \geq \frac{1}{\sqrt{d-2}} \,,
\end{equation}
as a function of the moduli of $\mathrm{QGT}_D$. There are two novelties in this equation. Firstly, although we are only varying the moduli of $\mathrm{QGT}_D$, the bound on $\alpha$ is stronger than the Sharpened DC applied directly to $\mathrm{QGT}_D$ ($\alpha \geq \frac{1}{\sqrt{D-2}}$) because we are considering the corresponding infinite distance limit of $\mathrm{QGT}_d$ arising from the toroidal compactification at fixed volume. Secondly, the bound applies to the set of particle towers and fundamental branes with $p \leq D-d+1$ spacetime dimensions, rather than just to particle towers, because fundamental branes fully wrapped on the torus can produce towers of winding modes.\footnote{Solitonic branes are excluded from consideration in the Brane DC as they do not produce infinite towers of winding modes.} Re-expressed in terms of $p_{\rm max} = D-d+1$ and $D$, we find the bound
\begin{equation}
\alpha \geq \frac{1}{\sqrt{D-p_{\rm max}-1}} \,,
\end{equation}
which is precisely the Brane DC applied to $\mathrm{QGT}_D$.

What about towers of type~\ref{case:osc}, i.e., the oscillation modes of fundamental branes that are partially wrapped on $T^n$? In all known examples where the leading light tower consists of such oscillation modes, the brane in question is a perturbative fundamental string. This observation has been formalized as part of the Emergent String Conjecture~\cite{Lee:2019wij}, which asserts that all infinite distance limits in moduli space are either decompactification limits or limits in which fundamental strings become tensionless, i.e., ``emergent string limits.'' Moreover, except in the superstring critical dimension $d=10$, in every emergent string limit there is a KK tower that becomes light at the same rate as the string oscillator tower~\cite{Klaewer:2020lfg}, where these extra dimensions hidden at the string scale restore the critical dimension $d=10$.

Thus, we do not need to consider type~\ref{case:osc} towers in the above argument because, apart from easily enumerated 10d theories, towers of string or brane oscillation modes are never required to satisfy the Sharpened DC. This is related to the fact that perturbative string oscillation modes saturate the Sharpened DC. Indeed, if we assume that type~\ref{case:osc} towers saturate the Sharpened DC, then they can only contribute non-trivially to satisfying the Sharpened DC if the moduli space is one dimensional (as 10d in Type IIA, Type I and Heterotic strings theories) or if such towers are dense in a finite neighborhood of some direction (as in 10d Type IIB string theory). However, the former is impossible in the situation at hand because we have at least one modulus in $\mathcal{M}_D$ (to be able to take an infinite distance limit) as well as the radion (overall volume modulus) of $T^n$. The latter is also impossible because the possible values of $\alpha_{\text{radion}}$ form a discrete, finite set (see Appendix~\ref{s.formulas}) which is incompatible with a dense set of Sharpened-DC-saturating towers. Thus, we again see that type~\ref{case:osc} towers cannot non-trivially contribute to satisfying the Sharpened DC in the infinite distance limit in question.

This dimensional reduction argument for the Brane DC is not without subtleties. In particular, as we take an infinite distance limit, the species scale will become exponentially lighter than the Planck scale. Since $T^n$ is held at fixed size in Planck units, this implies that our limit involves a small torus \emph{in units of the species scale}. This could lead to unforeseen consequences, which might invalidate our argument. However, we note that a small torus is much more benign than any other small geometry. For instance, the curvature is exactly zero and the entire effect of the torus can be represented by a period identification in space. Nonetheless, we do not pretend to understand the ways in which our argument might fail in general, viewing it instead as a very strong but not fully rigorous argument for the Brane DC that will be validated in many examples.\footnote{Our argument is somewhat similar to the dimensional reduction argument for the Tower WGC~\cite{Heidenreich:2015nta}, but likely stronger because the torus can be held at fixed, arbitrarily large size in Planck units.}

As a final remark, we note that the asymptotically tensionless branes required by this dimensional reduction argument do not need to be stable. As pointed out in the particle ($p=1$) case in \cite{Heidenreich:2015nta}, objects which are unstable in higher dimensions may be stable after reduction. Furthermore, the thorny issues involved when discussing unstable ``resonances'' tend to disappear in infinite-distance limits of field space, where decay widths tend to vanish asymptotically.
In Section \ref{s.sugrahalf} below, we will encounter examples in which unstable, non-BPS branes play a role in satisfying the Brane DC. 

\subsection{Exotic branes} \label{subsec:exotic}

By contrast, in the case of compactification to $d=3$---corresponding to the Brane DC with $p_{\text{max}} = D-2$, i.e., involving branes up to codimension-2---there are specific, identifiable subtleties that play an important role even in maximal SUGRA. These subtleties are the reason we have excluded the $d=3$ case above, even though as we will see they do not necessarily invalidate the $p_{\text{max}} = D-2$ Brane DC.

Firstly, note that there has yet been little research into the validity of the Distance Conjecture---let alone the Sharpened DC---in three dimensions (see however \cite{Perlmutter:2020buo, Baume:2020dqd, Baume:2023msm,Ooguri:2024ofs}). As a result, it is not entirely clear that the Sharpened DC can be applied after reduction to 3d. However, in this paper we provide some preliminary evidence for the Sharpened DC in 3d, suggesting that this issue is surmountable.

More important subtleties arise from the interesting, non-trivial behavior of low-codimension branes. For instance, codimension-2 branes carrying monodromies may force the moduli to trace out a non-trivial loop in moduli space upon circling the brane. This loop does not necessarily ``fit'' in a given duality frame, in which case we say that the brane is ``exotic'' from the perspective of that duality frame. Exotic branes have be extensively studied from various perspectives in \cite{Elitzur:1997zn, Blau:1997du, Hull:1997kb, Obers:1997kk, Obers:1998fb, Eyras:1999at, Lozano-Tellechea:2000mfy, Kleinschmidt:2011vu, Bergshoeff:2011se, Kikuchi:2012za, Hellerman:2002ax, deBoer:2012ma}.

For instance, the D7-brane of 10d type IIB string theory has monodromy $\tau \to \tau+1$. This brane is not exotic in the perturbative duality frame, $g_s \ll 1$, because the monodromy $C_0 \to C_0 + 1$ can be realized without leaving the frame. However, the S-dual ``NS7'' brane has monodromy $\tau \to \frac{\tau}{1-\tau}$. Starting from any point with $g_s \ll 1$, one must cross into the region $g_s \gg 1$ to reach the point $\frac{\tau}{1-\tau}$ that the NS7 monodromy relates this to. Equivalently, moving a fundamental string around the NS7-brane transforms it into an $(1,-1)$ string, which can never be perturbative in the same region as the original $(1,0)$ string is perturbative. Thus, the NS7 is exotic from the perspective of perturbative type IIB string theory. Note that whether a brane is exotic or not depends on the duality frame from which we view it, which explains why the NS7 seems ``ordinary'' from the S-dual perspective.

Returning to the argument given above, the danger is that a brane that is exotic in the large-volume duality frame that is easily accessed using the $D$-dimensional EFT becomes an ordinary brane in the duality frame that we reach by taking the infinite distance limit described previously. In particular, the tower of light particles that satisfy the Sharpened DC in this limit may be exotic branes from the large volume perspective.

For example, upon compactifying type IIA string theory on $S^1$, it is tempting to conclude that there is only one type of BPS 6-brane in the resulting nine-dimensional theory, namely an unwrapped D6-brane of the type IIA theory. However, this is not the case, as can be seen from the S-dual description in terms of M-theory on $T^2$. In this description, the unwrapped D6-brane is described by a particular elliptic fibration with the M-theory circle $S^1_{\text{M}} \subset T^2$ degenerating over the 6-brane. However, an equally good M-theory background involves the type IIA circle $S_{\text{IIA}}^1 \subset T^2$ degenerating. Similar to the discussion of the NS7-brane above, this 6-brane is exotic from the 10d type IIA perspective because the monodromy around it implies that the KK scale cannot be brought below the string scale simultaneously in a complete loop around the brane.\footnote{More generally, the nine-dimensional theory contains many other 6-branes, e.g., the $(p,q)$ 6-branes related to the unwrapped D6 by modular transformations. Other than the $(1,0)$ 6-brane arising from an unwrapped D6, these are all exotic from the 10d type IIA string theory perspective.}

This explains why these other 6-branes are missed by a na\"ive, large-volume analysis of type IIA string theory compactified on a circle. If similar exotic particle towers appear after toroidal compactification and these towers play an essential role in satisfying the Sharpened DC in the type of infinite distance limits considered above, then the dimensional reduction argument for the Brane DC given in Section \ref{subsec:dimred} could fail.

Fortunately, exotic branes---as we have defined them---do not occur at codimension three or higher. In particular, such exotic branes require non-trivial moduli gradients at infinity, whereas the gradient energy required for given moduli profile surrounding a $(p-1)$-brane grows as $R^{d-p-2}$ as the radius $R$ of the profile is increased. Thus, for a finite tension brane of codimension three or higher, the moduli approach a constant value at infinity, precluding these branes from being ``exotic'' in the sense described above.

Technically, the moduli profile around, e.g., a D7-brane also creates a total gradient energy in the region $r\le R$ that diverges logarithmically as $R \to \infty$. This is related to the fact that the backreaction of a D7 creates a singularity at finite distance from the brane. However, if the string coupling is $g_s \ll 1$ at a distance $r_0$ from the brane, the singularity lies at an \emph{exponentially} large radius $r_\infty \sim e^{\frac{2\pi}{g_s}} r_0$, e.g., for $g_s \simeq 0.1$, $r_\infty \sim 10^{27} r_0$. Thus, for all practical purposes the D7-brane is a finite tension object at weak string coupling.\footnote{More correctly, the D7-brane of type IIB string theory is ``log confined'' just like a charged particle in 3d Maxwell theory.} Similar considerations apply to other codimension-2 branes with moduli profiles at infinity.

Thus, while our dimensional reduction argument is unaffected by exotic branes for $p_{\rm max} \le D-3$, the case $p_{\rm max} \le D-2$ is more subtle, as exotic particles can and do appear upon toroidal compactification to $d=3$. Nonetheless, we will find evidence that that the $p_{\rm max} \le D-2$ Brane DC \emph{is} satisfied, at least in the best-understood examples, and that while exotic particle towers play an important role in satisfying the Sharpened DC in 3d, they do not remove the need for the $p_{\rm max} \le D-2$ Brane DC to be satisfied in higher dimensions. Nonetheless, these issues place the $p_{\rm max} \le D-2$ Brane DC on shakier ground than its higher-codimension relatives.

\subsection{Saturation conditions} \label{subsec:saturation}

When the Brane DC is saturated in a given infinite-distance limit for a given $p_{\rm max}$, the Sharpened DC is saturated in the corresponding infinite distance limit after compactification on $T^{p_{\rm max} - 1}$ in the manner described above. As discussed in Section \ref{s.introduction}, conjecturally the Sharpened DC can only be saturated in emergent string limits. Thus, for any direction in which it is saturated there is a fundamental string with $\vec{\alpha}$ vector pointing in that direction with $|\vec{\alpha}_{\text{str}}| = \frac{2}{\sqrt{d-2}}$.

Since the constraint on the direction $\hat{\alpha}_{\rm str}$ will prove critical below, we give another argument that this constraint is necessary. Suppose we assume only a leading stringy tower in some infinite distance limit where the Sharpened DC is satisfied. The species scale cannot be parametrically above the string scale, hence conjectural bounds on the species scale~\cite{vandeHeisteeg:2023ubh, vandeHeisteeg:2023uxj,Calderon-Infante:2023ler, Etheredge:2024tok} place the constraint $|\vec{\alpha}_{\rm str}|/2 \le \frac{1}{\sqrt{d-2}}$. However, for the stringy tower to satisfy the Sharpened DC we need $(\vec{\alpha}_{\rm str}/2) \cdot \hat{t} \ge \frac{1}{\sqrt{d-2}}$ where $\hat{t}$ is the direction of the infinite distance limit. The combination of these two inequalities forces $|\vec{\alpha}_{\rm str}|=\frac{2}{\sqrt{d-2}}$ and $\hat{\alpha}_{\rm str} = \hat{t}$, as stated above.

With this in mind, consider the $\mathrm{QGT}_d$ fundamental string whose existence and properties we have inferred from the saturation of the Brane DC. Assuming that this string is not an exotic brane, it must arise from a fundamental $(P-1)$-brane of $\mathrm{QGT}_D$ wrapping a $(P-2)$-cycle on the torus. In particular, since the infinite-distance limit we are considering lies within $\mathcal{M}_D$, the Sharpened DC saturation condition requires the string $\vec{\alpha}$ vector to be tangent to $\mathcal{M}_D$. This means that the string tension cannot depend on the size and shape of the torus. There are only two options that are independent of the shape moduli:
\begin{enumerate}
\item
Fundamental strings of $\mathrm{QGT}_D$, not wrapping the torus,
\item
Fundamental $p_{\rm max}$-branes of $\mathrm{QGT}_D$, wrapping the entire torus $T^{p_{\rm max} - 1}$,
\end{enumerate}
where the two options are equivalent when $p_{\rm max} = 1$. In addition, the radion component $\alpha_{\rm rad}$ of $\vec{\alpha}$ must vanish. Referring to \eqref{eqn:alpharad}, we find:
\begin{align}
\alpha_{\rm rad}^{(1)} &= 2 \sqrt{\frac{p_{\rm max} - 1}{(D-2)(d-2)}}\,, & \alpha_{\rm rad}^{(2)} &= -(D-3-p_{\rm max}) \sqrt{\frac{p_{\rm max} - 1}{(D-2)(d-2)}}\,,
\end{align}
in the two cases. Thus, non-exotic strings with suitable $\vec{\alpha}$ vectors can only exist when $p_{\rm max} = 1$ or $p_{\rm max} = D-3$. The strings arise from $p_{\rm max}$-branes of $\mathrm{QGT}_D$ with $\vec{\alpha}$ vectors asymptotically tangent to the infinite-distance geodesic, with length
\begin{equation}
|\vec{\alpha}_{p_{\rm max}+1}| = \frac{2}{\sqrt{D-p_{\rm max}-1}} \,.
\end{equation}
This is the Brane DC saturation condition stated in Section \ref{s.introduction}.

What if this string is an exotic brane in the large volume, $\mathrm{QGT}_D$ description? For $p_{\rm max} \le D-4$, the string is codimension three or higher, so it cannot be exotic. In particular, this validates the conclusion that the Brane DC cannot be saturated for $p_{\rm max} \in \{2, \ldots, D-4\}$. For $p_{\rm max} = D-3$, on the other hand, the string is a codimension-2 object in the compactified $\mathrm{QGT}_4$ theory. Even in very simple examples, exotic strings \emph{do} appear after compactification to 4d, so in principle there is a real danger that the required string is exotic in this case.

However, general reasoning suggests that the tension of a codimension-2 brane cannot be exponentially small in Planck units near a region in moduli space where it is exotic. This is because part of the tension of the brane originates from the moduli gradient energy necessitated by its monodromy. As explained in Section \ref{subsec:exotic}, exotic codimension-2 branes have monodromies that do not ``fit'' inside duality frames in which they are exotic. Thus as the moduli are adjusted closer to such a duality frame, the gradient energy around the brane larger and larger as the moduli are forced to traverse an increasingly long path in and out of the duality frame.

Now consider the low-tension string that results from an infinite distance limit within $\mathcal{M}_D$ that saturates the Sharpened DC in $\mathrm{QGT}_4$. The tension of this string can be exponentially reduced by traveling along the infinite-distance geodesic within $\mathcal{M}_D$ while holding the radion fixed. Moreover, since the string must have $\alpha_{\rm rad} = 0$ as argued above, increasing the size of the torus does not change its tension in 4d Planck units, meaning that we can approach the large-volume duality frame while keeping the tension of the string exponentially small. Per the above argument, this strongly suggests that the string cannot be exotic at large volume, hence it must arise from a $\mathrm{QGT}_D$ fundamental vortex wrapping the entire torus, as argued previously.

It is interesting to contrast this situation with what happens when we saturate the $p_{\rm max} = D-2$ Brane DC. Na\"ively, this should not be possible by the above arguments, since no $\mathrm{QGT}_D$ brane can produce a $\mathrm{QGT}_3$ string with $\alpha_{\rm rad} = 0$. However, we will see in examples that the $p_{\rm max} = D-2$ Brane DC \emph{can} be saturated. How is this possible?

The resolution is that, since the required string is now codimension-1 rather than codimension-2, it \emph{can} be an exotic brane. In particular, a codimension-1 exotic brane is simply a domain wall connecting two vacua that cannot both be brought inside the duality frame simultaneously. Unlike a codimension-2 brane with an exotic monodromy, moving such a brane towards a duality frame in which it is exotic does not necessarily incur a gradient energy penalty, hence nothing prevents a codimension-1 brane that is exotic at large volume from having $\alpha_{\rm rad} = 0$. Indeed, in a few cases we have checked in the landscape, we find that such exotic strings \emph{do} appear in $\mathrm{QGT}_3$ in directions in which the $p_{\rm max} = D-2$ Brane DC is saturated.

\section{Examples\label{s.examples}}

In this section, we show how the Brane DC is satisfied in various string theory examples, and saturated in many of them. We focus on type II and heterotic string theories and their toroidal compactifications, restricting our attention to flat slices of the moduli space obtained by setting the axions to various convenient values.

In type II toroidal compactifications with all axions set to zero, we verify that the Brane DC satisfied for all $d \in \{3 , \dots, 10\}$ and all $p_\text{max}\in\{1,\dots,d-2\}$, and is only saturable for $p_\text{max}\in\{1,d-3,d-2\}$ with suitable strings (vortices) invariably appearing in $p_\text{max}=1$ ($p_\text{max}=d-3$) saturation directions. In all but one case, the Brane DC is satisfied by \halfBPS\ branes, with the exception being the $p_\text{max}=1$ case in 10d, where the Brane DC sometimes requires non-BPS string oscillator modes. When $p_\text{max}=d-2$, the branes needed to satisfy the Brane DC are frequently ``exotic branes" \cite{Elitzur:1997zn, Blau:1997du, Hull:1997kb, Obers:1997kk, Obers:1998fb, Eyras:1999at, Lozano-Tellechea:2000mfy, Kleinschmidt:2011vu, Bergshoeff:2011se, Kikuchi:2012za, Hellerman:2002ax, deBoer:2012ma}, see Section~\ref{subsec:exotic}.

For heterotic compactifications, we consider various slices with different convenient values of the axions, verifying the $p_\text{max}\in\{1,\dots,d-3\}$ Brane DC in 10d and 9d and the $p_\text{max}=d-3$ Brane DC in 8d and 7d.\footnote{We defer an exhaustive treatment for other dimensions, slices, and values of $p_\text{max}$ to future work.} Unlike the type II case, this frequently requires non-BPS branes, each of which has a well-known origin in the duality frame where it is required. However, in some directions where the $p_\text{max}=d-3$ Brane DC is apparently saturated we are unable to identify the vortex required by the Brane DC saturation condition. This suggests that novel non-BPS branes exist, either the missing vortices or other branes that remove the saturation. Such branes, if they exist, may also help to satisfy the $p_\text{max}=d-2$ Brane DC, which appears to fail in a similar set of directions. This situation is reminiscent of the Cobordism Conjecture \cite{McNamara:2019rup}, which also requires the existence of certain mysterious branes in string theory. We speculate that the same novel branes may satisfy both conjectures.

\subsection{Maximal supergravities in diverse dimensions \label{s.sugramax}}

We begin by testing our conjectures in toroidal compactifications of type II string theory in diverse dimensions. For simplicity, we take all tori to be rectangular and set all axions to zero, thereby restricting to a flat $(11-d)$-dimensional slice of moduli space consisting of the radions for each circle and the dilaton (where the distinction between radion and dilaton depends on the duality frame).

We identify the \halfBPS\ particle towers / fundamental branes and their $\alpha$-vectors in such a compactification iteratively, starting with a known list of \halfBPS\ objects in either 11d M-theory or 10d Type IIB, then successively compactifying on a series of circles, where in terms of QGT$_{d+1}$, QGT$_{d}$ contains \halfBPS\ particle towers / fundamental branes of the following types:
\begin{enumerate}
\item \halfBPS\ particle towers / fundamental branes of QGT$_{d+1}$ either wrapping or not wrapping the circle,
\item Graviton KK modes and KK monopoles,
\item Codimension-2 exotic branes.
\end{enumerate}
This does not generate a complete list of the \halfBPS\ particle towers and fundamental branes, but the ones that are missed, such as the KK modes of a \halfBPS\ particle tower of QGT$_{d+1}$, typically have $\alpha$-vectors that depend on the axions. We find that these other branes do not play an important role in satisfying the Brane DC within the radion-dilaton slice that we are considering.

The radion/dilaton dependence of non-exotic branes listed above can be calculated using the formulas in Appendix \ref{s.formulas}, and these branes are sufficient to satisfy the Brane DC with $p_\text{max}\in\{1,\dots,d-3\}$ in dimensions $d\geq 4$. Verifying the $p_\text{max}=d-2$ Brane DC, as well as the saturation condition for the $p_\text{max}=d-3$ Brane DC and generally any of our conjectures in $d=3$ requires an understanding of exotic branes. We characterize these branes using U-duality in Appendix~\ref{s.cod2branes} (see also~\cite{Eyras:1999at, deBoer:2012ma}). In particular, we show that for every non-exotic \halfBPS\ codimension-2 brane of the type described above, there is an exotic codimension-2 brane with equal and opposite $\alpha$-vector. This $\vec{\alpha} \to - \vec{\alpha}$ reflection ``symmetry'' of the \halfBPS\ codimension-2 branes turns out to be sufficient to identify the exotic branes needed to satisfy our conjectures.

\subsubsection{10d}
We begin with type IIA and IIB string theory, and analyze the dilaton-dependence of the branes. In IIB string theory, we set the axion $C_0$ to zero.

\begin{figure}
\begin{subfigure}{\linewidth}
\begin{center}
\includegraphics[width = .8\linewidth]{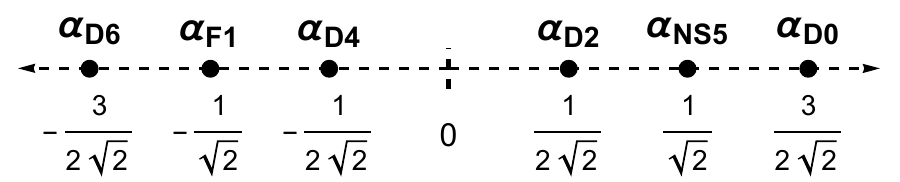}
\caption{IIA.}\label{f.10dII.IIA}
\end{center}
\end{subfigure}
\begin{subfigure}{\linewidth}
\begin{center}
\includegraphics[width = .8\linewidth]{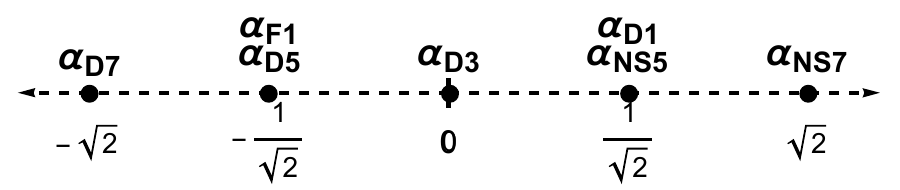}
\caption{IIB.}
\label{f.10dII.IIB}
\end{center}
\end{subfigure}
\caption{Dilaton-components of $\alpha$-vectors for various branes in 10d IIA and IIB string theories, satisfying the Brane DC in the dilaton direction.}
\label{f.10dII}
\end{figure}

In 10d IIA string theory, the D0, F1, D2, D4, and NS5-branes have tensions/masses that scale with the canonically normalized dilaton $\phi$ via
\begin{subequations}
\begin{align}
	T_\text{D$(p-1)$} &\sim \exp\biggl(\frac{p-4}{\sqrt{8}}\phi \biggr), &
	T_\text{F1} &\sim \exp\biggl(\frac{1}{\sqrt{2}}\phi\biggr), &
	T_\text{NS5} &\sim \exp\biggl(-\frac{1}{\sqrt{2}}\phi\biggr),
\end{align}
\end{subequations}
and the corresponding $\alpha$-vectors are
\begin{subequations}
\begin{align}
	\vec\alpha_\text{D$(p-1)$}= \frac{4-p}{\sqrt 8}\hat \phi,\qquad
	\vec\alpha_\text{F1}= -\frac{1}{\sqrt 2}\hat \phi,\qquad
	\vec\alpha_\text{NS5}= \frac{1}{\sqrt{2}}\hat \phi.
\end{align}
\end{subequations}
The convex hull is depicted in Figure \ref{f.10dII.IIA}.

The Brane DC for $p_\text{max}\in\{1,\dots d-2\}$ is satisfied in 10d IIA string theory. For $p_\text{max}=1$, this follows from the fact that the convex hull of fundamental string oscillators and D0-branes contains the origin-centered interval of radius $1/\sqrt{8}$. The fact that the string oscillators have $|\vec \alpha| = 1/\sqrt{8}$ implies that the $p_\text{max}=1$ Brane DC is saturated by the string oscillators in the tensionless string limit, as expected.
In addition, the convex hull of fundamental strings and D0-branes contains the origin-centered interval of radius $1/\sqrt{2}$, and so the Brane DC with $p_\text{max}\in\{2,\dots,d-4\}$ is satisfied with room to spare. The convex hull of D6-branes and D0-branes contains the ball of radius $3/\sqrt{8}>1$, so the Brane DC with $p_\text{max}\in \{ d-3 , d-2 \}$ is also satisfied but not saturated. 

In 10d IIB string theory, the tensions of the fundamental strings, D-branes, and NS5-branes are given by the formulas above, while the tensions of the $(1,0)$ and $(0,1)$ 7-branes are given by
\begin{align}
	T_8&\sim e^{\pm\frac{8-4}{\sqrt{d-2}}\phi}=e^{\pm \sqrt{2}\phi}.
\end{align}
The $\alpha$-vectors of all of these branes are depicted in Figure \ref{f.10dII.IIB}.

The Brane DC is satisfied for $p_\text{max}\in\{1,\dots,d-2\}$. The convex hull of fundamental and D-string oscillators contains the origin-centered interval of radius $1/\sqrt{8}$, and thus the $p_\text{max}=1$ Brane DC is satisfied and saturated by string oscillators. The convex hull of strings contains the origin-centered interval of radius $1/\sqrt{2}$, and thus the $p_\text{max}\in\{2,\dots,d-3\}$ Brane DC is satisfied, and it is saturated in the $p_\text{max}=d-3$ case. The 7-branes enclose the interval of radius $5/\sqrt{8}>1$, so the Brane DC with $p_\text{max}=d-2$ is satisfied but not saturated.

\subsubsection{9d }

\begin{figure}
\begin{subfigure}{.5\linewidth}
\centering
\includegraphics[width = \linewidth]{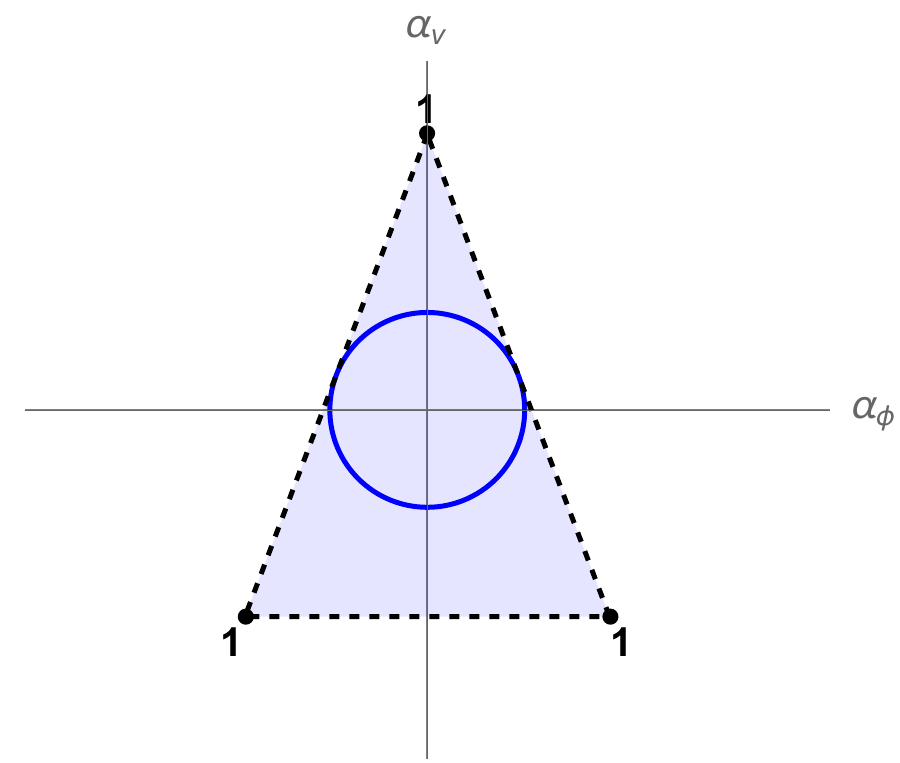}
\caption{Convex hull of particle towers.}\label{f.9dIIdvpmax1}
\end{subfigure}
\begin{subfigure}{.5\linewidth}
\centering
\includegraphics[width = \linewidth]{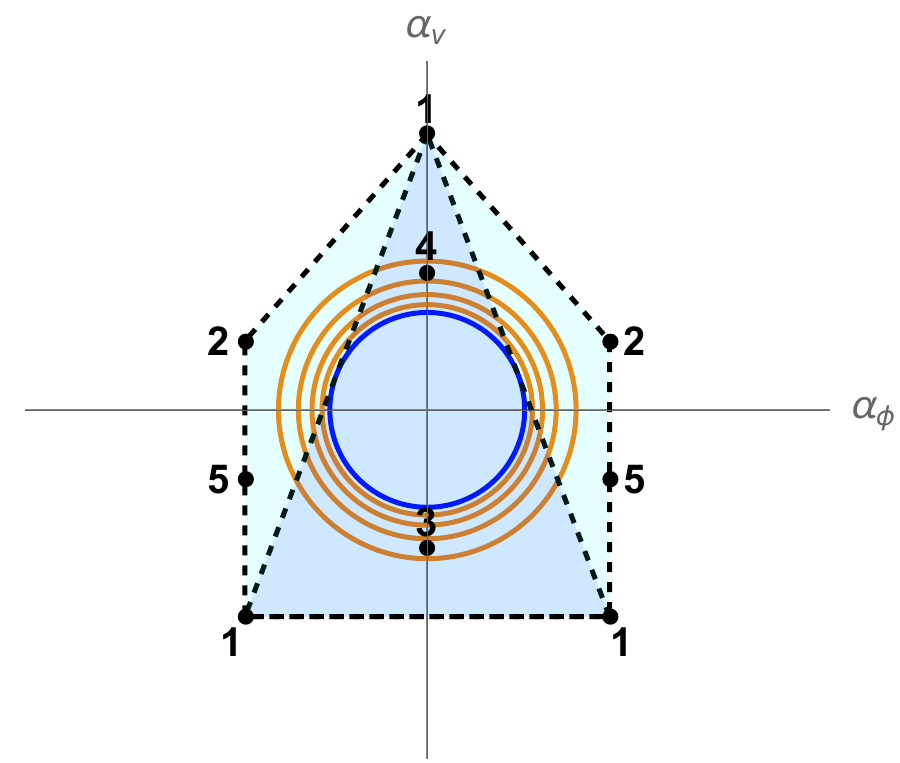}
\caption{Convex hull of $(p-1)$-branes up to spacetime dimension $p_\text{max}\in\{2,3,4,5\}$.}\label{f.9dIIdvpmax5}
\end{subfigure}
\begin{subfigure}{.5\linewidth}
\centering
\includegraphics[width = \linewidth]{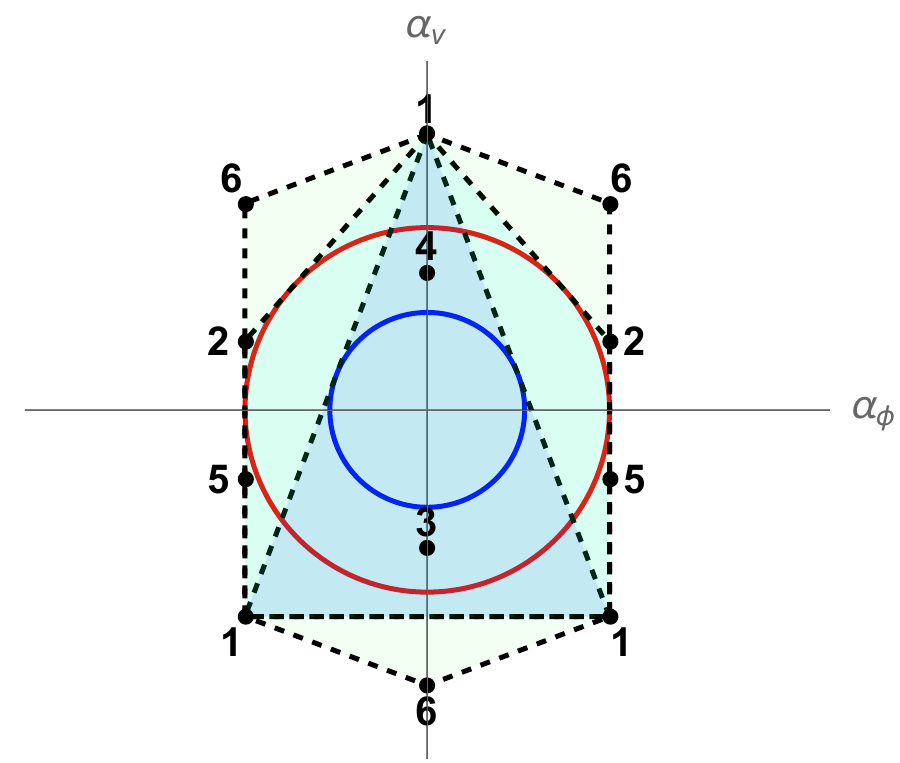}
\caption{Convex hull of $(p-1)$-branes up to spacetime dimension $p_\text{max}=6$.}\label{f.9dIIdvpmax6}
\end{subfigure}
\begin{subfigure}{.5\linewidth}
\centering
\includegraphics[width = \linewidth]{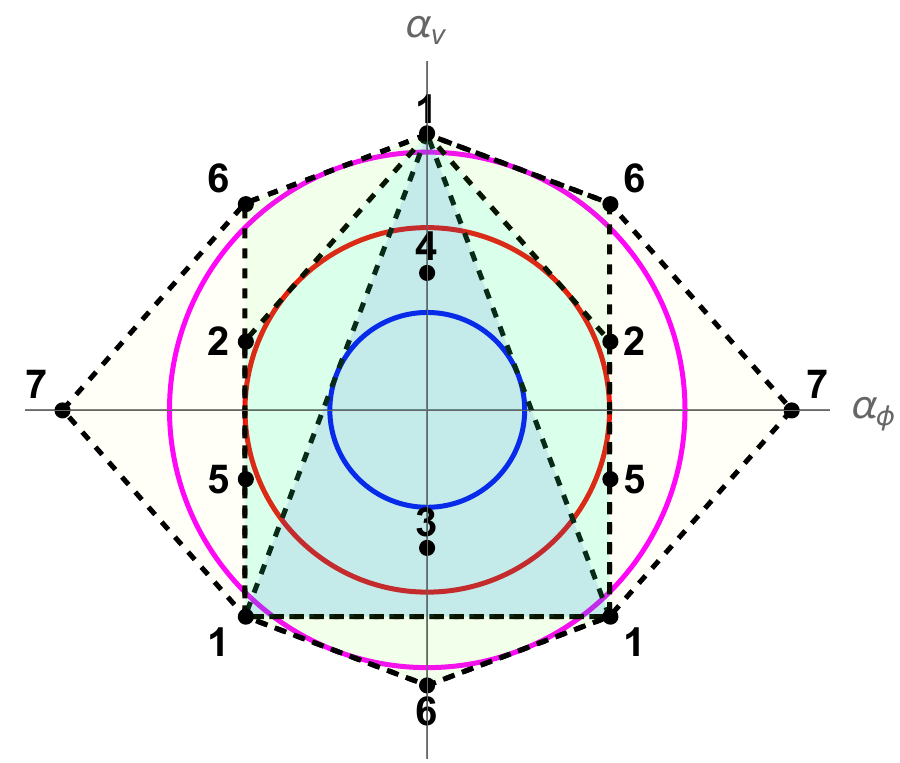}
\caption{Convex hull of $(p-1)$-branes up to spacetime dimension $p_\text{max}=7$.}
\label{f.9dIIdvpmax7}
\end{subfigure}

\caption{Radion-radion components of $\alpha$-vectors of \halfBPS\ $(p-1)$-branes from IIB string theory on a circle, with various brane dimensions $p$ and convex hulls depicted. The blue, orange, red, and magenta circles have radii $1/\sqrt 7$, $\{1/\sqrt{6},1/\sqrt{5}, 1/\sqrt4,1/\sqrt{3}\}$, $1/\sqrt{2}$, and $1$, and thus Figures \ref{f.9dIIdvpmax1}, \ref{f.9dIIdvpmax1}, and \ref{f.9dIIdvpmax7} depict saturation of the $p_\text{max}\in\{1,d-3,d-4\}$ Brane DC, and Figure \ref{f.9dIIdvpmax5} depicts satisfaction but not saturation of the $p_\text{max}\in\{2,\dots,d-4\}$ Brane DC. In Figure \ref{f.9dIIdvpmax7}, there are codimension-2 branes in the directions where the $p_\text{max}=6$ Brane DC is saturated.}
\label{f.9dII}
\end{figure}

We next investigate the dilaton-radion components of \halfBPS\ $\alpha$-vectors from IIB string theory on a circle with the axion set to zero.  The \halfBPS\ branes here are just the reductions of the branes from the 10d case, as well as the KK-modes and KK-monopoles. The $\alpha$-vectors of \halfBPS\ branes are depicted in Figure \ref{f.9dII}.

The Brane DC is satisfied, and it is saturated in only the $p_\text{max}\in\{1,d-3,d-2\}$ cases, as expected by our arguments above. The convex hull of particle towers contains the origin-centered ball of radius $1/\sqrt{7}$, and thus the $p_\text{max}=1$ Brane DC is satisfied and saturated by string oscillators. The convex hull of particle towers and strings contains the origin-centered ball of radius $1/\sqrt{2}$, and the $p_\text{max}\in\{2,\dots,5\}$ Brane DC is satisfied but not saturated (see Figure \ref{f.9dIIdvpmax5}). The $p_\text{max}=6=d-3$ Brane DC is satisfied and saturated, since the convex hull contains the ball of radius $1/\sqrt 2$ (see Figure \ref{f.9dIIdvpmax6}). The $p_\text{max}=7=d-2$ Brane DC is satisfied and saturated, since the convex hull contains the ball of radius $1$ (see Figure \ref{f.9dII}). Additionally, as expected from our argument in Section~\ref{subsec:saturation}, in each directions where the $p_\text{max}=d-3$ Brane DC is saturated there is a \halfBPS\ vortex (6-brane) with twice the $\alpha$-vector.

It is interesting to note that in Figure \ref{f.9dII}, the branes lie on a lattice. Also, as can be seen in Figure \ref{f.9dIIdvpmax6}, there is reflection symmetry $\vec{\alpha} \to -\vec{\alpha}$, $p \to d-p-2$, relating the tensions of $(p-1)$-branes and $(d-p-3)$-branes.\footnote{If one includes instantons, there is also a reflection symmetry between codimension-2 branes and instantons.} This ``brane Hodge duality'' appears because the related branes are electrically and magnetically charged under the same gauge field, where since the depicted branes are \halfBPS\ their tensions are fixed by their charges, which are inversely related by Dirac quantization. 

Genuine dualities fixing the radion-dilaton plane that we are studying manifest themselves as symmetries that map $(p-1)$-branes to $(p-1)$-branes. In particular, the reflection symmetry across the vertical axis corresponds to an S-duality of type IIB string theory. Note also that the $\alpha$-vectors of the \halfBPS\ codimension-2 branes form the $A_1$ root system, which are the ``roots'' of the duality group $SL(2,\mathbb{Z})$ (more precisely, of its supergravity analog $SL(2,\mathbb{R})$), where S-duality can be understood as the $\mathbb{Z}_2$ Weyl reflection of this root system.

All of these patterns persist in lower dimensions.

\subsubsection{8d}

\begin{figure}
\begin{subfigure}{.5\linewidth}
\centering
\includegraphics[width = \linewidth]{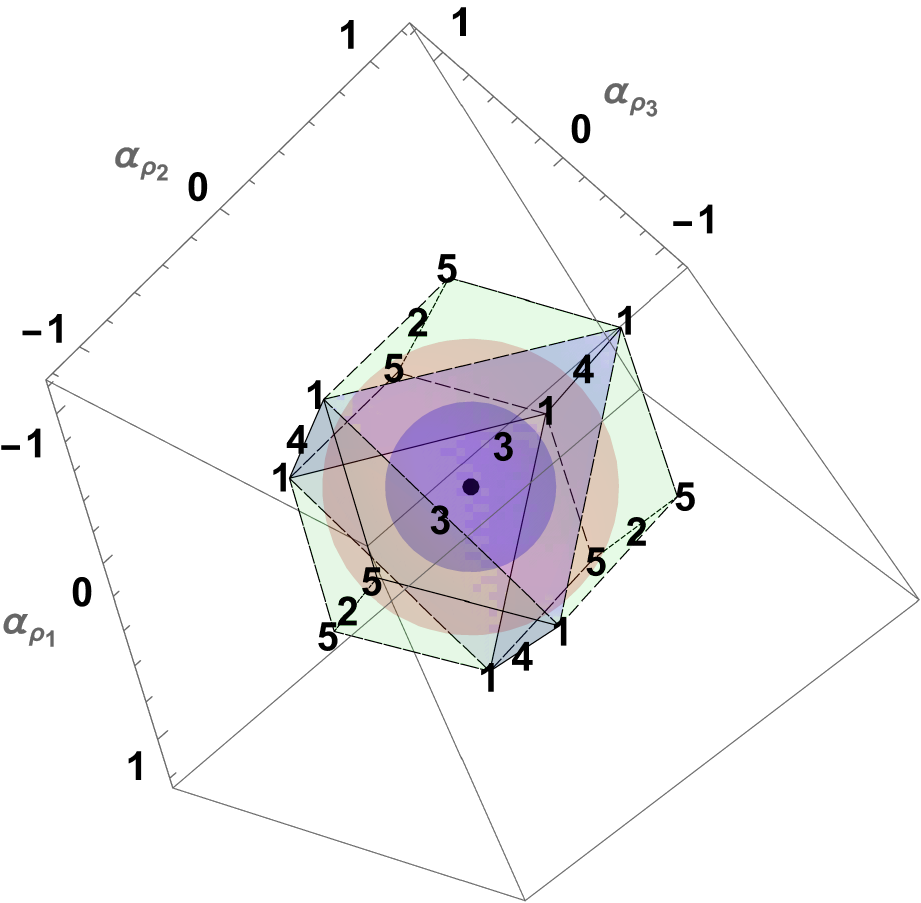}
\caption{The radion-radion-radion slice.}\label{f.8dIIrads}
\end{subfigure}
\begin{subfigure}{.5\linewidth}
\centering
\includegraphics[width = \linewidth]{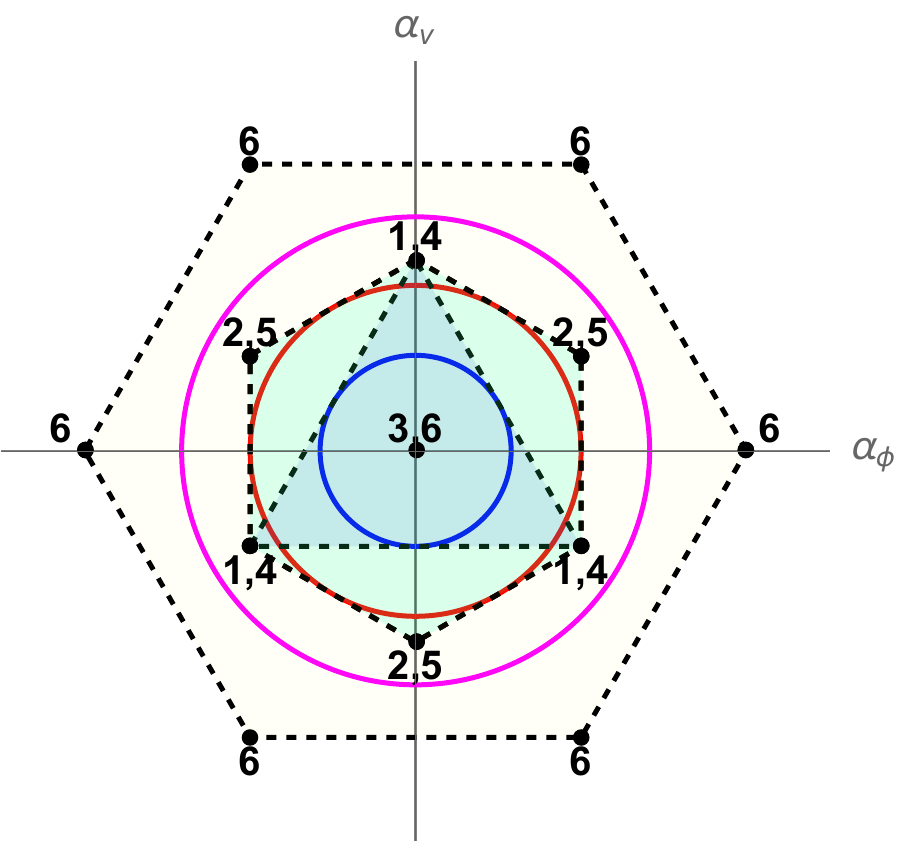}
\caption{Projection onto dilaton-volume plane.}
\label{f.8dIIdv}
\end{subfigure}
\caption{$\alpha$-vectors of \halfBPS\ branes in 8d maximal supergravity. Figure \ref{f.8dIIrads} depicts the radion-radion-radion slice from the perspective of M-theory on $T^3$, with codimension-2 branes omitted to reduce clutter. The triangular prism is the convex hull generated by particle towers and the hexagonal prism is the convex hull generated by particle towers and branes up to $p_\text{max}=d-3$. Figure \ref{f.8dIIdv} depicts the dilaton-overall volume slice from the perspective of Type IIB string theory on $T^2$. Balls with radii of $1/\sqrt{6}$, $1/\sqrt{2}$, and 1 (Figure~\ref{f.8dIIdv} only) are shown to illustrate that the Brane DC is satisfied for all $p_\text{max}$, with saturation for $p_\text{max}=1$, $p_\text{max}=d-3$, and $p_\text{max}=d-2$ (see the main text discussion). As can be seen in Figure \ref{f.8dIIdv}, there are vortex (5-brane) $\alpha$-vectors pointing in every direction where the $p_\text{max}=d-3$ Brane DC is saturated.}
\label{f.8dII}
\end{figure}

We now consider the 8d theory obtained by compactifying M-theory on $T^3$, or equivalently by compactifying Type IIB string theory on $T^2$. As can be seen in Figure \ref{f.8dII}, the Brane DC is satisfied in the radion-radion-dilaton slice for all $p_\text{max}\in\{1,\dots,d-2\}$, and it is saturated in only the $p_\text{max}\in\{1,d-3,d-2\}$ cases. The convex hull of particle towers contains the origin-centered ball of radius $\frac{1}{\sqrt{6}}$, and thus the Brane DC with $p_\text{max}=1$ is satisfied, and saturated by string oscillators. The convex hull of particle towers and strings contains the origin-centered ball of radius $\frac{1}{\sqrt{2}}$, and thus the $p_\text{max}\in\{2,\dots,d-3\}$ Brane DC is satisfied, with saturation in the $p_\text{max}=d-3$ case. 

In particular, the $p_\text{max} = d-3$ convex hull is the hexagonal prism depicted in Figure~\ref{f.8dIIrads}, which touches the origin-centered ball of radius $\frac{1}{\sqrt{2}}$ in the center of each of its faces. One can check that there is indeed a \halfBPS\ vortex (5-brane) with twice the $\alpha$-vector in each of these 8 directions (see, e.g., the end-on view in Figure~\ref{f.8dIIdv}). The full set of vortices includes exotic branes, but each is U-dual to either an unwrapped M5-brane in the M-theory description or an unwrapped D5-brane in the type IIB description, where the required U-dualities are generated by type IIB S-duality as well as double T-duality on $T^2$. Note that the U-duality orbit of the unwrapped D5 generates the 6 roots of the $A_2$ root system (as seen in Figure~\ref{f.8dIIdv}), whereas the U-duality orbit on the unwrapped M5 generates the 2 roots of the $A_1$ root system. Once again, these are the ``roots'' of the duality group $SL(3,\mathbb{Z}) \times SL(2,\mathbb{Z})$, and U-duality acts on the radion-radion-dilaton plane via the Weyl group $S_3 \times \mathbb{Z}_2$ of this root system.

Having mapped out these \halfBPS\ vortices via U-duality, one can check that the $p_\text{max}=d-2$ Brane DC is satisfied, and indeed it is saturated in some directions,\footnote{In particular, there are twelve directions of saturation, lying at the center of each $1$--$5$ edge of the hexagonal prism in Figure~\ref{f.8dIIrads}.} albeit not ones that are visible in the Type IIB dilaton-volume slice shown in Figure \ref{f.8dII}.

\subsubsection{$7\geq d\geq 3$} \label{subsec:maximaSUGRA73} \nopagebreak

\begin{figure}
\begin{subfigure}{.5\linewidth}
\centering
\includegraphics[width = \linewidth]{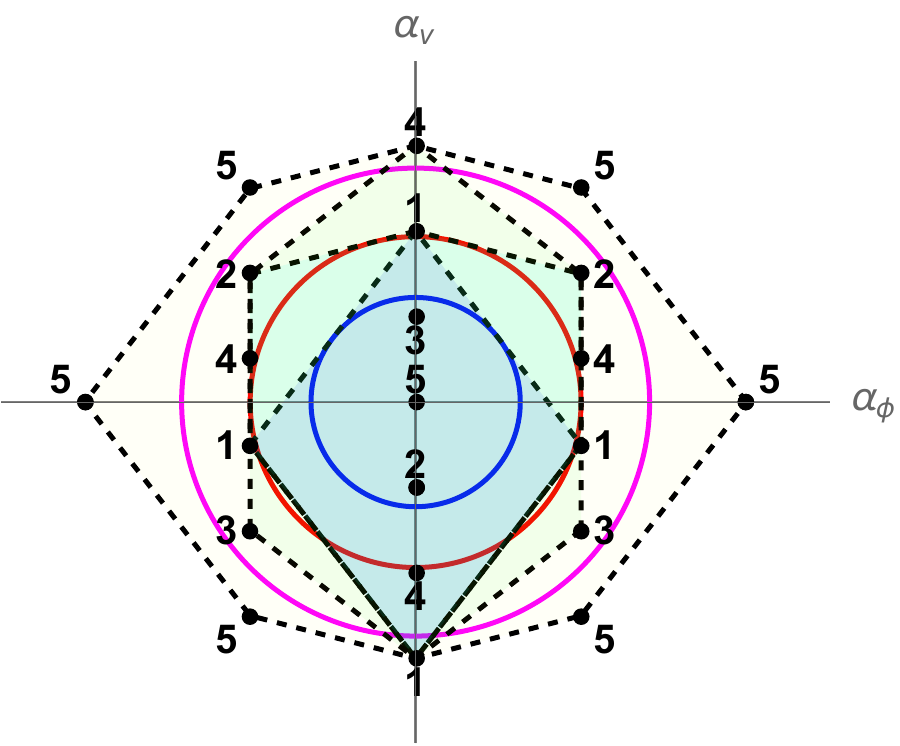}
\caption{7d maximal supergravity.}\label{f.7dIIdv}
\end{subfigure}
\begin{subfigure}{.5\linewidth}
\centering
\includegraphics[width = \linewidth]{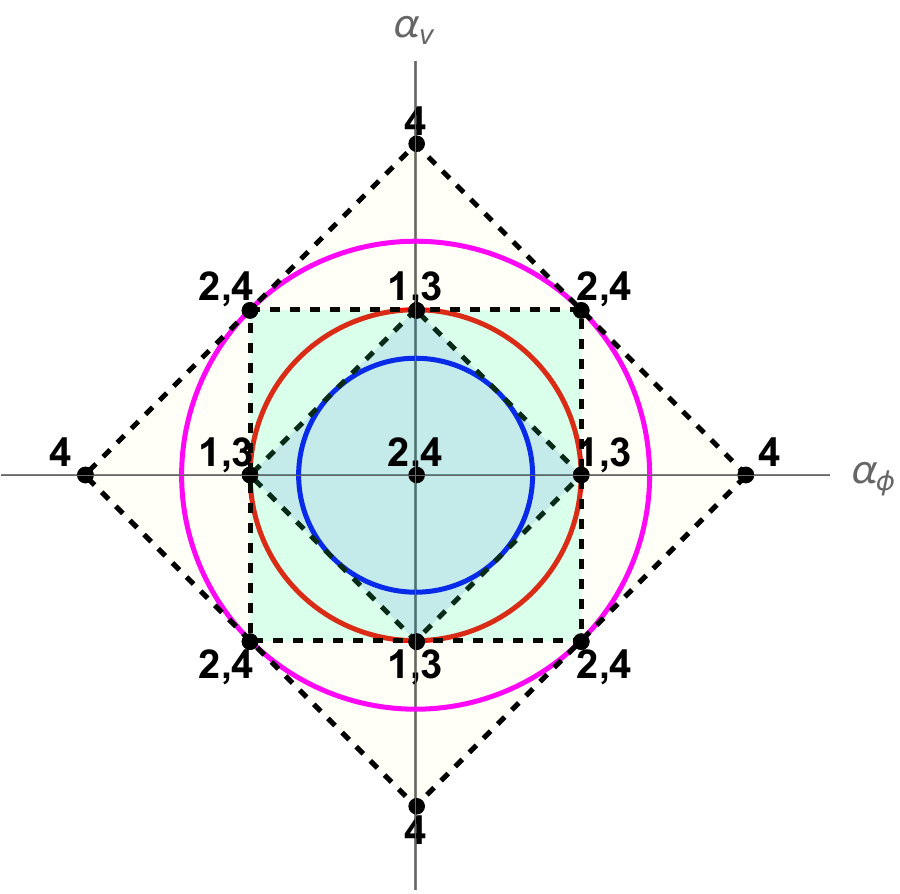}
\caption{6d maximal supergravity.}\label{f.6dIIdv}
\end{subfigure}
\begin{subfigure}{.5\linewidth}
\centering
\includegraphics[width = \linewidth]{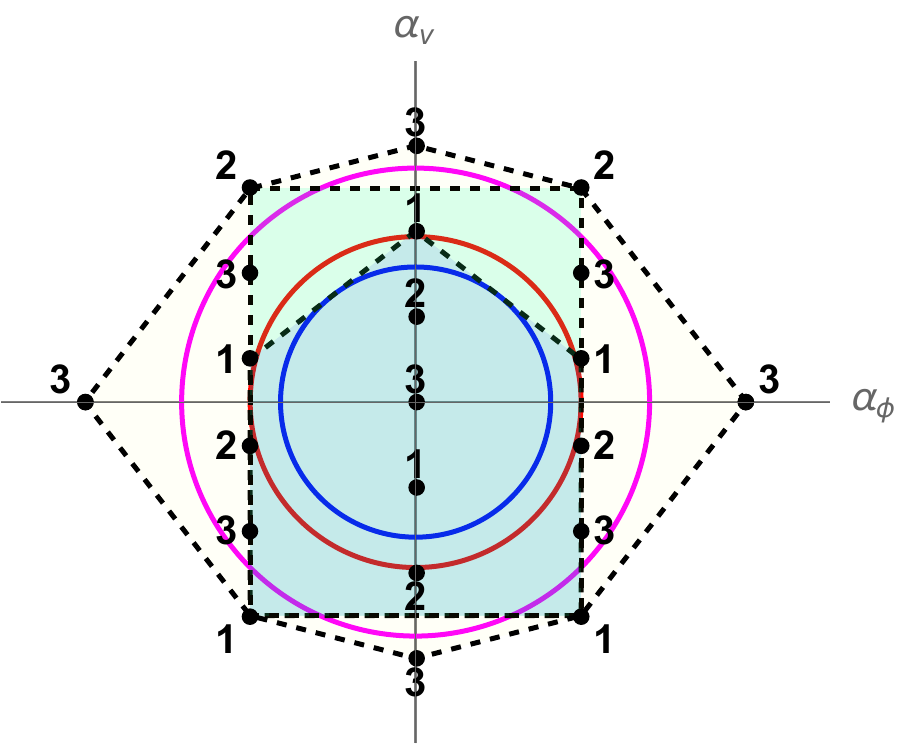}
\caption{5d maximal supergravity.}\label{f.5dIIdv}
\end{subfigure}
\begin{subfigure}{.5\linewidth}
\centering
\includegraphics[width = \linewidth]{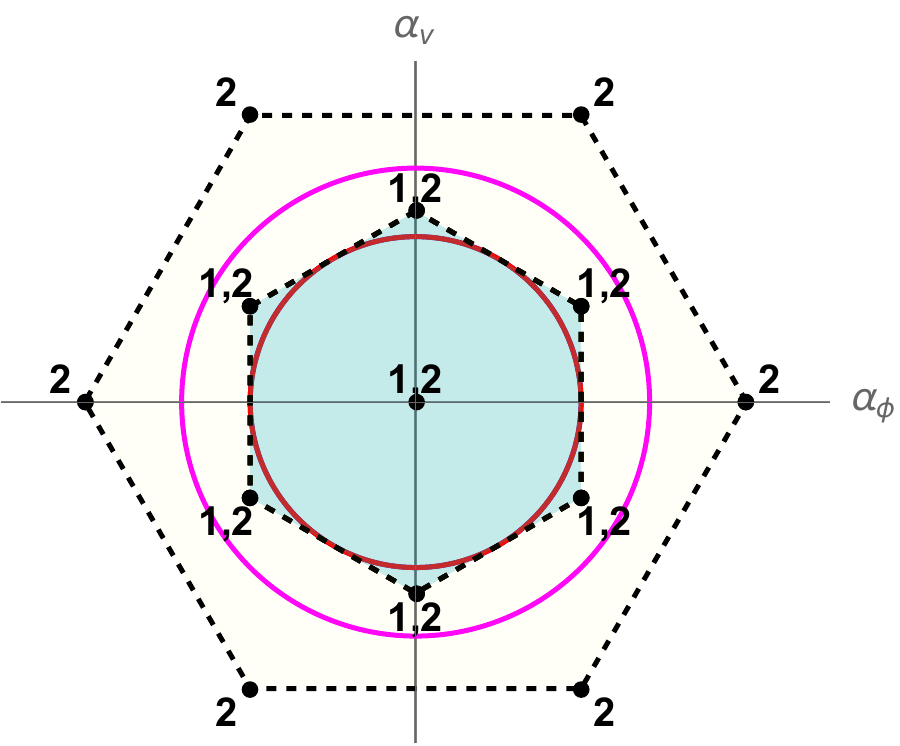}
\caption{4d maximal supergravity.}\label{f.4dIIdv}
\end{subfigure}
\caption{Dilaton-volume components of $\alpha$-vectors of \halfBPS\ branes from IIB on rectangular tori of increasing dimension, with axions set to zero. The blue, red, and magenta circles have radii $1/\sqrt{d-2}$, $1/\sqrt{2}$, and 1. For every saturation of the $p_\text{max}=1,d-3$ Brane DC, there is an $\alpha$-vector of a string or a codimension-2 brane.
}
\label{f.7654dII}
\end{figure}

\begin{figure}
\centering
\includegraphics[width = .5\linewidth]{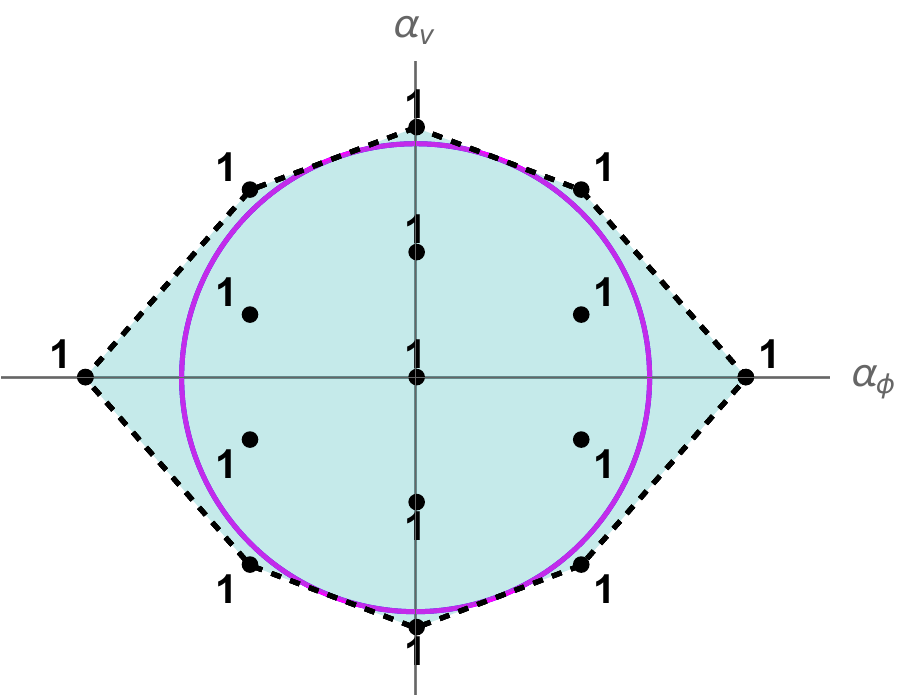}
\caption{Dilaton-volume components of $\alpha$-vectors of \halfBPS\ particles from IIB on a rectangular 7-torus, with axions set to zero. The blue circle has radius $1/\sqrt{d-2}=1$. The Brane DC and Sharpened DC is satisfied and saturated. }
\label{f.3dIIdv}
\end{figure}

One can continue this calculation to lower dimensions. For $d \le 7$, the dilaton/radion slice we have been considering is four-dimensional or higher, and becomes difficult to visualize. For illustration, we depict the dilaton-overall volume plane from the perspective of Type IIB string theory on a torus in Figures \ref{f.7654dII} and \ref{f.3dIIdv}. As can be seen in these figures, the Brane DC is always satisfied within this plane, with the $p_{\rm max} = 1$, $p_{\rm max} = d-3$ and sometimes also the $p_{\rm max} = d-2$ Brane DC being saturated, and with appropriate strings/vortices invariably appearing when the $p_{\rm max} = 1$ or $p_{\rm max} = d-3$ Brane DC is saturated.

However, as we saw already in 8d, some features are missed by restricting to this plane. For instance, the $p_{\rm max} = d-2$ Brane DC turns out to be saturated in \emph{every} dimension, even though this saturation is not visible in Figures~\ref{f.8dIIdv}, \ref{f.7dIIdv}, \ref{f.5dIIdv}, or \ref{f.4dIIdv}.

To check the Brane DC throughout the entire $(11-d)$-dimensional dilaton/radion slice, consider the polyhedron formed by the convex hull of the brane $\alpha$-vectors for each value of $d$ and $p_{\rm max}$. By finding the vertices of the dual polyhedron using SageMath~\cite{sagemath}, we determine the point of closest approach to the origin (the ``pericenter'') $\vec{\alpha}_{\rm pc}$ of each facet of this convex hull. The results are summarized in Table~\ref{tab:PCchecks}.

\begin{table}
\newcommand{\of}{\hspace{-0.3em}:}
$\begin{array}{|c|c|c|c|c|}
\hline
d & p_{\rm max} & \text{\#facets} & \text{facet $|\vec{\alpha}_{\rm pc}|^2$s} & |\vec{\alpha}_{\rm pc}|^2_{\rm min} \\ \hline
9 & 1 & 3 & 2\of \frac{1}{7}, 1\of \frac{9}{14} & \cellcolor{SatColor}\frac{1}{7} \\ 
 & \text{2--5} & 5 & 4\of \frac{1}{2}, 1\of \frac{9}{14} & \frac{1}{2} \\ 
 & 6 & 6 & 2\of \frac{1}{2}, 4\of 1 & \cellcolor{SatColor}\frac{1}{2} \\ 
 & 7 & 8 & 4\of 1, 4\of \frac{9}{8} & \cellcolor{SatColor}1 \\ \hline
8 & 1 & 5 & 3\of \frac{1}{6}, 2\of \frac{1}{2} & \cellcolor{SatColor}\frac{1}{6} \\
 & \text{2--4} & 14 & 14\of \frac{1}{2} & \frac{1}{2} \\
 & 5 & 8 & 8\of \frac{1}{2} & \cellcolor{SatColor}\frac{1}{2} \\
 & 6 & 24 & 12\of 1, 12\of \frac{9}{8} & \cellcolor{SatColor}1 \\ \hline
7 & 1 & 10 & 5\of \frac{1}{5}, 5\of \frac{9}{20} & \cellcolor{SatColor}\frac{1}{5} \\
& 2 & 45 & 5\of \frac{9}{20}, 40\of \frac{1}{2} & \frac{9}{20} \\
& 3 & 40 & 40\of \frac{1}{2} & \frac{1}{2} \\
& 4 & 20 & 20\of \frac{1}{2} & \cellcolor{SatColor}\frac{1}{2} \\
& 5 & 80 & 30\of 1, 40\of \frac{9}{8}, 10\of \frac{5}{4} & \cellcolor{SatColor}1 \\ \hline
\end{array}$ \hspace{0.2em}
\raisebox{23pt}{$\begin{array}{|c|c|c|c|c|}
\hline
d & p_{\rm max} & \text{\#facets} & \text{facet $|\vec{\alpha}_{\rm pc}|^2$s} & |\vec{\alpha}_{\rm pc}|^2_{\rm min} \\ \hline
6 & 1 & 26 & 10\of \frac{1}{4}, 16\of \frac{9}{20} & \cellcolor{SatColor}\frac{1}{4} \\
 & 2 & 136 & 16\of \frac{9}{20}, 120\of \frac{1}{2} & \frac{9}{20} \\
  & 3 & 40 & 40\of \frac{1}{2} & \cellcolor{SatColor}\frac{1}{2} \\
  & 4 & 250 & 90\of 1, 160\of \frac{9}{8} & \cellcolor{SatColor}1 \\ \hline
5 & 1 & 99 & 27\of \frac{1}{3}, 72\of \frac{1}{2} & \cellcolor{SatColor}\frac{1}{3} \\
  & 2 & 72 & 72\of \frac{1}{2} & \cellcolor{SatColor}\frac{1}{2} \\
  & 3 & 1134 & 270\of 1, 864\of \frac{9}{8} & \cellcolor{SatColor}1 \\ \hline
4 & 1 & 702 & 126\of \frac{1}{2}, 576\of \frac{9}{14} & \cellcolor{SatColor}\frac{1}{2} \\
  & 2 & 5274 & 753\of 1, 3950\of \frac{9}{8}, 571\of \frac{8}{7} & \cellcolor{SatColor}1 \\ \hline
3 & 1 & 19440 & 2160\of 1, 17280\of \frac{9}{8} & \cellcolor{SatColor}1 \\ \hline 
\end{array}$}
\caption{Properties of the convex hull of \halfBPS\ brane $\alpha$-vectors in maximal SUGRA for various $d$ and $p_{\rm max}$, including the total number of facets as well as the number with a given squared-distance to the facet pericenter $|\vec{\alpha}_{\rm pc}|^2$. The minimum value of this quantity $|\vec{\alpha}_{\rm pc}|^2_{\rm min}$ across all facets is the squared-radius of the largest enclosed origin-centered ball, which is constrained to be greater than $\frac{1}{d-p_{\rm max} - 1}$ by the Brane DC. All listed $|\vec{\alpha}_{\rm pc}|^2_{\rm min}$'s satisfy the Brane DC and the highlighted ones saturate it.} \label{tab:PCchecks}
\end{table}

From the table, one can see that the Brane DC is satisfied in each dimension $d \ge 3$ for all $1 \le p_{\rm max} \le d-2$, with saturation for $p_{\rm max} = 1$, $d-3$ and $d-2$. Note that this also confirms that the Sharpened DC holds in 3d maximal SUGRA, which to our knowledge is one of the first checks of this conjecture in 3d.

One can check that for $d\ge 4$ the directions in which the $p_{\rm max}=1$ and $p_{\rm max} = d-3$ Brane DCs are saturated are precisely the directions of \halfBPS\ string and vortex $\alpha$-vectors, respectively, satisfying the Brane DC saturation condition. Note that in any given duality frame, some of these vortices are exotic branes. Even without the power of U-duality, we could have predicted their existence and their precise $\alpha$-vectors by requiring the Brane DC saturation condition to be satisfied.

Checking the $p_{\rm max}=1$ saturation condition in 3d is more difficult\footnote{There is no $p_{\rm max} = d-3$ saturation condition in 3d since $d-3=0 < 1$.} because the required strings are codimension-1. While these strings should not be exotic branes in the infinite distance limit in which they are required to be appear due to the Sharpened DC being saturated, they are frequently exotic in the large volume duality frame that we naturally work with when constructing the theory via compactification. This is the same problem that we faced with vortices, but in the codimension-1 case we have not found a simple criterion (analogous to the $\vec{\alpha} \to -\vec{\alpha}$ reflection ``symmetry'' of the \halfBPS\ vortices) that would enable us to identify all \halfBPS\ codimension-1 exotic branes.

Fortunately, as shown in Appendix~\ref{subsec:3dsat}, U-duality is sufficiently powerful to ensure that a \halfBPS\ string exists for every direction in which the Sharpened DC is saturated in 3d. Note that this also implies that for each instance of $p_{\rm max} = d-2$ Brane DC saturation in higher dimensions, there is a corresponding \halfBPS\ string in 3d. However, as noted in Section~\ref{subsec:saturation}, these 3d strings are invariably exotic at large volume, hence we cannot easily lift to a purely higher-dimensional $p_{\rm max} = d-2$ Brane DC saturation condition.

Finally, note that the \halfBPS\ vortex $\alpha$-vectors always form a root system, respectively $A_1$, $A_2 A_1$, $A_4$, $D_5$, $E_6$, $E_7$ and $E_8$ in dimension $d=9, 8, 7, 6, 5, 4$ and $3$. For $d \le 8$, this matches the exceptional series $E_{11-d}$ that plays a well-known role in U-duality. It would be interesting to further explore the links between brane $\alpha$-vectors and duality groups.\footnote{For instance, in maximal SUGRA one can show that the \halfBPS\ brane $\alpha$-vectors are all integral weights of the Lie algebra $E_{11-d}$. This is doubtless closely related to U-duality, but it would be interesting to explore whether similar structures appear in less supersymmetric theories.}

\subsection{Half-maximal supergravity in diverse dimensions \label{s.sugrahalf}}

We next turn our attention to half-maximal supergravity examples. In the 10d and 9d theories, we investigate the Brane DC with $p_\text{max}\in\{1,\dots,d-3\}$, but in lower dimensions we focus on the $p_\text{max}=d-3$ Brane DC for simplicity. All our examples satisfy the Brane DC, with non-BPS branes now playing an essential role.

By contrast, we will see that the $p_\text{max}=d-3$ Brane DC saturation condition, as well as the $p_\text{max}=d-2$ Brane DC, fail to be satisfied by known branes in some directions. While an apparent violation of the $p_\text{max}=d-2$ Brane DC might be explained by the appearance of exotic branes upon compactification, see the discussion in Section~\ref{subsec:exotic}, we argued in Section~\ref{subsec:saturation} that exotic branes cannot be used to satisfy the $p_\text{max}=d-3$ Brane DC saturation condition. This suggests the existence of novel, non-BPS branes in these theories, whose properties we will constrain.

\subsubsection{10d}
\begin{figure}
\begin{subfigure}{\linewidth}
\begin{center}
\includegraphics[width = .8\linewidth]{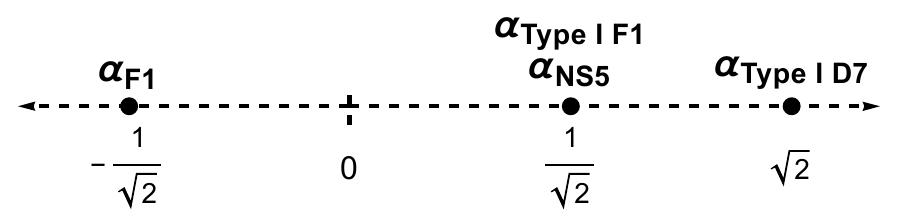}
\caption{10d heterotic SO$(32)$ $\alpha$-vectors.}\label{f.10hetalphas.SO32}
\end{center}
\end{subfigure}
\begin{subfigure}{\linewidth}
\begin{center}
\includegraphics[width = .8\linewidth]{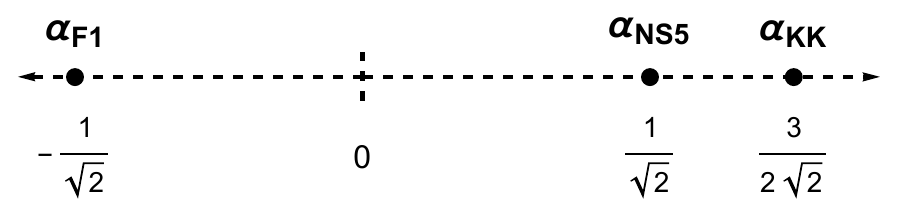}
\label{f.10dhet.E8E8}
\caption{10d heterotic $E_8\times E_8$ $\alpha$-vectors.}
\end{center}
\end{subfigure}
\caption{$\alpha$-vectors for various branes in 10d heterotic string theories. The non-BPS type I strings and KK-modes are only valid in the limits of strong coupling. The $p_\text{max}\leq d-3$ Brane DC is satisfied.}
\label{f.10dhet}
\end{figure}

In 10d heterotic string theory, fundamental strings and the NS5-branes are \halfBPS, with dilaton-dependent tensions
\begin{subequations}
\begin{align}
	T_\text{F1}\sim \exp\left(\frac{1}{\sqrt{2}}\phi\right),\qquad
	T_\text{NS5}\sim \exp\left(-\frac{1}{\sqrt{2}}\phi\right).
\end{align}
\end{subequations}
Thus, their $\alpha$-vectors are
\begin{align}
	\vec \alpha_\text{F1}=-\frac 1{\sqrt{2}}\hat \phi,\qquad \vec \alpha_\text{NS5}=\frac{1}{\sqrt{2}}\hat \phi.
\end{align}
These two branes satisfy and saturate the $p_\text{max}=d-3$ Brane DC, see Figure \ref{f.10dhet}. However, for $p_\text{max}\in\{1,d-2\}$, non-BPS branes are necessary. These branes\footnote{When discussing these non-BPS branes, we will assume that we are in a regime of moduli space where these branes are light enough to be stable.} differ depending on whether the theory is SO$(32)$ or $E_8\times E_8$ heterotic string theory.

In the strong coupling regime $\phi\gg 1$ of SO$(32)$ heterotic string theory, non-BPS type I strings become light, with $\alpha$-vector
\begin{align}
	\vec \alpha_\text{I1}=\frac{2}{\sqrt{d-2}}\hat \phi=\frac 1{\sqrt{2}}\hat \phi.
\end{align}

Meanwhile, in the strong-coupling limit of $E_8\times E_8$ heterotic string theory, the theory is described by M-theory on an interval. The interval KK modes are non-BPS, with $\alpha$-vector
\begin{align}
	\vec \alpha_\text{KK}=\sqrt{\frac{d-1}{d-2}}\hat \phi.
\end{align}

The $\alpha$-vectors for each of these 10d heterotic theories are depicted in Figure \ref{f.10dhet}.
In the $p_\text{max}=1$ case, the Brane DC is saturated by fundamental string oscillators in the $-\hat \phi$ direction, and in the $+\hat \phi$ direction it is either saturated by type I string oscillators (heterotic $SO(32)$) or satisfied but not saturated by KK-modes (heterotic $E_8 \times E_8$). In the $p_\text{max}\in\{2,\dots,7\}$ cases, the Brane DC is satisfied by fundamental strings in the $-\hat \phi$ direction, and by either type I strings, NS5-branes, or KK-modes in the $+\hat \phi$ direction, where for $p_\text{max} = d-3=7$ the Brane DC is saturated in some cases.

The codimension-2 case is more interesting, as the Brane DC with $p_\text{max}=d-2$ is not satisfied in the $-\hat{\phi}$ direction (the perturbative heterotic limit) by the branes discussed above. This leaves two possibilities; either (i) the Brane DC with $p_\text{max}=d-2$ is not satisfied in heterotic string theories, or else (ii) these theories contain additional non-supersymmetric branes. Per the discussion in Section \ref{s.dimred}, a violation of the $p_\text{max}=d-2$ Brane DC could indicate either a dimensional reduction subtlety (such as an exotic brane) or a violation of the Sharpened DC in three dimensions. Such a violation might not be too surprising, as gravity does not propagate in 3d, and thus one might expect some swampland criteria that hold in higher dimensions to fail in 3d.

However, we already found evidence in favor of the Sharpened DC in 3d in Section~\ref{subsec:maximaSUGRA73}. Moreover, even without considering three-dimensional QGTs, there is good reason to suspect that (ii) is correct, i.e., that there are additional low-tension branes in the perturbative heterotic limit that we have not considered. To see this, note that the $p_\text{max}=d-3$ Brane DC is saturated by heterotic strings in the weakly-coupled heterotic direction $-\hat{\phi}$. By the argument of Section~\ref{subsec:saturation}, this saturation requires the existence of a light vortex (7-brane) with $\vec \alpha_\text{8}=- \sqrt{2} \hat \phi$. Such a 7-brane would ensure that the $p_\text{max}=d-2$ Brane DC is satisfied. If this 7-brane does not exist then the $p_\text{max}=d-3$ Brane DC should not be saturated, implying the existence of some other $(p-1)$-brane, $p \leq 7$, with $\alpha$-vector $\alpha_p = - \alpha \hat \phi$, $\alpha > \frac{1}{\sqrt{2}}$. This hypothetical brane would also satisfy the $p_\text{max}=d-2$ Brane DC if it satisfies the slightly stronger constraint $\alpha \geq 1$.

Indeed, the former possibility is realized in the strong coupling limit of $\mathrm{SO}(32)$ heterotic string theory,\footnote{Likewise, the strong coupling limit of the heterotic $E_8 \times E_8$ theory realizes the latter possibility. A very naive analysis of the \halfBPS\ branes would suggest that the $p_\text{max}=d-3$ Brane DC is saturated by NS5 branes in this direction, but in fact the non-BPS KK modes on the M-theory interval remove this saturation.} where the $p_{\rm max} = 7$ Brane DC is saturated by NS5-branes and type I strings. By our argument above, we expect a codimension-2 brane with $\vec \alpha_8 = \sqrt{2} \hat \phi$, and in this case such a brane is known: the non-BPS D7-brane of type I string theory \cite{Witten:1998cd} exactly fits the bill. This brane ensures that after reduction to 4d, the $\phi \rightarrow \infty$ limit is consistent with the Emergent String Conjecture, and it ensures that the Brane DC with $p_{\rm max} = 8$ is satisfied in this limit in 10d.

Thus, we see that the Sharpened DC in 4d, via dimensional reduction of the $p_\text{max}=d-3$ Brane DC in 10d, implies the existence of new non-supersymmetric branes in $\mathrm{SO}(32)$ and $E_8 \times E_8$ heterotic string theory. We speculate that some of these branes may be the novel branes discussed in \cite{McNamara:2019rup, Kaidi:2023tqo} (see also \cite{Polchinski:2005bg}), which were recently predicted using the Cobordism Conjecture \cite{McNamara:2019rup}. It would be interesting to explore further a possible connection between this conjecture and the Brane DC.

\subsubsection{9d}

Consider heterotic string theory on a circle. Here, the moduli space is composed of the dilaton, radion, and sixteen axions. We now examine the dilaton-radion components of $\alpha$-vectors for SO(32) or $\mathrm{E_8\times E_8}$ string theory reduced on a circle with the axions set to zero.

\begin{figure}
\begin{center}
\includegraphics[width = .65\linewidth]{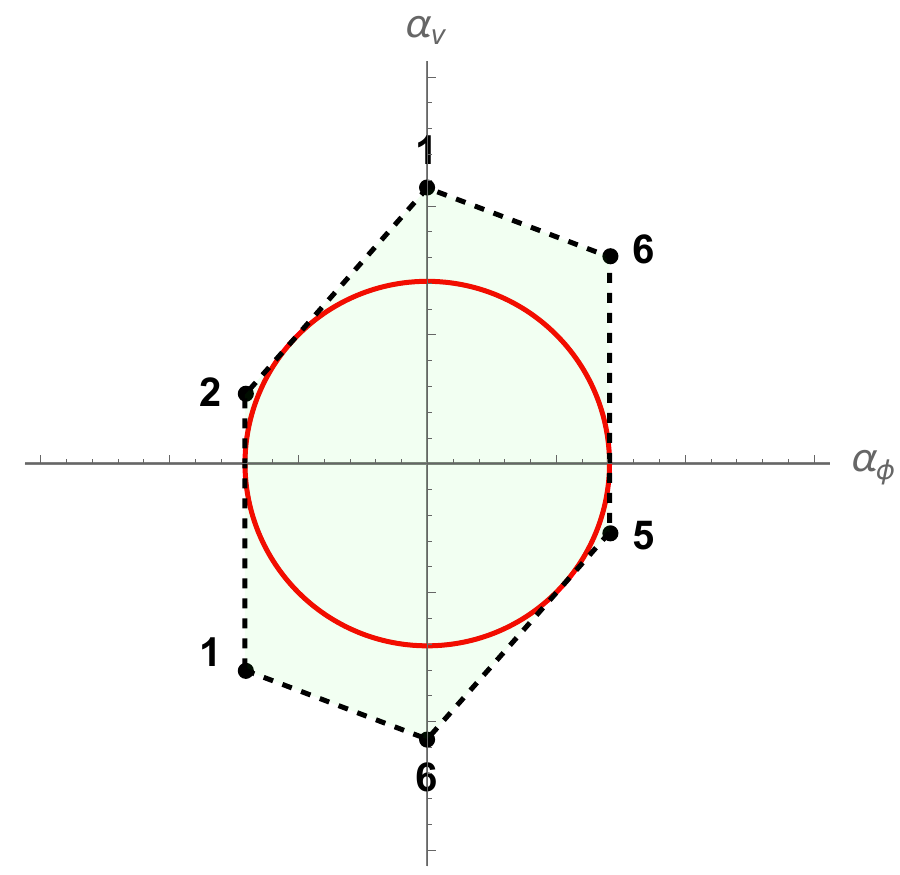}
\end{center}
\caption{{Dilaton-radion components of $\alpha$-vectors of \halfBPS\ $(p-1)$-branes from either SO$(32)$ or $\mathrm{E_8\times E_8}$ heterotic string theory on a circle. The circle has radius $1/\sqrt 2$, and thus the $p_\text{max}=d-3$ Brane DC is satisfied and saturated.}}
\label{f.9dhet}
\end{figure}

The $p_\text{max}=1$ Brane DC, or Sharpened DC, was tested in these theories in \cite{Etheredge:2023odp}. For the SO(32) or $\mathrm{E_8\times E_8}$ theories, the $p_\text{max}=d-3$ Brane DC is satisfied, and saturated, by \halfBPS\ states, as depicted in Figure \ref{f.9dhet}. The Brane DC for $p_\text{max} \neq 1, d-3$ requires non-BPS branes. Note also that, although there are four directions of saturation in Figure \ref{f.9dhet}, some of these saturations may be removed by non-BPS branes, as we will see in an example below.

Analyzing the spectrum of non-BPS branes is nontrivial. In general, such branes have $\alpha$-vectors that are non-constant functions of the moduli and ``slide" \cite{Etheredge:2023odp}. Additionally, these branes may become highly unstable in some duality frames (e.g., those in which their tensions are large), meaning that we need to study each duality frame in turn in order to identify the light, long-lived non-BPS branes that may play a role in satisfying the Brane DC.

A simple example where some of these complications are removed is the heterotic circle compactification with Wilson lines that break the gauge group to $\mathrm{SO(16)\times SO(16)}$. This compactification is well known as a simple way to connect the SO$(32)$ and $\mathrm{E_8\times E_8}$ heterotic theories via T-duality. Moreover, it turns out that in this case the brane $\alpha$-vectors do not slide, simplifying our analysis.

\begin{figure}
\begin{center}
\includegraphics[width = .65\linewidth]{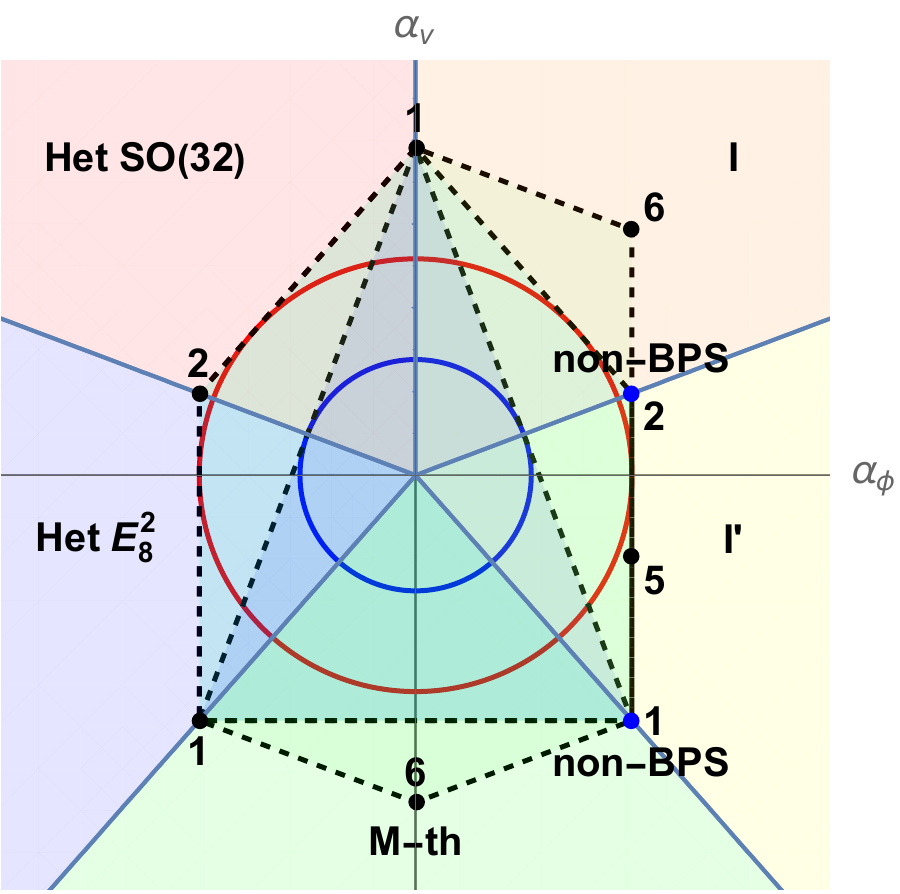}
\end{center}
\caption{Dilaton-radion components of $\alpha$-vectors of branes in 9d heterotic string theory with Wilson lines chosen to preserve the gauge group $\mathrm{SO(16)\times SO(16)}$. The black dots are \halfBPS\ brane $\alpha$-vectors, and the blue dots are non-BPS brane $\alpha$-vectors, where the latter are evaluated in the phase(s) in which they are light. The blue and red circles have radii $1/\sqrt{d-2}$ and $1/\sqrt2$, and thus the $p_\text{max}=d-3$ Brane DC is satisfied and saturated. The non-BPS Type I wrapped and unwrapped strings are crucial for the $p_\text{max}\in\{1,\dots,5\}$ Brane DC.}
\label{f.9dhetframes}
\end{figure}

As in 10d, two non-BPS branes turn out to be sufficient: the type I fundamental string and the M-theory interval KK modes, themselves dual to wrapped type I strings. The brane $\alpha$-vectors are plotted in Figure \ref{f.9dhetframes} together with the applicable duality frame in each asymptotic direction. The $p_\text{max}=1$ Brane DC (i.e., the Sharpened DC) is satisfied and saturated~\cite{Etheredge:2023odp}, where the non-BPS interval KK modes are crucial to satisfying the conjecture and the Type I strings are also required by the Sharpened DC saturation condition. For $p_\text{max}\in\{2,\dots,5\}$ both the interval KK modes and the Type I strings are required, satisfying but not saturating the Brane DC. As already discussed, for $p_\text{max}=6$ the \halfBPS\ branes are sufficient to satisfy the Brane DC, but the non-BPS interval KK modes remove one apparent direction of saturation from Figure~\ref{f.9dhet}.

There are three remaining directions in Figure \ref{f.9dhetframes} where the $p_\text{max}=d-3$ Brane DC is saturated. In one of these directions, $+\hat{\phi}$, a suitable vortex (6-brane) satisfying the Brane DC saturation condition is known to exist. This can be described as either the type I non-BPS 7-brane wrapping the circle, or an ordinary Type IIA D6 brane not wrapping the interval in the T-dual type I$^\prime$ description. However, two apparent directions of saturation remain unaccounted for. Thus, we predict that in asymptotic limits in these three directions of moduli space, there either exist novel non-BPS branes of spacetime dimension $p\leq d-3$ that spoil this saturation, or else novel codimension-2 branes that become light. In either case, we see that our arguments above require the existence of additional non-BPS branes.

\subsubsection{8d}

\begin{figure}
\begin{subfigure}{.5\linewidth}
\centering
\includegraphics[width = \linewidth]{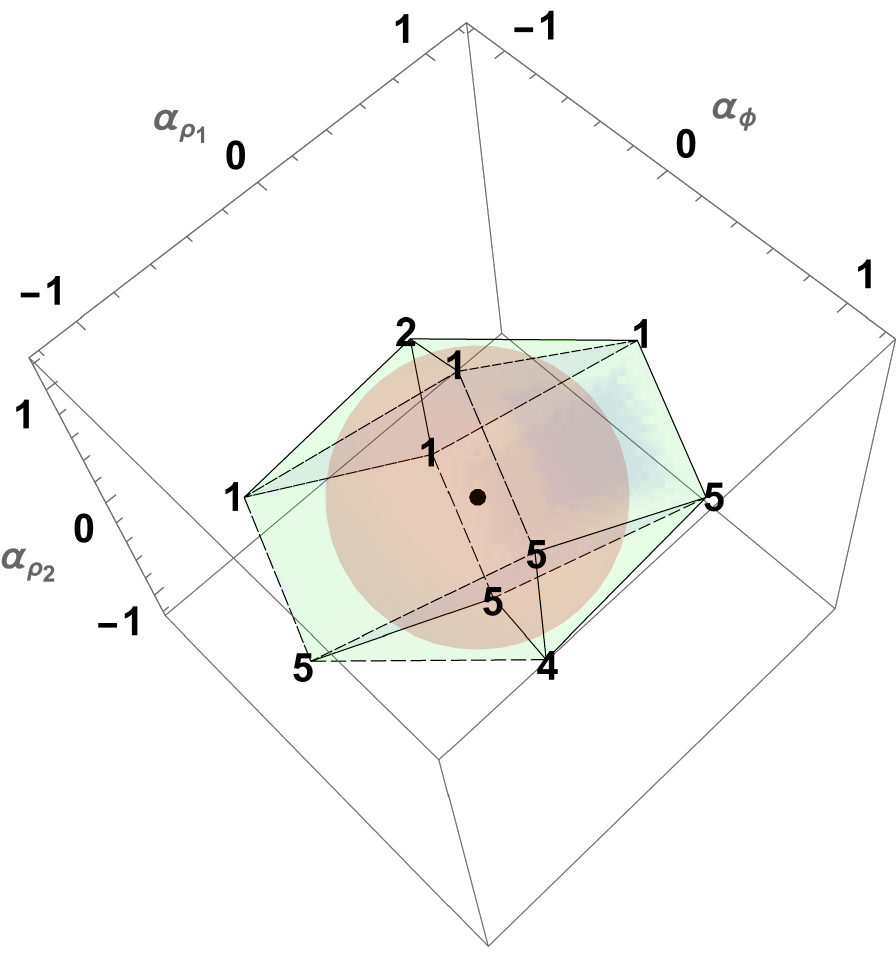}
\caption{Dilaton-radion-radion.}\label{f.8dhetrads}
\end{subfigure}
\begin{subfigure}{.5\linewidth}
\centering
\includegraphics[width = \linewidth]{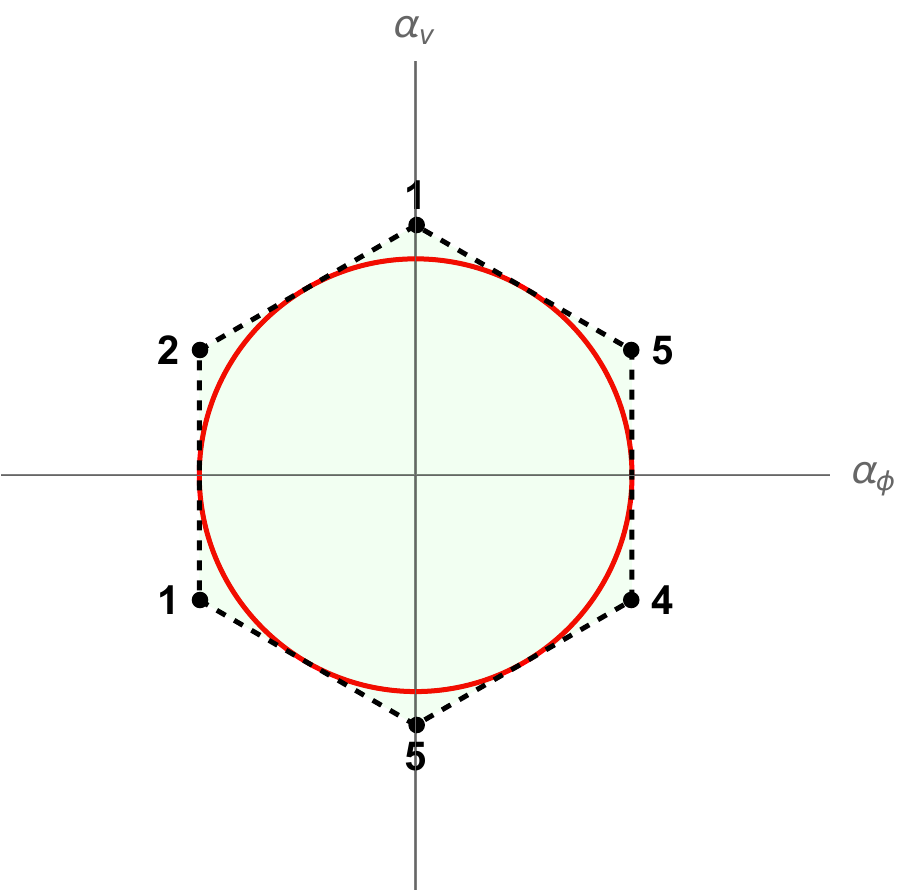}
\caption{Dilaton-volume.}
\label{f.8dhetdv}
\end{subfigure}
\caption{Dilaton-radion-radion (left) and dilaton-volume (right) components of $\alpha$-vectors of \halfBPS\ $(p-1)$-branes from either SO$(32)$ or $\mathrm E_8\times E_8$ heterotic string theory on a circle. The spheres and circles have radii $1/\sqrt{2}$. The $p_\text{max}=d-3$ Brane DC is satisfied and saturated.}
\label{f.8dhet}
\end{figure}

Next, we consider the dilaton-radion-radion components of $\alpha$-vectors for SO(32) or $\mathrm{E_8\times E_8}$ string theory reduced on a rectangular torus with the axions set to zero. The $p_\text{max}=d-3$ Brane DC is satisfied, and saturated, by \halfBPS\ states, as depicted in Figure \ref{f.8dhet}. The Brane DC with $p_\text{max} \neq 1, d-3$ requires non-BPS branes.

For the branes we have considered, the $p_\text{max}=d-3$ Brane DC is saturated in twelve directions in this dilaton-radion-radion slice of moduli space. Thus, by our dimensional reduction argument above, we predict that in asymptotic limits in these twelve directions of moduli space, there either exist non-BPS branes of spacetime dimension $p\leq d-3$ that spoil this saturation, or else there exist light codimension-2 branes with $\alpha$-vectors pointing in the directions of saturation.\footnote{Some of these branes may be well-known objects, as in the 9d example above. We defer a detailed analysis to future work.}

\subsubsection{7d}
Unlike in 10d, 9d, and 8d half-maximal supergravity, in 7d half-maximal supergravity the \halfBPS\ branes are not sufficient to satisfy the $p_\text{max}=d-3$ Brane DC. For the branes of spacetime dimension  $ p \leq p_\text{max}=d-3$, the ball of radius $1/\sqrt{10-d}$ is the largest origin-centered ball enclosed in the convex hull of the $\alpha$-vectors.\footnote{Similarly, in dimensions 4, 5, and 6, the ball of radius $1/\sqrt{10-d}$ is the largest origin-centered ball enclosed in the convex hull of \halfBPS\ branes with $p_\text{max}=d-3$.} However, the $p_\text{max}=d-3$ Brane DC requires the larger ball of radius $1/\sqrt{2}$ to be enclosed. Thus, non-BPS branes are necessary for the $p_\text{max}=d-3$ Brane DC in this example. We now demonstrate in detail for a slice of moduli space that such non-BPS branes do exist, and consequently the $p_\text{max}=d-3$ Brane DC is satisfied and saturated.

\begin{figure}
\center
\includegraphics[width = .7\linewidth]{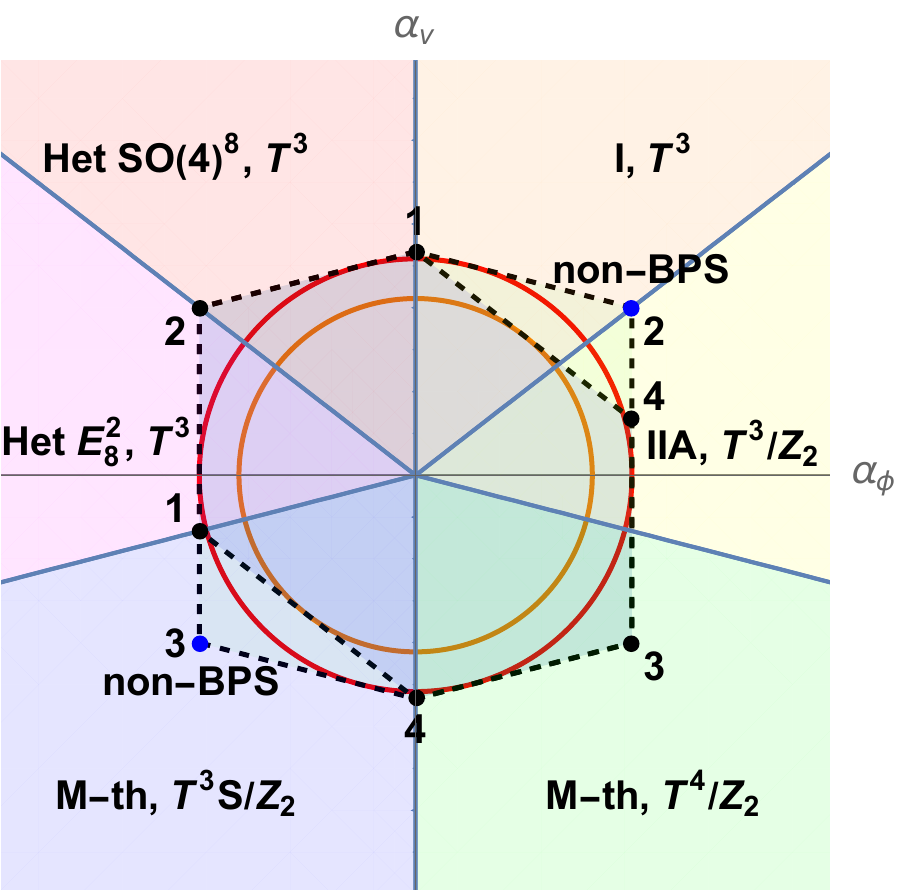}
\caption{Dilaton-volume components of $\alpha$-vectors of branes in 7d heterotic string theory in various phases dual to SO$(4)^8$ heterotic string theory. All branes have at most four spacetime dimensions. The black dots represent BPS branes, and the blue dots represent non-BPS branes. The non-BPS state on the top right is the type I fundamental string in the type I phase, and the non-BPS state on the bottom left is the unwrapped M2-brane in the M-theory phase. The non-BPS branes are understood to be evaluated when their respective phases are valid. The orange circle, enclosed by only the \halfBPS\ branes, has radius $1/\sqrt{3}$, and thus the $p_\text{max}=d-3$ Brane DC is not satisfied by BPS branes alone. But, when non-BPS branes are included, the red circle of radius $1/\sqrt{2}$ is enclosed, and thus the $p_\text{max}=d-3$ Brane DC is satisfied and saturated.}
\label{f.7dhetSO48}
\end{figure}

The non-BPS branes required for the $p_\text{max}=d-3$ Brane DC in the volume-dilaton plane depend on the values of the axions. In this section, we fix the axions so that the weak coupling limit corresponds to SO$(4)^8$ heterotic string theory in 7d. For the other duality frames, we will investigate separately the non-BPS states in each frame of the volume-dilaton moduli space, since the existence and stability of such branes is questionable in regimes where their tensions are super-Planckian. Thus, we must approach the Brane DC in these theories one duality frame at a time. (At this point, the reader who is less interested in the computational details and more interested in the final result may wish to skip straight to Figure \ref{f.7dhetSO48}.)

Let us begin with  SO$(32)$ heterotic string theory compactified on a rectangular $T^3$ with Wilson lines that break the gauge group to SO$(4)^8$. Let $R_{1,2,3}$ be the three radii in the heterotic theory. The heterotic SO$(4)^8$ phase is valid when the string coupling is small, and also the torus radii are large compared with the string length,
\begin{align}
g_s \ll 1, \qquad R_{1, 2, 3}(\alpha')^{-1/2} \gg 1.
\end{align}

Before we dualize to obtain another phase, we first establish how these terms are related to the canonically normalized dilaton $\phi$ and volume modulus $v$. The canonically normalized dilaton $\phi$ is related to the string coupling $g_s$ by
\begin{align}
	\phi=\frac 2{\sqrt{D-2}}\log g_s=\frac 1{\sqrt2}\log g_s\quad \Rightarrow\quad  g_s=\exp\left(\sqrt2\phi\right).
\end{align}
Determining the canonically normalized volume modulus requires some more steps. In terms of the radii $R_i$, the volume $V$ of the torus is
\begin{align}
	V=(2\pi R_1)\cdots (2\pi R_k).
\end{align}
Meanwhile, from equation (2.12) of \cite{Etheredge:2022opl}, a KK-mode from compactification of $k$ multiple rectangular dimensions has a mass that scales as
\begin{align}
	m_\text{KK}^d\sim \exp\left(-\sqrt{\frac{k+d-2}{k(d-2)}}v\right)\sim V^{-1/k},
\end{align}
where $v$ is the canonically normalized volume modulus. Thus
\begin{align}
	(2\pi R_1)\cdots (2\pi R_k)\sim V\sim \exp \left(\sqrt{\frac{k(k+d-2)}{d-2}}v\right)=\exp \left(\sqrt{\frac{24}{5}}v\right).
\end{align}
Let us fix the ratios of the radii so that
\begin{align}
	R_i\sim V^{1/k}\sim \exp \left(\sqrt{\frac{k+d-2}{k(d-2)}}v\right)=\exp \left(\sqrt{\frac{8}{15}}v\right).
\end{align}
The string scale, 10d Planck scale, and 7d Planck scale are then related by
\begin{align}
	\frac{1}{2}  (2 \pi)^7 g^2_s {\alpha'}^4=\kappa_{10}^2=V\kappa_7^2\,.
\end{align}
So, for $d=7$,
\begin{align}
	\alpha' \sim g_s^{-1/2}V^{1/4}\sim \exp\left(-\frac 1{\sqrt2}\phi+\frac 14\sqrt{\frac{k(k+d-2)}{d-2}}v \right)=\exp\left(-\frac 1{\sqrt2}\phi+\sqrt{\frac{3}{10}}v \right).
\end{align}
Thus, the range of validity in terms of the canonically normalized moduli is
\begin{align}
\exp(\phi)  \ll 1, \qquad \exp \left(\phi+\sqrt{\frac{5}{3}}v\right) \gg1 .
\end{align}

We next investigate the Type I phase. This is obtained by S-dualizing the heterotic SO$(4)^8$ theory. The Type I phase is valid when
\begin{align}
 g_s^{\Iota} \ll 1, \qquad R^{\Iota}_{1, 2, 3} (\alpha'_\Iota)^{-1/2}\gg 1.
\end{align}
The parameters here are related to the heterotic parameters via
\begin{align}
  g_s^{\Iota} =  1 / g_s,\qquad
  R_{1, 2, 3}^{\Iota} =  R_{1, 2, 3}, \qquad
  \kappa_{10, \Iota}^2 =  \kappa_{10 }^2 ,
\end{align}
where absence of indices indicates that the quantity is the quantity of the heterotic SO$(4)^8$ theory. Also,
\begin{align}
	\kappa_{10,\Iota} g^2_{s, \Iota} {\alpha'}^4_{\Iota} =\kappa_{10} g^2_{s}
   {\alpha'}^4 \qquad \Rightarrow \qquad \alpha'_{\Iota} =
   \sqrt{\frac{g_{s}}{g_{s, \Iota}}} \alpha'_{\tmop{het}} = g_{s} \alpha'.
\end{align}
In terms of the heterotic moduli,
\begin{subequations}
\begin{align}
	g_s^\Iota &=1/g_s=\exp(-\sqrt 2\phi),\\
	R^\Iota_{1,2,3}&\sim \exp \left(\sqrt{\frac{8}{15}}v\right),\\
	\alpha'_\Iota &\sim \exp\left(\frac 1{\sqrt 2}\phi+\sqrt{\frac{3}{10}}v \right).
\end{align}
\end{subequations}
Thus, this phase of moduli space is valid when
\begin{align}
	\exp \phi \gg 1,\qquad \exp \left(-\phi+\sqrt{\frac53}v \right)\gg 1.
\end{align}
In this phase of moduli space, the \halfBPS\ branes are not sufficient for the $p_\text{max}=d-3$ Brane DC. However, there is also the non-BPS Type I string, which has tension
\begin{align}
	T_{\text{F1, }\Iota}\sim \frac 1{\alpha'_\Iota}\sim \exp\left(-\frac 1{\sqrt 2}\phi-\sqrt{\frac{3}{10}}v \right),
\end{align}
and so
\begin{align}
	\vec \alpha_\text{F1, I}=\frac 1{\sqrt 2}\hat \phi+\sqrt{\frac{3}{10}}\hat v,
\end{align}
which suffices to satisfy the Brane DC in this phase (see Figure \ref{f.7dhetSO48} below).

We next T-dualize to obtain type IIA string theory on an orbifolded three-torus $T^3/\mathbb Z_2$. This phase is valid when
\begin{align}
 g_s^{\text{IIA}} \ll 1, \qquad R^{\text{IIA}}_{1, 2, 3}(\alpha')^{-1/2} \gg 1.
\end{align}
These terms are related to the type I theory, and thus the SO(4)$^8$ theory, by
\begin{subequations}
\begin{align}
  g_s^{\text{IIA}} &=  g_s^{\Iota}  \frac{{\alpha'_{\Iota}}^{3 /
  2}}{R_1^{\Iota} R_2^{\Iota} R_3^{\Iota}} \sim  \exp\left(-\frac 1{\sqrt 8}\phi-\sqrt{\frac{15}{8}}v \right),\\
  R_{1, 2, 3}^{\text{IIA}} &=  \frac{\alpha'_{\Iota}}{R_{1, 2, 3}^\Iota} \sim \exp\left(\frac 1{\sqrt 2}\phi-\frac1{30}v \right),
 \\
  \alpha'_{\text{IIA}} &=  \alpha'_{\text{I}}\sim \exp\left(\frac 1{\sqrt 2}\phi+\sqrt{\frac{3}{10}}v \right).
\end{align}
\end{subequations}
Thus, the range of validity for this phase is
\begin{align}
\exp\left(\phi+\sqrt{15}v \right) \gg 1, \qquad \exp\left( \phi-\sqrt{\frac{5}{3}}v \right)\gg 1.
\end{align}
The \halfBPS\ branes are not sufficient for the $p_\text{max}=d-3$ Brane DC. But, in this phase the fundamental type II string, which is non-BPS, ensures that the $p_\text{max}=d-3$ Brane DC is satisfied. The moduli-dependence of this  string matches that of the type I fundamental string.

We next take the strong coupling limit of type IIA string theory, thereby decompactifying to M-theory. Here, let the M-theory radius be radius $R_4^\Mu$. This theory is valid when all of the lengths of the 4-torus are greater than the 11-dimensional Planck length,
\begin{align}
	R^\Mu_{1,2,3,4}(\ell_{11}^\Mu)^{-1}\gg 1.
\end{align}
The three radii $R^\Mu_{1,2,3}$ are equal to the IIA three radii $R^\text{IIA}_{1,2,3}$ and the M-theory radius $R^\Mu_4$ is related to the IIA Planck length $\ell_s^\text{IIA}\sim \sqrt{\alpha'_\text{IIA}}$ and IIA string coupling $g_s^\text{IIA}$ by
\begin{subequations}
\begin{align}
	R^\Mu_{1,2,3}&=R^\text{IIA}_{1,2,3},\\
	R_4^\Mu &=g_s^{\text{IIA}} \ell_s^{\text{IIA}}=g_s^{\text{IIA}} {\alpha'}_\text{IIA}^{1/2},\\
	l^\Mu_{11}&\sim M_{11}^{-1}\sim    \left( g_s^{\text{IIA}} \right)^{1/3} {\alpha'}^{1 / 2}_{\text{IIA}}.
\end{align}
\end{subequations}
Thus, the validity of this phase is
\begin{align}
	\exp(\phi)\gg 1,\qquad \exp(\phi+\sqrt{15} v)\ll 1.
\end{align}
To obtain the neighboring phase of this phase in this slice of moduli space involves a chain of dualities. Rather than following this chain, we will instead T-dualize the original heterotic SO$(4)^8$ theory that we started with and explore the phases in the other direction through this volume-dilaton slice of moduli space.

Let us T-dualize the heterotic SO$(4)^8$ theory that we started with to obtain a heterotic $E_8\times E_8$ theory. This phase is valid when
\begin{align}
 g_s^{\text{E}_8^2} \ll 1, \qquad R^{\text{E}_8^2}_{1, 2, 3}(\alpha')^{-1/2} \gg 1.
\end{align}
These terms are related to the terms of the previous phase by
\begin{subequations}
\begin{align}
  g_s^{\text{E}_8^2} &=  g_s  \frac{{\alpha'}^{3 /
  2}}{R_1 R_2 R_3} \sim  \exp\left(\frac{1}{4} \left(\sqrt{2} \phi -\sqrt{30} v\right) \right),\\
  R_{1, 2, 3}^{\text{E}_8^2} &=  \frac{\alpha'}{R_{1, 2, 3}} \sim \exp\left(-\frac{\phi }{\sqrt{2}}-\frac{v}{\sqrt{30}} \right),\\
  \alpha'_{\text{E}_8^2} &=  \alpha'\sim \exp\left(\frac 1{\sqrt 2}\phi+\sqrt{\frac{3}{10}}v \right).
\end{align}
\end{subequations}
Thus, the range of validity for this phase is
\begin{align}
\exp\left(\phi-\sqrt{15}v \right) \ll  1, \qquad \exp\left(  \phi+\sqrt{\frac 53} v \right)\ll  1.
\end{align}

We next take the strong coupling limit of the $\mathrm{E_8\times E_8}$ heterotic string theory to decompactify to M-theory on an interval. Here, the radius $R_4^\text{HW}$ is the M-theory interval lengths. This theory is valid when all of the lengths of the 4-torus are greater than the 11-dimensional Planck length,
\begin{align}
	R^\text{HW}_{1,2,3,4}(\ell_{11}^\text{HW})^{-1}\gg 1.
\end{align}
The three radii $R^\text{HW}_{1,2,3}$ are equal to the IIA three radii $R^\mathrm{E_8^2}_{1,2,3}$ and the M-theory radius $R^\text{HW}_4$ is related to the IIA Planck length $\ell_s^\mathrm{E_8^2}\sim \sqrt{\alpha'_\mathrm{E_8^2}}$ and Heterotic $\mathrm{E_8\times E_8}$ string coupling $g_s^\mathrm{E_8^2}$ by
\begin{subequations}
\begin{align}
	R^\text{HW}_{1,2,3}&=R^\mathrm{E_8^2}_{1,2,3}\sim \exp\left(-\frac{\phi }{\sqrt{2}}-\frac{v}{\sqrt{30}}\right),\\
	R_4^\text{HW} &=g_s\mathrm{E_8^2}\ell_s^\mathrm{E_8^2}=g_s^\mathrm{E_8^2} {\alpha'}_\mathrm{E_8^2}^{1/2}\sim \exp\left(-\sqrt{\frac{6}{5}} v\right),\\
	l^\text{HW}_{11}&\sim (M^{\text{HW}}_{11})^{-1}\sim    \left( g_s^\mathrm{E_8^2} \right)^{1/3} {\alpha'}^{1 / 2}_\mathrm{E_8^2}\sim \exp\left(-\frac{\sqrt{15} v+5 \phi }{15 \sqrt{2}}\right).
\end{align}
\end{subequations}
Thus, the validity of this phase is
\begin{align}
	\exp(\phi)\ll 1,\qquad \exp(\phi-\sqrt{15} v)\gg 1.
\end{align}
The Brane DC is not satisfied by \halfBPS\ branes in this phase. However, in this phase, there is a light, non-BPS, unwrapped M2-brane. This M2-brane has a tension
\begin{align}
	T_3\sim (M^\text{HW}_{11})^3\sim (l^\text{HW}_{11})^{-3}\sim \exp\biggl(3\frac{\sqrt{15} v+5 \phi }{15 \sqrt{2}}\biggr),
\end{align}
and thus an $\alpha$-vector of
\begin{align}
	\vec \alpha_p=-\frac{1}{\sqrt{2}}\hat \phi-\sqrt{\frac{3}{10}}\hat v,
\end{align}
which ensures that the $p_\text{max}=d-3$ Brane DC is satisfied in this phase.

All of these phases are depicted in Figure \ref{f.7dhetSO48}, and the $p_\text{max}=d-3$ Brane DC is satisfied and saturated in this example.

For the branes we have considered, the $p_\text{max}=d-3$ Brane DC is saturated in six directions in this dilaton-volume slice of moduli space. As before, this should either imply the existence of specific light codimension-2 branes, or else imply that there are additional light non-branes of spacetime dimension $p\leq d-3$ that we have not identified above. It would be interesting to analyze this further.

\section{Observations\label{s.observations}}

\subsection{Connection with the species polytope}

So far, our analysis has dealt with the tensions of branes, but it is interesting to analyze the energy scale associated with each brane. A $(p-1)$-brane has a characteristic energy scale that is proportional to the $p$-th root of its tension,
\begin{align}
	\Lambda_{p} \sim T_p^{1/p}.
\end{align}
Above this energy scale, the brane can be readily created from the vacuum. Unless it is a soliton, this process is not readily described in a local effective field theory, hence we expect that this characteristic energy scale must be no lighter than the UV cutoff $\Lambda_{\rm QG}$, sometimes known as the ``species scale'' \cite{Donoghue:1994dn, Dvali:2007hz, Dvali:2007wp}, at which gravity becomes strongly coupled:
\begin{align}
\Lambda_{\rm QG} \lesssim \Lambda_{p} \,.
\end{align}
This bound translates to constraints on the $\alpha$-vectors of the $(p-1)$-brane at hand. Let us define 
\begin{equation}
 \vec \lambda_p \equiv \frac1p \vec \alpha_p \,.
 \end{equation}
to be the ``$\lambda$-vector" of the $(p-1)$-brane, and let us define
\begin{align}
\vec \lambda_{\rm QG} \equiv - \vec \nabla \log \Lambda_{\rm QG}
\end{align}
to be the ``species vector'' 
\cite{Calderon-Infante:2023ler, Etheredge:2024tok}, which captures the rate at which the species scale $\Lambda_{\rm QG}$ becomes light.
 Then, in a generic asymptotic limit in the unit $\hat t$ direction, we must have that
\begin{align}
	\frac{1}{p} \hat t  \cdot \vec \alpha_p \equiv \hat t \cdot \vec \lambda_p \leq \hat  t \cdot \vec \lambda_{\rm QG} \leq \frac{1}{\sqrt{d-2}}\,, 
\label{speciesbound}
\end{align}
where the last inequality, originally proposed in \cite{vandeHeisteeg:2023ubh, vandeHeisteeg:2023uxj,Calderon-Infante:2023ler}, can be justified in generic infinite-distance limits using the results of \cite{Etheredge:2024tok}.

\begin{figure}
\begin{subfigure}{.32\linewidth}
\centering
\includegraphics[width = \linewidth]{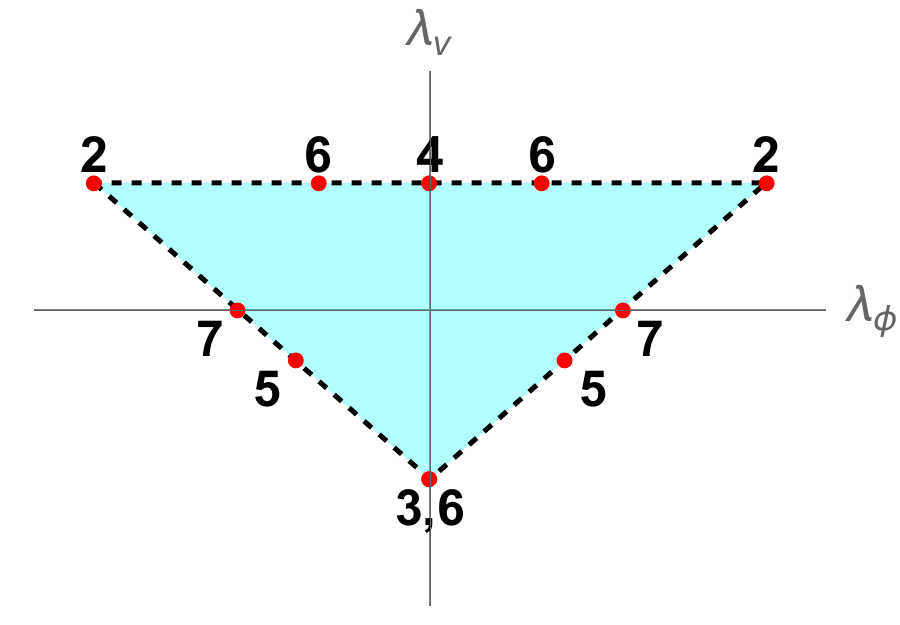}
\caption{9d maximal supergravity.}\label{f.9dIIdvspecies}
\end{subfigure}
\begin{subfigure}{.32\linewidth}
\centering
\includegraphics[width = \linewidth]{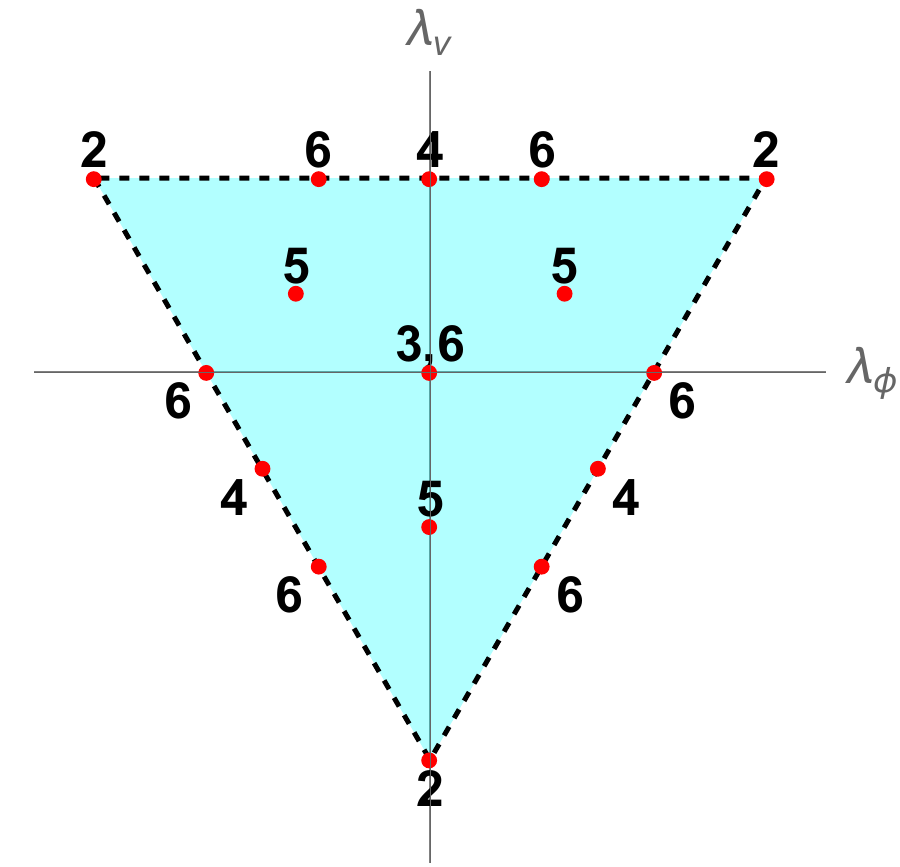}
\caption{8d maximal supergravity.}\label{f.8dIIdvspecies}
\end{subfigure}
\begin{subfigure}{.32\linewidth}
\centering
\includegraphics[width = \linewidth]{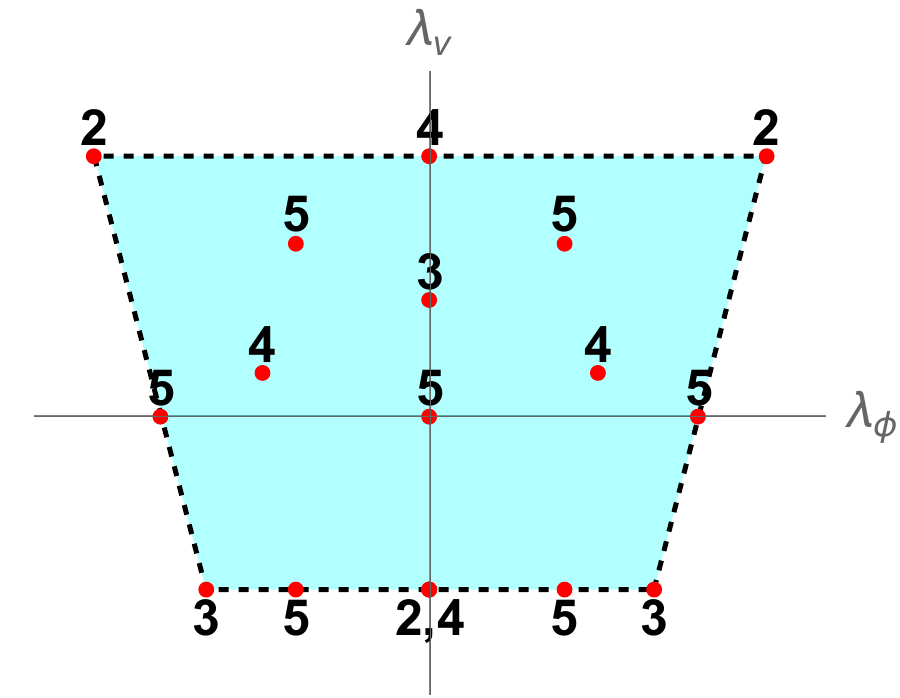}
\caption{7d maximal supergravity.}\label{f.7dIIdvspecies}
\end{subfigure}
\begin{subfigure}{.32\linewidth}
\centering
\includegraphics[width = \linewidth]{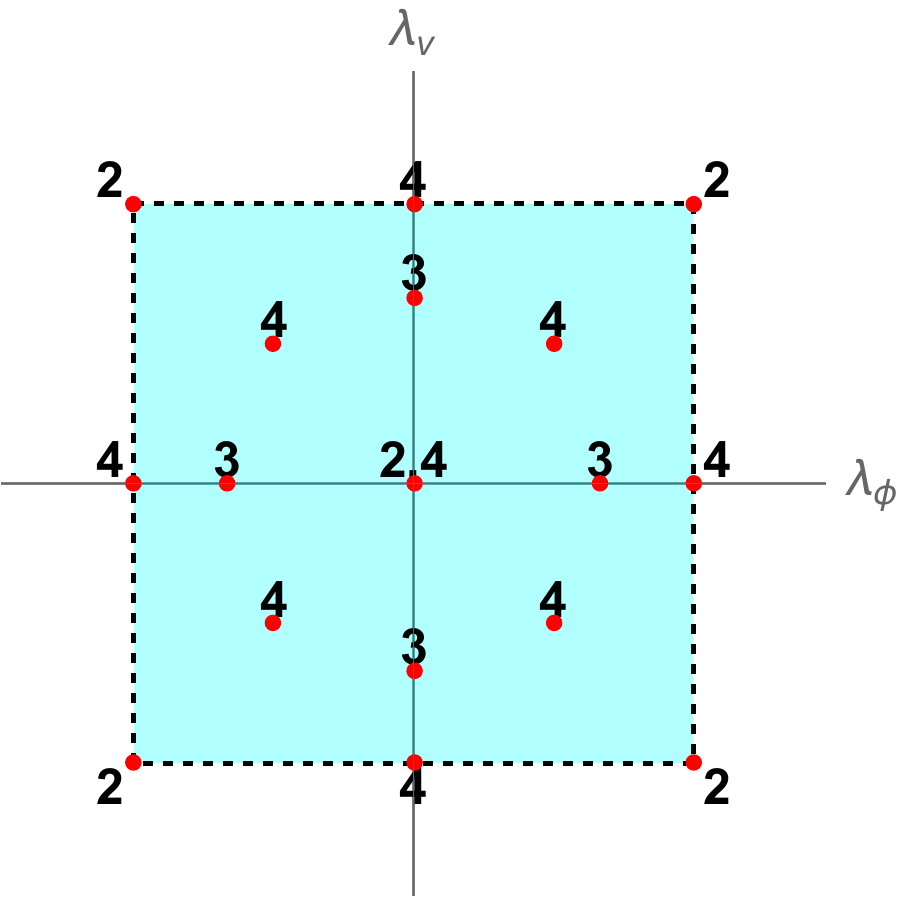}
\caption{6d maximal supergravity.}\label{f.6dIIdvspecies}
\end{subfigure}
\begin{subfigure}{.32\linewidth}
\centering
\includegraphics[width = \linewidth]{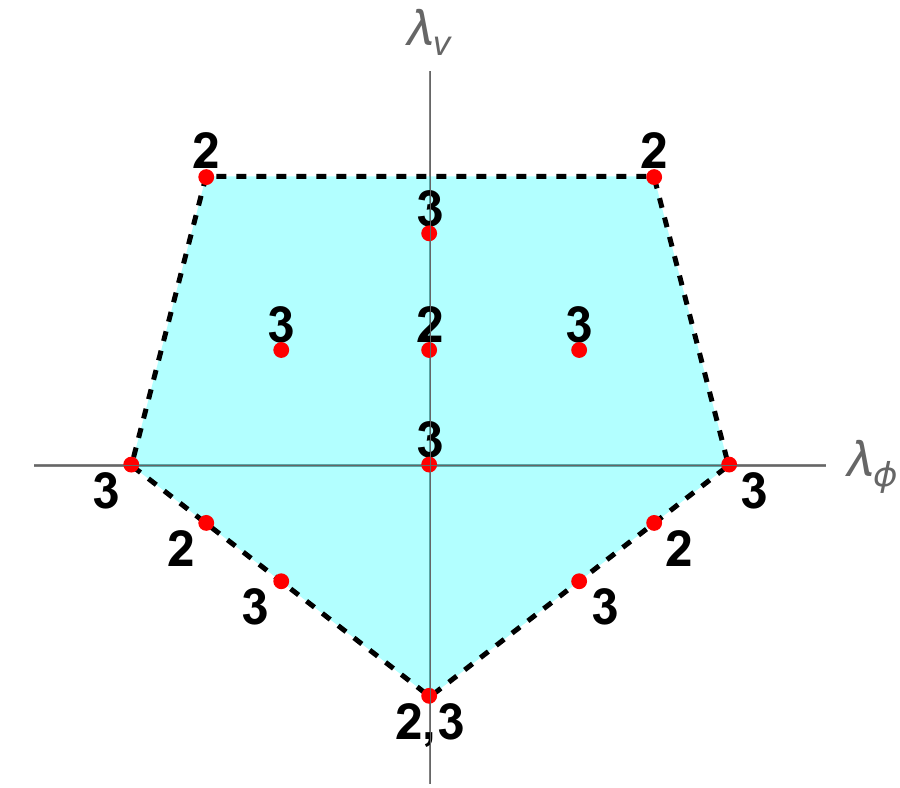}
\caption{5d maximal supergravity.}\label{f.5dIIdvspecies}
\end{subfigure}
\begin{subfigure}{.32\linewidth}
\centering
\includegraphics[width = \linewidth]{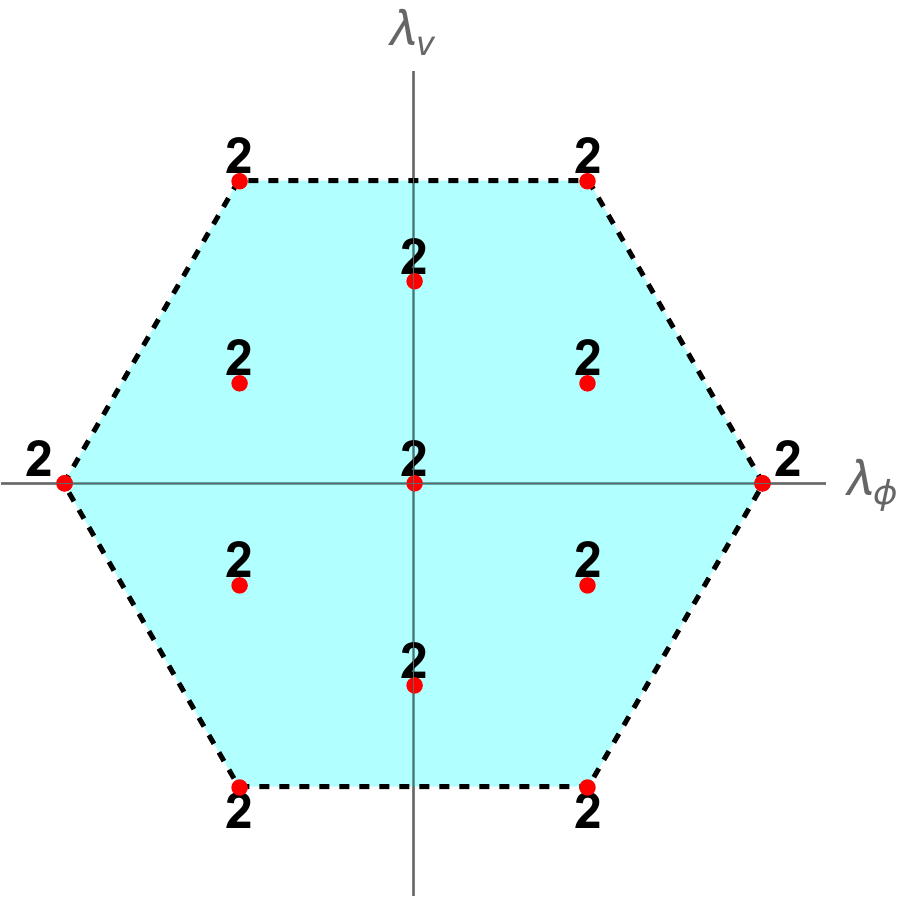}
\caption{4d maximal supergravity.}\label{f.4dIIdvspecies}
\end{subfigure}
\caption{Dilaton-volume components of $\lambda$-vectors and species polytopes of \halfBPS\ in maximal supergravity in various dimensions. The cyan region is the species polytope, and the red dots are $\lambda$-vectors for $p>1$. Note that these $\lambda$-vectors are always contained in the species polytopes and often reside on its boundaries. (In fact, many $\lambda$-vectors that appear to lie strictly in the interior of the species polytope in this planar projection actually lie on the boundary when considering the full polytope.)}
\label{f.speciespattern}
\end{figure}

The result \eqref{speciesbound} can be understood graphically by comparison with the ``species polytopes" \cite{Calderon-Infante:2023ler, Etheredge:2024tok}, which capture the rate at which the species scale becomes light as a function of direction in moduli space. In flat slices of maximal supergravity, these species polytopes are constructed simply by taking the convex hull of the species vectors $\vec \lambda_{\rm QG}$ across the various duality frames of the theory. The result is shown in Figure \ref{f.speciespattern}, plotted along with the $\lambda$-vectors of the BPS branes.

As shown in Figure \ref{f.speciespattern}, many of the branes responsible for the Brane DC have $\lambda$-vectors that reside on the boundary of the species polytopes. The fact that these $\lambda$-vectors lie inside the species polytope follows from \eqref{speciesbound}. The more surprising feature of Figure \ref{f.speciespattern} is the fact that, for many branes, the $\lambda$-vectors lie precisely on the boundary of the species polytope.

As explained in \cite{Etheredge:2024tok}, in maximal supergravity the species polytope is (up to a normalization) the dual of the convex hull generated the $\alpha$-vectors of the particle towers of the theory. This implies that for any point $\vec \lambda$ on the boundary of the convex hull, there exists a particle tower with $\alpha$-vector $\vec\alpha_1$ that satisfies
\begin{equation}
\vec\alpha_1 \cdot \vec \lambda = \frac{1}{d-2}\,.
\end{equation}
Traveling in a direction $\hat t$ for which $\vec \alpha_1 = \vec \alpha_{\rm lightest}$ is the lightest particle tower, this gives precisely the tower-species pattern of \cite{Castellano:2023jjt, Castellano:2023stg}:
\begin{equation}
\vec \alpha_{\rm lightest} \cdot \vec \lambda_{\rm QG} = \frac{1}{d-2}\,.
\end{equation}
Here, taking $\vec \lambda = \vec\lambda_p$ to be a $\lambda$-vector of a brane that lies on the boundary of the species polytope in the direction of travel $\hat t$, we find a generalization of this tower-species pattern:\footnote{We thank Irene Valenzuela for discussions on this point.}
\begin{equation}
\vec \alpha_{\rm lightest} \cdot \vec \lambda_{p} = \frac{1}{d-2}\,.
\end{equation}

More generally, the $\lambda$-vectors of fundamental branes satisfy $\vec \alpha_{\rm lightest} \cdot \vec \lambda_{p} \leq  \frac{1}{d-2}$, which is closely related to the bound \eqref{speciesbound} and ensures that no brane tension scale becomes light more quickly than the species scale.\footnote{We thank Jos\'e Calder\'on-Infante for discussions on this point.}

\subsection{Taxonomy rules for particles and branes \label{s.observations.ESC}}

Another interesting pattern in our analysis above is that the $\alpha$-vectors of light $(p-1)$-branes all have lengths given by 
\begin{align}
	\vec \alpha_p^2=2-\frac{p(d-p-2)}{d-2}.
	\label{e.speciallength}
\end{align}
This relation can be justified using the results of the taxonomy program of \cite{Etheredge:2024tok} (see also \cite{Etheredge:2022opl}), which argued on the basis of the Emergent String Conjecture that in a generic infinite-distance limit in moduli space, the light towers of particles must have $\alpha$-vectors with lengths given by
\begin{equation}
\vec \alpha_1^2 = \frac{n+d-2}{n(d-2)}  \,,
\end{equation}
for some integer $n$. Physically, this result follows by demanding that the tower of particles in question represents either a KK-mode for a decompactification of $n$ dimensions or else a string oscillator mode (in which case we formally have $n = \infty$). The result \eqref{e.speciallength} follows from demanding that, upon compactifying $(p-1)$ dimensions and having a $(p-1)$-brane fully wrap the compactification manifold, the wound state is a KK-mode in the lower-dimensional theory with length
\begin{align}
	\vec\alpha_1^2=\frac{d-1}{d-2}\,,
\end{align}
corresponding to the KK-mode of a single decompactifying dimension, $n=1$. Formula \eqref{e.speciallength} thus provides a generalization of the taxonomy rules of \cite{Etheredge:2024tok} from particles to branes.

By similar reasoning, \cite{Etheredge:2024tok} found constraints not only on the lengths of $\alpha$-vectors of light particles, but also on the dot products (or equivalently, the angles) between pairs of $\alpha$-vectors. Using a similar dimensional reduction argument, we can uplift these constraints from particles to branes, as we now explain.

To begin, we consider a $D$-dimensional theory with $(p-1)$-branes and $(q-1)$-branes, where without loss of generality we assume $p\geq q$. We then compactify this theory on a rectangular $(p-1)$-torus, wrapping the branes around the cycles of the torus to produce a collection of particle species in the resulting theory in $d=D-p+1$ dimensions. In particular, we allow the $(p-1)$-brane to wrap all of the $p-1$ one-cycles of the torus, while the $(q-1)$-brane wraps only $q-1$ one-cycles of the torus. 
The relevant moduli in the lower-dimensional theory are given by
\begin{align}
	(\vec \phi_D,\rho_{q-1},\rho_{p-q}),
\end{align}
where $\vec \phi_D$ are the moduli of the $D$-dimensional theory, $\rho_{q-1}$ controls the volume of the cycles that the $(q-1)$-brane wraps, and $\rho_{p-q}$ controls the volume of the cycles that the $(q-1)$-brane does not wrap. With respect to these moduli, the $\alpha$-vectors of the particles from the wrapped $(p-1)$ and $(q-1)$ branes are given by
\begin{align}
\begin{aligned}
	\vec \alpha_{p\rightarrow1}^{(d)}&=\left (\vec \alpha_p^{D},-\frac{(D-p-2)\sqrt{q-1} }{\sqrt{(D-2)(D-q-1)}},-\frac{(D-p-2) \sqrt{p-q}}{\sqrt{(D-p-1)(D-q-1)}}\right ),\\
	\vec \alpha_{q\rightarrow1}^{(d)}&=\left (\vec \alpha_q^{D},-\frac{(D-q-2)\sqrt{q-1} }{\sqrt{(D-2) (D-q-1)}},\frac{\sqrt{p-q}}{\sqrt{(D-p-1)(D-q-1)}}\right ).
\end{aligned}
\label{e.redap1aq1alphas}
\end{align}
One can make progress if the $\alpha$-vectors of these wrapped branes satisfy the dot-product rules from \cite{Etheredge:2024tok},\footnote{Understanding precisely when pairs of branes result in $\alpha$-vectors that satisfy the product rules in \cite{Etheredge:2024tok} will be the subject of future work \cite{Etheredge:BraneTaxonomy}.}:
\begin{align}
	\vec \alpha_{p\rightarrow 1}^{(d)}\cdot \vec \alpha_{q\rightarrow 1}^{(d)}=\frac 1{d-2}=\frac 1{D-p-1}.\label{e.ddimdotprod}
\end{align}
By \eqref{e.redap1aq1alphas}, this dot product is also equal to
\begin{align}
	\vec \alpha_{p\rightarrow 1}^{(d)}\cdot \vec \alpha_{q\rightarrow 1}^{(d)}=\vec \alpha_{p}^{(D)}\cdot \vec \alpha_{q}^{(D)}+(D-p-2) \left(\frac{1}{-D+p+1}+\frac{q}{D-2}\right).
\end{align}
Thus, when the dot product \eqref{e.ddimdotprod} is satisfied in the lower-dimensional theory, the dot product in the parent $D$-dimensional theory is,
\begin{align}
	\vec \alpha_p^{(D)}\cdot \vec \alpha_q^{(D)}=1-\frac{q (D-p-2)}{D-2}\,,
\end{align}
Combining this with \eqref{e.speciallength}, we have the brane generalization of the particle taxonomy rules of \cite{Etheredge:2024tok}:
\begin{align}
	\vec \alpha_i \cdot \vec \alpha_j = 1+\delta_{i j} -\frac{p_i (d-p_j-2)}{d-2}, \qquad p_i \le p_j, \label{e.taxonomy}
\end{align}
where $i$, $j$ index the different branes and $p_i$, $p_j$ count their spacetime dimensions.

The above rules apply when the two branes can be fully wrapped on the torus to produce $n=1$ winding modes that become light in the same duality frame. The conditions for this to occur are not obvious from the higher-dimensional perspective. In particular, note that~\eqref{e.taxonomy} does not apply to every pair of branes, and sometimes \emph{does} apply pairs that are not even light in the same higher-dimensional duality frame. A more general rule that applies to any pair is
\begin{align}
	\vec \alpha_i \cdot \vec \alpha_j = a_{i j} -\frac{p_i (d-p_j-2)}{d-2}, \qquad p_i \le p_j,
\end{align}
where $a_{i j}$ is some number that depends on the pair. Indeed, this can be viewed as a \emph{definition} of $a_{i j}$. However, as shown in Appendix \ref{s.taxonomy} it turns out that $a_{i j}$ behaves simply under dimensional reduction on tori and in particular for \halfBPS\ branes in maximal SUGRA $a_{i j}$ is always an integer between $-2$ and $+2$.

We defer a more complete analysis of these rules to future work \cite{Etheredge:BraneTaxonomy}.

\subsection{Lattices}

In our above studies of maximal and half-maximal supergravity theories in the landscape, we found that the dilaton-radion-\dots-radion components of $\alpha$-vectors of \halfBPS\ resided on lattices. For example, in ten dimensions, the $\alpha$-vectors reside on the dilaton line in increments of $1/\sqrt{d-2}=1/\sqrt{8}$ (see Figures \ref{f.10dII} and \ref{f.10dhet}). For M-theory on a rectangular two-torus, the radion-dilaton components of \halfBPS\ states reside on a lattice (see Figure \ref{f.9dII}). For heterotic string theory on a circle, the radion-dilaton components of the \halfBPS\ states lie on the same lattice that the \halfBPS\ states of M-theory on the rectangular two-torus. Similar patterns occur in lower-dimensional examples as well.

It seems likely that the ``taxonomy'' results of the previous subsection may offer further justification for this observation. We leave a more thorough investigation to future work.

\subsection{Instanton Distance Conjecture}

Heretofore, we have restricted the value of $p_{\rm max}$ in $D$ dimensions to the set $\{ 1, ..., d-2 \}$. We have seen that, after toroidal reduction, these versions of the Brane DC are related to the Sharpened DC in $d-p_{\rm max}+1 \geq 3$ dimensions.

However, it is natural to consider the extrapolation of the Brane DC bound to the case $p_{\rm max} =0$ as well.\footnote{We thank Cumrun Vafa and Timo Weigand for discussions on this point.} In this case, in any infinite-distance limit, the Brane DC would na\"ively require an instanton of action $S$ which decays as
\begin{align}
		S\sim \exp\left( -\alpha \Delta \right), ~~  ~~~~~~\alpha \geq \frac{1}{\sqrt{d-1}}\,. \label{e.IDCbound}
	\end{align}
	However, this na\"ive statement seems to be the opposite of what we observe. In the weak coupling limit of the type IIB string, for instance, the action of the D($-1$)-brane diverges as $S \sim \exp(\sqrt{2} \Delta)$. Likewise, compactifying type IIA string theory on a circle of radius $R$ to 9d, instantons come from D0-branes wound around the circle, and the action of these instantons diverges in the decompactification limit $R \rightarrow \infty$. Instantons with small action do exist in the weak coupling limit of type IIB string theory, but these are exotic branes in this duality frame, related to the D($-1$)-brane by $SL(2, \mathbb{Z})$ duality. Furthermore, whereas the precise notion of a (non-BPS) charged particle becomes well-defined only in a weak coupling limit $g \rightarrow 0$, where the tower Weak Gravity Conjecture \cite{Arkanihamed:2006dz, Heidenreich:2015nta, Andriolo:2018lvp} predicts a tower of light particles with $m \rightarrow 0$, the precise notion of an instanton becomes well-defined only in the large action limit $S \rightarrow \infty$. 
	
	All of this supports a modification of the na\"ive statement of the $p_{\rm max} = 0$ version of the Brane DC in which the sign of the exponential is flipped:
	\begin{align}
		S\sim \exp\left( \alpha \Delta \right), ~~  ~~~~~~\alpha \geq \frac{1}{\sqrt{d-1}}\,. \label{e.IDCboundII}
	\end{align}
This statement, which one might refer to as the ``Instanton Distance Conjecture,'' is satisfied in the strong coupling limit of 10d type IIB string theory by the D($-1$)-brane, which has $\alpha = \sqrt{2} > 1/3 = 1/\sqrt{d-1}$. In any other infinite-distance limit of this theory, then, there will exist instantons related to the D($-1$)-brane via $SL(2, \mathbb{Z})$ duality transformations.

However, even this improved version of the Instanton DC seems not to be completely universal.
In type IIA string theory and the heterotic string theories in 10d, for example, the instantons needed to satisfy this conjecture are not immediately apparent. It is possible that there exist presently unknown, non-BPS instantons which satisfy the conjecture, but it is also plausible that the Brane DC simply does not extend to the $p_{\rm max} = 0$ case, or perhaps it extends only under certain conditions. Clearly, more work is needed to put an instanton version of the Brane DC on firmer footing, which we leave for the future.

\section{Conclusions \label{s.conclusions}}

We have proposed the Brane DC, a series of precise bounds on the exponential decay rates of brane tensions in infinite-distance limits of moduli space. We have seen that the Brane DC is necessary to ensure that the Sharpened DC is satisfied after toroidal reduction, and we have verified that it is satisfied in a number of theories with maximal and half-maximal supersymmetry. We have further argued that saturation of the Brane DC is related to the presence of emergent string limits, which occurs only for the Brane DC with $p_{\rm max} \in \{ 1, d-3, d-2 \}$.

A number of future directions have presented themselves throughout the course of this work. For instance, in Section \ref{s.observations}, we observed several patterns involving the $\alpha$-vectors of light branes in asymptotic limits of maximal supergravity. Some of these patterns are clearly related to the Emergent String Conjecture, and it would be worthwhile to analyze more carefully the consequences of the Emergent String Conjecture for higher-dimensional branes.

Our dimensional reduction arguments focused exclusively on toroidal reductions, but compactifications on more general Calabi-Yau manifolds may lead to stronger consistency conditions (but potentially many caveats). Relatedly, the examples we studied above dealt exclusively with theories with maximal and half-maximal supersymmetry, but it would interesting to study the Brane DC in more general settings with less supersymmetry.

The Brane DC is not the first swampland conjecture to postulate the existence of certain branes. The absence of global symmetries \cite{Hawking:1974sw, Banks:2010zn}, the Completeness Hypothesis \cite{polchinski:2003bq}, the Cobordism Conjecture \cite{McNamara:2019rup}, and the Weak Gravity Conjecture \cite{Arkanihamed:2006dz} all motivate the existence of branes in string theory (see e.g. \cite{Gaiotto:2014kfa, McNamara:2019rup, Montero:2020icj, Rudelius:2020orz, Heidenreich:2020pkc, Heidenreich:2021tna, Kaidi:2023tqo}). It could be profitable to explore connections between these conjectures and the Brane DC.

One of the major leaps in the history of string theory was the discovery and appreciation of the importance of branes. In the end, a graduation from particles and strings to branes may prove to be a crucial step in the development of the swampland program as well.

\section*{Acknowledgements}

We are grateful for conversations with Alek Bedroya, Jos\'e Calder\'on-Infante, Bernardo Fraiman, Clifford Johnson, Dieter L\"ust, Jacob McNamara, Miguel Montero, Salvatore Raucci, Matthew Reece, Ignacio Ruiz, Savdeep Sethi, Cumrun Vafa, Irene Valenzuela, Timo Weigand, and Max Weisner, and we thank Jos\'e Calder\'on-Infante, Jacob McNamara, Miguel Montero, and Irene Valenzuela for comments on a draft. We are grateful for ``The Landscape vs.\ the Swampland'' workshop at the Erwin Schr\"odinger Institute where some of this work was completed. ME was supported in part by the Heising-Simons Foundation, the Simons Foundation, and grant no. PHY-2309135 to the Kavli Institute for Theoretical Physics (KITP). ME and BH received support from NSF grant PHY-2112800. The work of TR was supported in part by STFC through grant ST/T000708/1.

\appendix

\section{Brane Formulas \label{s.formulas}}

In this appendix, we review useful formulas related to the tensions of various particles and branes.

\subsection{D-branes, NS5-branes, and fundamental strings \label{s.formulas.tensions}}

In 10d string units,
\begin{align}
	T_{\text{D}p}\sim 1/g_s=e^{-\Phi}=e^{-\frac{4}{\sqrt{d-2}}\phi},
\end{align}
where $\phi=\frac{\sqrt{d-2}}{4}\Phi$ is the canonically normalized dilaton. One also has
\begin{align}
	T_\text{F1}\sim \text{constant},\qquad 	T_\text{NS5}\sim 1/g_s^2\sim e^{-2\Phi}=e^{-\frac 8{\sqrt{d-2}}\phi}.
\end{align}
In 10d Planck units,
\begin{subequations}
\begin{align}
	T_{\text{D}p}&\sim g_s^{-1}\cdot g_s^{\frac{p+1}4}=g_s^{\frac{p-3}{4}}=e^{\frac{p-3}{4}\Phi}=e^{\frac{p-3}{\sqrt{d-2}}\phi},\\
	T_\text{F1}&\sim 1\cdot g_s^{\frac{1}{2}}=e^{\frac 12\Phi}=e^{\frac{2}{\sqrt{d-2}}\phi},\\
	T_\text{NS5}&\sim g_s^{-2}\cdot g_s^{\frac32}=g_s^{-\frac{1}{2}}=e^{-\frac{1}{2}\Phi}=e^{-\frac{2}{\sqrt{d-2}}\phi}.
\end{align}
\end{subequations}
Additionally, oscillator modes of $(p-1)$-branes scale with the tensions via
\begin{align}
	m_\text{osc}\sim T_p^{\frac 1p},
\end{align}
and thus the $\alpha$-vectors of oscillator modes of $(p-1)$-branes are related to the $\alpha$-vectors of $(p-1)$-branes by
\begin{align}
	\vec \alpha_\text{osc}=\frac 1p\vec\alpha_p.
\end{align}

\subsection{KK-modes and monopoles}

For a circle reduction, the mass of the KK-modes scales like the gauge coupling of the graviphoton:
\begin{align}
	|e_\text{KK}|  \sim m_\text{KK}\sim \ept{-\sqrt{\frac{d-1}{d-2}}\rho } .
\end{align}
KK-monopoles are $(d-4)$-branes that are EM-dual to the KK-modes. Their associated gauge coupling is given by
\begin{align}
	g_\text{KK} \sim  \frac 1	{e_\text{KK}}.
\end{align}
The tension of a BPS KK-monopole scales as
\begin{align}
	T_\text{KK}\sim \frac 1{|g_\text{KK}|}\sim \frac 1{m_\text{KK}}\sim \ept{\sqrt{\frac{d-1}{d-2}}\rho }.
\end{align}
Thus, the $\alpha$-vector for a KK-monopole is opposite that for a KK-mode:
\begin{align}
	\vec \alpha_\text{KK mode}=-\vec \alpha _\text{KK mon}.
\end{align}

\subsection{Reduction formulas}

Consider reducing a $D$-dimensional theory with various branes with a compact $n$-manifold $M_n$ of dimension $n$. Let $\rho$ be the volume modulus of this manifold.

Reduction from $D=d+n$ dimensions to $d$ dimensions with a $(P-1)$-brane wrapping $k$-dimensions of the internal geometry, yields a $(p-1)$-brane with $P=p+k$. The tension of the $(p-1)$-brane scales with the volume of the $n$-manifold $M_n$ and the $D$-dimensional tension by \cite{Etheredge:2022opl}
\begin{align}
	T_p^{(d)}\sim \exp\left(-\frac{p n-k(d-2)}{\sqrt{n(n+d-2)(d-2)}}\rho\right)T_P^{(D)},
	\label{dimredform}
\end{align}
where $\rho$ is the canonically-normalized radion controlling the volume of the $n$-manifold $M_n$.

Let $\vec \phi_D$ be the moduli of the $D$-dimensional theory. The $\alpha$-vectors in the $d$-dimensional theory depend on the $\alpha$-vectors of the $D$-dimensional theory with respect to the moduli $\vec \phi_D$. To see this, consider the ($\vec \phi_D$-$\rho $)-slice of moduli space. The components in this slice of $\alpha$-vectors of these wrapped branes in the $d$-dimensional theory are
\begin{align}
	\vec \alpha_p^{(d)}=\left(\vec \alpha_P^{(D)},\frac{p n-k(d-2)}{\sqrt{n(n+d-2)(d-2)}}\right). \label{eqn:alpharad}
\end{align}
There are also KK-modes and monopoles, with $\alpha$-vectors of
\begin{align}
	\vec \alpha_{\text{KK},1,d-3}^{(d)}=\left (0,\pm \sqrt{\frac{n+d-2}{n(d-2)}}\right). 
\end{align}
The particles in the $d$-dimensional theory are either from fully-wrapped branes,
\begin{align}
	\vec \alpha_1^{(d)}=\left(\vec \alpha_P^{(D)},\frac{ n-(P-1)(d-2)}{\sqrt{n(n+d-2)(d-2)}}\right),
\end{align}
or from oscillations of non-fully wrapped branes
\begin{align}
	\vec \alpha_1^{(d)}=\frac 1p\left(\vec \alpha_P^{(D)},\frac{p n-k(d-2)}{\sqrt{n(n+d-2)(d-2)}}\right),
\end{align}
or from KK-modes or KK-monopole oscillators.

\section{Exotic branes in maximal supergravity \label{s.cod2branes}}

In this appendix we use U-duality to prove several facts about low-codimension exotic branes in maximal supergravity.

\subsection{$\vec\alpha \leftrightarrow -\vec \alpha$ closure}
First, we show that in maximal supergravity, for every \halfBPS\ vortex (codimension-2 brane) that is light in some duality frame, there is another \halfBPS\ vortex with opposite $\alpha$-vector. 

The argument is as follows. The duality frame in which the vortex is light lies in some principal plane~\cite{Etheredge:2024tok} in the moduli space, consisting of dilaton/radion directions with the axions held fixed. Henceforward, we focus exclusively on \halfBPS\ branes with $\alpha$-vectors parallel to this plane, ignoring, e.g., multicharged towers or multicharged fundamental branes which invariably have axion components to their $\alpha$-vectors. We show that these branes satisfy the following properties\footnote{Property (\ref{step1}) extends to 3d, as shown in Section~\ref{subsec:3dsat}, but this relies on the results of the present argument. Likewise, property (\ref{step1}) extends to $0 \leq p \leq d-2$ (including instantons and vortices), but this also relies on the results of the present argument. The present statements are restricted to ensure that our argument is not circular.}
\begin{enumerate}
	\item Each \halfBPS\ particle tower in $d\geq4$ is U-dual to an $S^1$ KK tower.\label{step1}
	\item The \halfBPS\ $(p-1)$-brane $\alpha$-vectors and the \halfBPS\ $(d-p-3)$-brane $\alpha$-vectors are related by $\vec\alpha \rightarrow - \vec\alpha$ for $1 \leq p \leq d-3$.\label{step2}
	\item The \halfBPS\ codimension-2 brane $\alpha$-vectors have $\vec\alpha \rightarrow - \vec\alpha$ symmetry.\label{step3}
\end{enumerate}

Statement (\ref{step3}) is ultimately what we want to prove, but our argument will depend on statements (\ref{step1}) and (\ref{step2}), so we begin by establishing these. These statements are 10d Type IIB and 11d M-theory, so let us consider either theory compactified on a torus. In particular, after compactifying on $S^1$ to $d\geq 4$, we obtain \halfBPS\ towers / branes of codimension $\geq 3$ in the following ways:
\begin{enumerate}[label=(\alph*)]
	\item The $S^1$ KK tower,
	\item $S^1$ KK monopoles,
	\item Wound and unwound \halfBPS\ fundamental branes.
\end{enumerate}
From the M-theory viewpoint, this gives explicit $S^1$ KK towers, wound M2 and M5 branes, and $S^1$ KK monopoles that wrap the remaining compact directions. Wound M2 and M5 branes are dual to F1 strings and D4 branes from the type IIA perspective. The first is T-dual to a KK tower, whereas the second is T-dual to a D0 tower, which is S-dual to a KK tower. Likewise, all the $S^1$ KK monopoles are S-dual to each other, and in particular they are S-dual to a fully wrapped D6 brane, which is T-dual to a D0 tower, which is S-dual to an $S^1$ KK tower. Thus, point (\ref{step1}) is established.

For point (\ref{step2}), let us refer to branes related in this way as ``Hodge dual''. We proceed by induction: assuming that (\ref{step2}) holds in D-dimensions, we show that it is true in $d=D-1$ dimensions after compactification on $S^1$. First, observe that the $S^1$ KK tower and the $S^1$ KK monopole are Hodge dual. 

Now consider a ``Hodge-dual" pair of $P$ and $(D-P-2)$-dimensional towers / fundamental branes in the $D$-dimensional theory, with $\vec\alpha_{D-P-2} = -\vec \alpha_P$ by assumption. From this pair we obtain four different towers / fundamental branes with $1 \leq p \leq d-3$ in the $d=D-1$ dimensional theory, depending on whether we wrap or do not wrap each object, unless $P = 1$ or $P=D-3$, in which case we only obtain two towers / fundamental branes within the stipulated $1 \leq p \leq  d-3$ range.

Wrapping / not-wrapping the $(P-1)$-brane is Hodge dual to not-wrapping / wrapping the $((D-P-2) - 1)$-brane, respectively. The only non-trivial point is to check the radion coupling~\eqref{eqn:alpharad}, which is $p/\sqrt{(d-1)(d-2)}$ for an unwrapped object resulting in a $(p-1)$-brane, and $-(d-p-2)/\sqrt{(d-1)(d-2)}$ for a wrapped object resulting in a $(p-1)$-brane. Then exchanging wrapped and unwrapped and taking $p\rightarrow d-p-2$ gives $\vec \alpha \rightarrow  - \vec\alpha$ as required. So point (\ref{step2}) is established.

To establish point (\ref{step3}), which we call ``vortex reflection symmetry,'' we again proceed by induction on the number of toroidally-compactified directions. Note that (\ref{step3}) is trivially true in 11d M-theory, and (non-trivially) true in 10d Type IIB string theory.
Now suppose that it holds in $D$-dimensions. We have argued that exotic codimension-2 branes in the $S^1$ compactification to $d=D-1$ dimensions will have $\alpha_\text{radion} < 0$, so there are no exotic codimension-2 branes within the $\alpha_\text{radion} = 0$ plane. Thus, the codimension-2 branes in this plane all come from wrapped codimension-2 branes of $\mathrm{QGT}_D$. These branes have $\vec \alpha \rightarrow -\vec\alpha$ symmetry in $\mathrm{QGT}_D$ by assumption, so they have it in $\mathrm{QGT}_d$ as a result.

The codimension-2 branes with $ \alpha_\text{radion} > 0$ (other than the KK tower itself in the $d=3$ case) come from codimension-3 branes not wrapping the $S^1$. So what we have to show is that for each such codimension-2 brane, there is another codimension-2 exotic brane with the opposite $\alpha$-vector.

\begin{figure}
\center
\includegraphics[width = .7\linewidth]{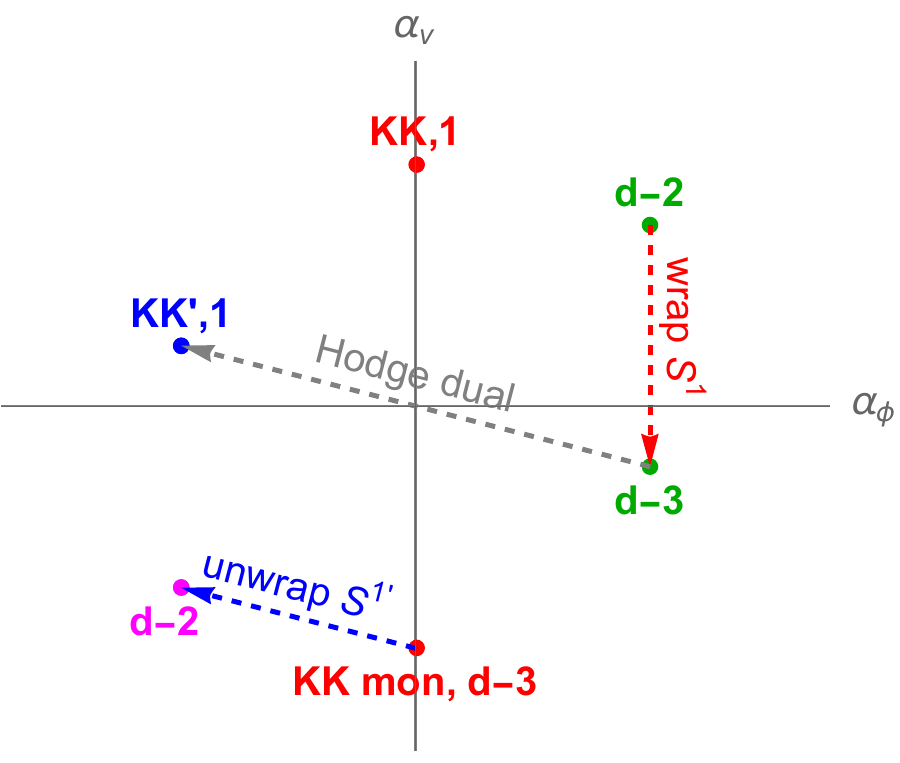}
\caption{The inductive argument for vortex reflection symmetry in $d\ge 4$. The top red point is $\vec\alpha_\text{KK}$ and the top-right green point is $\vec\alpha$ for the non-exotic vortex (coming from an unwrapped codimension-3 brane of $\mathrm{QGT}_D$) whose exotic reflected counterpart we are seeking. The figure depicts the plane containing these two points and the origin. The lower-right green point is the result of wrapping the vortex on $S^1$, with the upper-left blue point being the particle tower Hodge dual to this codimension-3 brane. Transforming to the U-dual frame where this is an $S^{1\prime}$ KK tower (as authorized by property (\ref{step1})), the KK monopole in the original frame (bottom red point) becomes a wrapped brane. Unwrapping this brane from the $S^{1\prime}$ gives the desired vortex (lower-left magenta point).}
\label{f.dgeq4}
\end{figure}

The proof of this fact for $d\geq 4$ is illustrated in Figure \ref{f.dgeq4}, depicting the plane in which the $\alpha$-vectors for the $S^1$ KK modes and the non-exotic vortex with $\alpha_\text{radion} > 0$ both lie. Wrapping the codimension-3 brane of $\mathrm{QGT}_D$ that gives rise to this vortex on the $S^1$ leads to a codimension-3 brane with an $\alpha$-vector within the same plane. The Hodge dual of this brane is a particle tower that also lies within this plane. By property (\ref{step1}), this tower is an KK tower for some U-dual $S^{1\prime}$ decompactification. From the perspective of this U-dual $S^{1\prime}$ compactification, the KK monopole of the original frame has a value of $\alpha_\text{radion}^{\prime}$ that implies that it comes from a wrapped codimension-3 brane. Unwrapping this brane from $S^{1\prime}$ generates another vortex which miraculously has precisely the opposite $\alpha$ vector from the vortex that we started with. Note that this vortex is exotic from the perspective of the $S^1$ compactification, but not from the perspective of the $S^{1\prime}$ compactification. Likewise, the original, non-exotic vortex we started with is exotic from the perspective of the new, $S^{1\prime}$ compactification duality frame.

\begin{figure}
\center
\includegraphics[width = .7\linewidth]{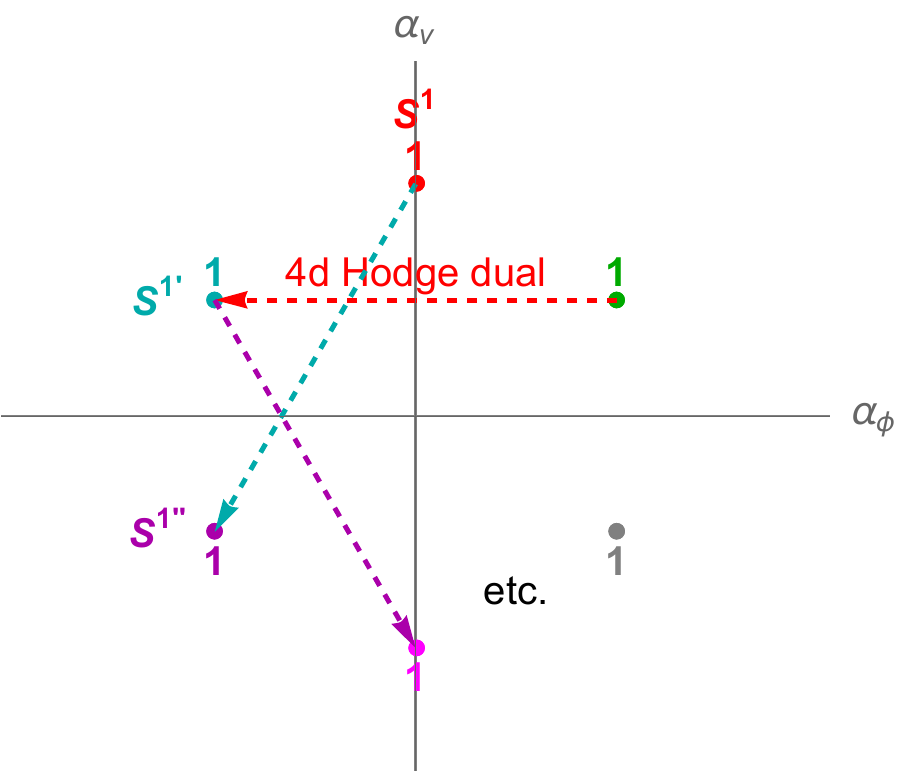}
\caption{The argument for vortex reflection symmetry in 3d. As before, the top red point is $\vec\alpha_\text{KK}$ and the top-right green point is $\vec\alpha$ for the non-exotic particle tower (vortex) arising from a 4d particle tower that we are interested in. Applying 4d Hodge duality to this tower and reducing on $S^1$, we find the cyan tower, which can be viewed as an $S^{1\prime}$ KK tower by property (\ref{step1}). Applying 4d Hodge duality in this $S^{1\prime}$ duality frame to the original KK tower yields the dark purple tower, which is the $\vec\alpha \to -\vec\alpha$ reflection of the non-exotic tower we started with. Note that iterating this argument gives all six $\alpha$-vectors in the figure.
}
\label{f.deq3}
\end{figure}

This completes the argument for $d \ge 4$. The proof for 3d is illustrated in Figure \ref{f.deq3}, depicting the same plane. For each 3d particle tower arising from a 4d particle tower, there is a Hodge dual 4d particle tower whose $\alpha$-vector is reflected across the line parallel to $\alpha_\text{KK}$. By property (\ref{step1}), this can be viewed as an $S^{1\prime}$ KK tower, and we can again consider 4d Hodge duality from the perspective of this $S^{1\prime}$ compactification. In particular, the original KK tower has a Hodge dual with respect to this new compactification which turns out to be a particle tower with opposite $\alpha$-vector to the non-exotic particle tower we started with, establishing vortex reflection symmetry in 3d as well.

In particular, from the perspective of the original, $S^1$ compactification, this establishes that there is an exotic particle tower with $\vec\alpha = - \vec\alpha_\text{KK}$ and also one with $\vec\alpha=-\vec\alpha_\text{part}$ for any non-exotic particle tower $\vec\alpha_{part}$ of interest.
So the $d=3$ argument uses the 4d Hodge duality property instead of the 3d one (which we have not shown yet) and does not require the wrapping/unwrapping steps that appeared in the $d\geq 4$ argument.

Note that the Hodge duality property applies to non-exotic \halfBPS\ codimension-2 branes and instantons by the same argument as given for property (\ref{step2}) above. Combined with the reflection property that establishes that we have \halfBPS\ instantons with the same $\alpha$-vectors as all the \halfBPS\ codimension-2 branes that we have been charting out.

\subsection{Saturation condition in 3d} \label{subsec:3dsat}

For the following argument we need to show that every \halfBPS\ particle tower in 3d is U-dual to an $S^1$ KK tower.
One way to prove this is to show that every \halfBPS\ particle tower / $(p-1)$-brane in maximal SUGRA in $d\leq7$ dimensions is U-dual to a D$(p-1)$ brane. 
(This is also true in 10d Type IIB, but not for M-theory on $T^k$, $k\leq3$.) 

One can prove this working downwards in dimension as before, starting with 10d Type IIB, where this is well known.
If this is true in $D$-dimensions, then after $S^1$ compactification it is true for all the unwrapped branes, and also for the wrapped ones by T-duality.
It is also true for the $S^1$ KK tower, because this is S-dual to a D0 tower.
For the $S^1$ KK monopole, this is T-dual to a type IIA NS5 brane not wrapping the $S^1$, but wrapping the remaining compact directions. Assuming $d<9$, we can T-dualize one of the directions it wraps to get back to type IIB and then S-dualize to a D5 brane and then T-dualize the wrapped directions to get a D$(d-4)$-brane as required.
Finally, for exotic codimension-2 branes, we argued before that in some U-dual frame they come from unwrapped codimension-3 branes of the $D=d+1$ dimensional theory, so the preceding considerations apply. 

However, one final wrinkle to deal with is that there is one \halfBPS\ brane in 9d maximal SUGRA that is not U-dual to a D-brane. It is the KK monopole of type IIB, i.e., an unwrapped NS5 of type IIA or an unwrapped M5 of M-theory.
When we compactify to 8d, with this brane wrapping the $S^1$, then we can T-dualize along this last $S^1$, starting from the type IIA description, giving a wrapped type IIB NS5 brane, which is S-dual to a wrapped D5, etc.
We are still in trouble when this brane does not wrap $S^1$. It is still not U-dual to a D-brane in the 8d theory.
In 8d maximal SUGRA there are two such codimension-2 branes (within the radion-radion-dilaton slice)
They are the ``roots" of the $\mathrm{SL}(2,\mathbb Z)$ part of the $\mathrm{SL}(2,\mathbb Z) \times  \mathrm{SL}(3,\mathbb Z)$ U-duality group,
i.e., their $\alpha$-vectors are parallel to the pericenters of the two triangular faces in the tower polytope.
When we compactify on $S^1$ to 7d, if we are only considering \halfBPS\ branes codimension-2 or higher, then we only need to think about what happens when these branes wrap the additional $S^1$.
And in that case, we can start with the type IIA description where they are NS5 branes, T-dualize on the new $S^1$ to get type IIB NS5 branes, S-dualize to D5 branes, etc.
So indeed in $d\leq 7$, our statement is true for all the \halfBPS\ branes of codimension $\geq 2$. In particular, in 3d the statement becomes that all \halfBPS\ particle towers are U-dual to D0 towers, which are of course $S^1$ KK modes, which is what we wanted to show.

Using this fact, we now argue that the Sharpened DC saturation condition holds in 3d. Since the Sharpened DC convex hull is generated by the \halfBPS\ particle towers, any direction of saturation lies on a facet with vertices that are \halfBPS\ particle towers. So let us focus on a plane including one of these vertices and the direction of saturation.

\begin{figure}
\center
\includegraphics[width = .7\linewidth]{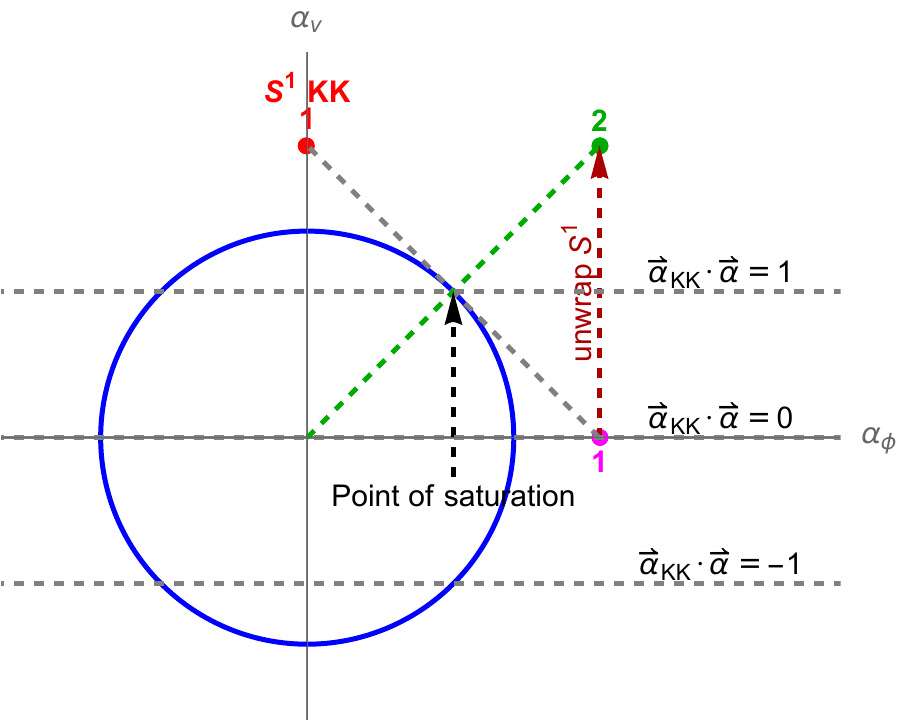}
\caption{Proving the Sharpened DC saturation condition in 3d maximal SUGRA. Here the red vector is the vertex we choose, and the dashed line is the plane of the facet, intersecting the radius 1-ball at the labelled point of saturation.
The possible values of $\vec \alpha \cdot \vec \alpha_\text{KK}$ are the integers from 2 to $-2$, from the following appendix \ref{s.taxonomy}. So they either project down to the point of saturation itself, or we get the magenta dot. One of them has to be the magenta dot, or else we violate the sharpened DC in 3d. Then, since $\vec \alpha \cdot \vec \alpha_{KK} = 0$ for the magenta dot, this is not an exotic tower, and it comes from a 4d string wrapping $S^1$. Considering the string not wrapped, we obtain the required string in 3d, labeled 2 in green, pointing in the direction of saturation.}
\label{f.deq3saturation}
\end{figure}

These vectors are depicted in Figure \ref{f.deq3saturation}. Here the red $\alpha$-vector is the vertex we choose, and the dashed line is the plane of the facet, intersecting the radius 1-ball at the labelled point of saturation. Now we use the fact that the red vector is an $S^1$ KK tower in some U-dual frame, and go to that frame.
The possible values of $\vec \alpha \cdot \vec \alpha_{KK}$ are the integers from $-2$ to $2$ (see also appendix \ref{s.taxonomy}).
This is because exotic branes have $\vec \alpha \cdot \vec \alpha_{KK} < 0$, and wrapped/unwrapped branes give us $0, 1, 2$, so by $\vec\alpha \rightarrow -\vec\alpha$ reflection symmetry we can only have $-2,-1,0,1,2$.
Also, the lengths of the \halfBPS\ alpha vectors are $\sqrt{2}$.
Thus the other vertices on this facet have to project down to one of the dashed lines in this plane.

They either project down to the point of saturation itself, or we get the yellow arrow.
One of them has to be the yellow arrow, or else we violate the sharpened DC in 3d.
Then, since $\vec \alpha \cdot \vec \alpha_{KK} = 0$ for the yellow arrow, this is not an exotic tower, and it comes from a 4d string wrapping $S^1$
Considering the string not wrapped, we obtain the required string in 3d, pointing in the direction of saturation! Thus, the saturation condition is satisfied in 3d maximal supergravity.

\section{Taxonomy rules and dimensional reduction \label{s.taxonomy}}

\subsection{Ansatz}

Suppose that we have a collection of branes $(p_i - 1)$-branes with alpha
vectors $\vec{\alpha}_i$ satisfying the taxonomy rules of \cite{Etheredge:2024tok}:
\begin{equation}
  \vec{\alpha}_i \cdot \vec{\alpha}_j = a_{i j} - \frac{p_i  (d - 2 - p_j)}{d
  - 2},
\end{equation}
for some constants $a_{i j}$. Note that $a_{i j}$ is \emph{not} symmetric:
\begin{equation}
  a_{i j} - a_{j i} = \frac{p_i  (d - 2 - p_j)}{d - 2} - \frac{p_j  (d - 2 -
  p_i)}{d - 2} = p_i - p_j .
\end{equation}
For this reason, it is convenient to also define:
\begin{equation}
  \tilde{a}_{i j} \equiv \min (a_{i j}, a_{j i}) = \left\{ \begin{array}{ccc}
    a_{i j}, &  & p_i \leq p_j,\\
    a_{j i}, &  & p_i > p_j,
  \end{array} \right.
\end{equation}
so that the taxonomy rule can alternately be written as:
\begin{equation}
  \vec{\alpha}_i \cdot \vec{\alpha}_j = \tilde{a}_{i j} - \frac{p_i  (d - 2 -
  p_j)}{d - 2}, \qquad \text{where} \quad p_i \leq p_j .
\end{equation}
Note that $\tilde{a}_{i j}$ can be defined even if there are no moduli, in
which case $a_{i j} = \frac{p_i  (d - 2 - p_j)}{d - 2}$. For instance, in
M-theory $a_{\text{M2}, \text{M2}} = a_{\text{M5}, \text{M5}} = 2$ and $\tilde{a}_{\text{M2}, \text{M5}} = 1$.

We now work out some $p_i$, $p_j$-independent constraints on $\tilde{a}_{i
j}$, assuming that $0 \leq p_i, p_j \leq d - 2$. Specifically, for
$p_i \leq p_j$ one finds
\begin{equation}
  | \vec{\alpha}_i - \vec{\alpha}_j |^2 = a_{i i} + a_{j j} - 2 \tilde{a}_{i
  j} - \frac{(p_j - p_i)  (d - 2 - (p_j - p_i))}{d - 2} .
\end{equation}
The last term is non-positive, so we conclude that
\begin{equation}
  \tilde{a}_{i j} \leq \frac{a_{i i} + a_{j j}}{2},
\end{equation}
with saturation only possible when $\vec{\alpha}_i = \vec{\alpha}_j$ and $p_i
= p_j$ or $p_i = 0$, $p_j = d - 2$. Similarly, one can place a lower bound on
$\tilde{a}_{i j}$ by noting that decreasing it will eventually violate the
triangle inequality:
\begin{equation}
  \vec{\alpha}_i^2  \vec{\alpha}_j^2 - (\vec{\alpha}_i \cdot \vec{\alpha}_j)^2
  \geq 0 .
\end{equation}
Thus,
\begin{equation}
  \tilde{a}_{i j}^{\min} = \frac{p_i  (d - 2 - p_j)}{d - 2} + (\vec{\alpha}_i
  \cdot \vec{\alpha}_j)^{\min} = \frac{p_i  (d - 2 - p_j)}{d - 2} - |
  \vec{\alpha}_i |  | \vec{\alpha}_j | .
\end{equation}
Let us hold $a_{i i}$, $a_{j j}$ fixed and vary $p_i, p_j$ subject to the
constraint $p_i \leq p_j$. In the region $p_j \leq \frac{d - 2}{2}$,
decreasing $p_i$ will increase $| \vec{\alpha}_i |$ since necessarily $p_i
\leq \frac{d - 2}{2}$, so that $\tilde{a}_{i j}^{\min}$ decreases
monotonically with $p_i$. Thus, the minimum possible value with have $p_i =
0$, where
\begin{equation}
  \tilde{a}_{i j}^{\min} = - \sqrt{a_{i i}}  \sqrt{a_{j j} - \frac{p_j  (d - 2
  - p_j)}{d - 2}} .
\end{equation}
We then set $p_j = d - 2$ to minimize this expression, which gives
$\tilde{a}_{i j}^{\min} = - \sqrt{a_{i i} a_{j j}}$. Likewise, in the region
$p_j \geq \frac{d - 2}{2}$, $\tilde{a}_{i j}^{\min}$ decreases
monotonically with \emph{increasing} $p_j$, so we set $p_j = d - 2$,
giving:
\begin{equation}
  \tilde{a}_{i j}^{\min} = - \sqrt{a_{i i} - \frac{p_i  (d - 2 - p_i)}{d - 2}}
  \sqrt{a_{j j}} .
\end{equation}
Now we can either set $p_i = 0$ or $p_i = d - 2$ to minimize this expression,
again giving $(a_{i j}^{\leq})^{\min} = - \sqrt{a_{i i} a_{j j}}$. Thus,
we conclude that:
\begin{equation}
  - \sqrt{a_{i i} a_{j j}} \leq \tilde{a}_{i j} \leq \frac{a_{i i} +
  a_{j j}}{2},
\end{equation}
where the upper bound corresponds to $\vec{\alpha}_i = \vec{\alpha}_j$ and
either $p_i = p_j$ or $p_i = 0$, $p_j = d - 2$, whereas the lower bound
requires $p_i, p_j = 0, d - 2$ and $\vec{\alpha}_i \propto - \vec{\alpha}_j$.

It is also interesting to note that if we restrict to $1 \leq p \leq
d - 3$ then by a similar approach we get a stronger lower bound:
\begin{equation}
  \tilde{a}_{i j} \geq \frac{1}{d - 2} - \sqrt{a_{i i} - \frac{d - 3}{d -
  2}}  \sqrt{a_{j j} - \frac{d - 3}{d - 2}},
\end{equation}
which can only be saturated for $p_i = 1$, $p_j = d - 3$ with $\vec{\alpha}_i
\propto - \vec{\alpha}_j$.

Before moving on we note that the value $a_{i i} = 2$ occurs naturally in
maximal SUGRA, which leads to the bounds $- 2 \leq \tilde{a}_{i j}
\leq 2$ for $0 \leq p_i, p_j \leq d - 2$ and also $\tilde{a}_{i
j} \geq - 1$ for $1 \leq p_i, p_j \leq d - 3$. The taxonomy
rules previously discussed in the Brane DC draft correspond to $\tilde{a}_{i
j} = 1$ for $i \neq j$.

\subsection{Dimensional reduction}

Now let us see how $a_{i j}$ changes under compactification. First, consider
reduction from $D = d + n$ to $d$ dimensions on an $n$-dimensional Ricci-flat
manifold $\mathcal{M}_n$, ignoring for the time being shape moduli, axions and
brane moduli and the KK modes of massive particles. Wrapping a brane with $P =
p + k$ spacetime dimensions on a $k$ cycle thereof, we find:
\begin{equation}
  \vec{\alpha}^{(d)} = \left( \vec{\alpha}^{(D)}, \frac{p n - k (d -
  2)}{\sqrt{n (n + d - 2) (d - 2)}} \right) .
\end{equation}
Thus,
\begin{align}
  \vec{\alpha}^{(d)}_i \cdot \vec{\alpha}^{(d)}_j &= \vec{\alpha}^{(D)}_i
  \cdot \vec{\alpha}_j^{(D)} + \frac{[p_i n - k_i  (d - 2)] [p_j n - k_j  (d -
  2)]}{n (n + d - 2) (d - 2)} \nonumber\\
  &= a_{i j}^{(D)} - \frac{P_i  (D - 2 - P_j)}{D - 2} + \frac{[p_i n - k_i 
  (d - 2)] [p_j n - k_j  (d - 2)]}{n (n + d - 2) (d - 2)} \nonumber\\
  &= a_{i j}^{(D)} - \frac{k_i  (n - k_j)}{n} + \frac{p_i  (d - 2 - p_j)}{d -
  2}, 
\end{align}
where we substitute $P_i = p_i + k_i$, $D = d + n$ and simplify. Thus,
\begin{equation}
  a_{i j}^{(d)} = a_{i j}^{(D)} - \frac{k_i  (n - k_j)}{n} .
\end{equation}
As a cross-check, this implies that:
\begin{equation}
  a_{i j}^{(d)} - a_{j i}^{(d)} = a_{i j}^{(D)} - a_{j i}^{(D)} - k_i + k_j =
  (P_i - k_i) - (P_j - k_j) = p_i - p_j,
\end{equation}
as required. Now consider the KK tower:
\begin{equation}
  \vec{\alpha}_{\text{KK}} = \left( 0, \sqrt{\frac{n + d - 2}{n (d - 2)}}
  \right) .
\end{equation}
Thus,
\begin{equation}
  \vec{\alpha}_{\text{KK}} \cdot \vec{\alpha}^{(d)}_j = \frac{p_j n - k_j 
  (d - 2)}{n (d - 2)} = \frac{n - k_j}{n} - \frac{d - 2 - p_j}{d - 2},
\end{equation}
and so
\begin{equation}
  a_{\text{KK}, j}^{(d)} = \frac{n - k_j}{n} .
\end{equation}
Likewise, when the geometry of $\mathcal{M}_n$ is favorable there is a KK
monopole with
\begin{equation}
  \vec{\alpha}_{\text{mon}} = \left( 0, - \sqrt{\frac{n + d - 2}{n (d - 2)}}
  \right) .
\end{equation}
Thus,
\begin{equation}
  \vec{\alpha}^{(d)}_i \cdot \vec{\alpha}_{\text{mon}} = - \frac{p_i n - k_i 
  (d - 2)}{n (d - 2)} = \frac{k_i}{n} - \frac{p_i}{d - 2},
\end{equation}
and so
\begin{equation}
  a_{i, \text{mon}}^{(d)} = \frac{k_i}{n} .
\end{equation}
Finally, from the pure KK sector, we find:
\begin{align}
  | \vec{\alpha}_{\text{KK}} |^2 = | \vec{\alpha}_{\text{mon}} |^2 &=
  \frac{n + d - 2}{n (d - 2)} = \frac{n + 1}{n} - \frac{d - 3}{d - 2} &
  &\Rightarrow & a_{\text{KK}, \text{KK}}^{(d)} &= a_{\text{mon},
  \text{mon}}^{(d)} = \frac{n + 1}{n}, \nonumber\\
  \vec{\alpha}_{\text{KK}} \cdot \vec{\alpha}_{\text{mon}} &= - \frac{n +
  d - 2}{n (d - 2)} = - \frac{1}{n} - \frac{1}{d - 2} & &\Rightarrow
  & a_{\text{KK}, \text{mon}}^{(d)} &= - \frac{1}{n} . 
\end{align}
Let us now specialize to $\mathcal{M}_n = S^1$ and express these results in
terms of $\tilde{a}_{i j}$. Firstly, in the KK sector we have:
\begin{equation}
  a_{\text{KK}, \text{KK}}^{(d)} = a_{\text{mon}, \text{mon}}^{(d)} =
  1, \qquad a_{\text{KK}, \text{mon}}^{(d)} = - 1 .
\end{equation}
In $d \geq 4$ the KK monopole has dimension at least as large as the KK
mode, and this gives:
\begin{equation}
  a_{\text{KK}, \text{KK}} = a_{\text{mon}, \text{mon}} = 1, \qquad
  \tilde{a}_{\text{KK}, \text{mon}} = - 1 .
\end{equation}
In $d = 3$, however, the order reverses, and we get instead that:
\begin{equation}
  a_{\text{KK}, \text{KK}} = a_{\text{mon}, \text{mon}} = 1, \qquad
  \tilde{a}_{\text{mon}, \text{KK}} = - 2 .
\end{equation}
Now consider the wrapped-brane sector. We find:
\begin{equation}
  a_{i j}^{(d)} = a_{i j}^{(D)} - k_i  (1 - k_j), \qquad k_i, k_j = 0, 1.
\end{equation}
Since there is only one direction to wrap or not wrap, it is not possible to
have $p_i < p_j$ together with $P_i > P_j$. Thus, by relabeling as needed, we
can take $p_i \leq p_j$ and $P_i \leq P_j$ WLOG, so that
\begin{equation}
  \tilde{a}_{i j}^{(d)} = \tilde{a}_{i j}^{(D)} - k_i  (1 - k_j),
\end{equation}
and $\tilde{a}_{i j}$ is preserved \emph{except} in the case where the
brane with initially equal or lower spacetime dimension wraps the circle and
the one initially with equal or higher spacetime dimension does not, in which
case $\tilde{a}_{i j}$ decreases by 1.

Now consider dot products between the two sectors. We have:
\begin{equation}
  a_{\text{KK}, j}^{(d)} = 1 - k_j .
\end{equation}
Thus, if the brane is \emph{not} an instanton in the reduced theory, then
we get $\tilde{a}_{\text{KK}, j} = 1$ for an unwrapped brane and
$\tilde{a}_{\text{KK}, j} = 0$ for a wrapped brane, whereas if the brane
is an instanton we get $\tilde{a}_{\text{KK}, j} = 0$ for an unwrapped
brane (descending from another instanton) and $\tilde{a}_{\text{KK}, j} =
- 1$ for a wrapped brane (descending from a particle).

Notice that this implies that codimension-2 branes can have either
$\tilde{a}_{\text{KK}, j} = 1$ or $\tilde{a}_{\text{KK}, j} = 0$ but
not $\tilde{a}_{\text{KK}, j} = - 1$. However, U-duality implies that
codimension-2 branes come in $\vec{\alpha}, - \vec{\alpha}$ pairs, which
implies that $\tilde{a}_{\text{KK}, j} = - 1$ should be realized by exotic
codimension-2 branes in the context of maximal SUGRA. Indeed, if we start
with a complete set of codimension-2 branes before the circle reduction then
these already come in $\vec{\alpha}, - \vec{\alpha}$ pairs so we see that none
of the exotic branes can have $\tilde{a}_{\text{KK}, j} = 0$, and they
certainly cannot have $\tilde{a}_{\text{KK}, j} = 1$ as these would
survive the as finite-tension codimension-2 branes in the decompactification
limit. Thus, the exotic codimension-2 branes are precisely those
codimension-2 branes with $\tilde{a}_{\text{KK}, j} = - 1$.

Likewise, the above implies that instantons have either $\tilde{a}_{\text{KK}, j} = 0$ or $\tilde{a}_{\text{KK}, j} = - 1$ but not
$\tilde{a}_{\text{KK}, j} = 1$. By analogous reasoning, the latter must
arise from exotic instantons in the maximal SUGRA case. Note that the exotic
instantons are those whose action \emph{decreases} in the
decompactification limit, whereas conventional instantons all have a constant
or increasing action in this limit.

The above assumes $d > 3$. In $d = 3$, the $\text{KK}$ modes themselves
are codimension-2, with $\tilde{a}_{\text{KK}, \text{KK}} = 2$, so
there should be one more codimension-2 exotic brane with $\tilde{a}_{\text{KK}, j} = - 2$. Likewise, the KK monopole is an instanton with
$\tilde{a}_{\text{mon}, \text{KK}} = - 2$ so there should be another
exotic instanton with $\tilde{a}_{i, \text{KK}} = 2$. This conforms to the
pattern that exotic instantons have an action that \emph{decreases} in the
decompactification limit, whereas exotic codimension-2 branes have a tension
that \emph{increases} in this limit, when measured in $d$-dimensional
Planck units.

Finally, consider dot products of branes and KK monopoles. We have:
\begin{equation}
  a_{i, \text{mon}}^{(d)} = k_i .
\end{equation}
Now the answer depends on the brane codimension and whether it is wrapped. If
the brane is codimension-3 or higher after reduction then we get
$\tilde{a}_{i, \text{mon}} = 1$ for a wrapped brane and $\tilde{a}_{i,
\text{mon}} = 0$ for an unwrapped brane. If, however, the brane is
codimension-2 after reduction then we get $\tilde{a}_{\text{mon}, i} = 0$
for a wrapped brane and $\tilde{a}_{\text{mon}, i} = - 1$ for an unwrapped
brane. Similar comments follow about exotic branes as above.

The preceding discussion implies that $\tilde{a}_{i j} \in \mathbb{Z}$ for
branes whose $\vec{\alpha}$ vectors lie in the same principal plane in
M-theory on $T^n$, where $- 2 \leq \tilde{a}_{i j} \leq 2$ and
saturation of these bounds has the consequences described in the previous
section.

\bibliographystyle{JHEP}
\bibliography{ref}
\end{document}